\documentclass{article}
\usepackage[utf8]{inputenc}
\usepackage{authblk}
\usepackage{amssymb}
\usepackage{amsmath} 
\usepackage[round]{natbib}
\usepackage{graphicx}
\usepackage[linesnumbered,ruled,vlined]{algorithm2e}
\usepackage{bm}
\usepackage[colorinlistoftodos]{todonotes}
\usepackage{comment}
\usepackage[colorlinks=true, allcolors=blue]{hyperref}
\usepackage{lineno}
\usepackage{ulem}
\usepackage{graphicx}
\newcommand*\patchAmsMathEnvironmentForLineno[1]{%
  \expandafter\let\csname old#1\expandafter\endcsname\csname #1\endcsname
  \expandafter\let\csname oldend#1\expandafter\endcsname\csname end#1\endcsname
  \renewenvironment{#1}%
     {\linenomath\csname old#1\endcsname}%
     {\csname oldend#1\endcsname\endlinenomath}}%
\newcommand*\patchBothAmsMathEnvironmentsForLineno[1]{%
  \patchAmsMathEnvironmentForLineno{#1}%
  \patchAmsMathEnvironmentForLineno{#1*}}%
\AtBeginDocument{%
\patchBothAmsMathEnvironmentsForLineno{equation}%
\patchBothAmsMathEnvironmentsForLineno{align}%
\patchBothAmsMathEnvironmentsForLineno{flalign}%
\patchBothAmsMathEnvironmentsForLineno{alignat}%
\patchBothAmsMathEnvironmentsForLineno{gather}%
\patchBothAmsMathEnvironmentsForLineno{multline}%
}
\usepackage{color}
\usepackage{verbatim}
\usepackage{comment,xcolor}

\newcommand{\shota}[1]{\textbf{\textcolor{green}{[SS2: #1]}}}
\newcommand{\RNum}[1]{\uppercase\expandafter{\romannumeral #1\relax}}
\newcommand{\Add}[1]{\textcolor{red}{#1}} 
\newcommand{\Erase}[1]{\textcolor{red}{\sout{\textcolor{black}{#1}}}} 
\usepackage[a4paper,top=2cm,bottom=2cm,left=2cm,right=2cm,marginparwidth=2cm]{geometry}
\usepackage{setspace}
\title{Exclusion of the fittest predicts microbial community diversity in fluctuating environments}

\author[1]{Shota Shibasaki}
\author[2$\dagger$]{Mauro Mobilia}
\author[1$\dagger$]{Sara Mitri}
\affil[1]{\small{Department of Fundamental Microbiology, University of Lausanne, Switzerland}}
\affil[2]{Department of Applied Mathematics, School of Mathematics, University of Leeds, United Kingdom}
\affil[$\dagger$]{these authors contributed equally}
\date{}
\begin{document}

\maketitle
E-mail addresses of all authors. \par
\quad SS: shota.shibasaki@unil.ch \par
\quad MM: M.Mobilia@leeds.ac.uk\par
\quad SM: sara.mitri@unil.ch
\par
Short running title: Exclusion of the fittest and diversity
\par
\par
Type of article: article
\par 
Numbers of words: 187 words in abstract, and 7159 words in main text.
\par
Number of references: 78.
\par
Numbers of figures: 6, tables: 1, text box 0.
\par
Corresponding authors: 
\begin{description}
\item[SS]Quartier UNIL-Sorge, Batiment Biophore, Office 2917.3 , CH-1015 Lausanne. Phone : + 41 21 692 56 00. E-mail: shota.shibasaki@unil.ch.
\item[SM]Quartier UNIL-Sorge,
Batiment Biophore, Office 2414, CH-1015 Lausanne. Phone : +41 21 692 56 12. E-mail: sara.mitri@unil.ch.
\end{description}
\par

\newpage
\begin{abstract}
Microorganisms live in environments that inevitably fluctuate between mild and harsh conditions. As harsh conditions may cause extinctions, the rate at which fluctuations occur can shape microbial communities and their diversity, but we still lack an intuition on how. Here, we build a mathematical model describing two microbial species living in an environment where substrate supplies randomly switch between abundant and scarce. We then vary the rate of switching as well as different properties of the interacting species, and measure the probability of the weaker species driving the stronger one extinct.
We find that this probability increases with the strength of demographic noise under harsh conditions and peaks at either low, high, or intermediate switching rates depending on both species' ability to withstand the harsh environment. This complex relationship shows why finding patterns between environmental fluctuations and diversity has historically been difficult. In parameter ranges where the fittest species was most likely to be excluded, however, the beta diversity in larger communities also peaked. 
In sum, how environmental fluctuations affect interactions between a few species pairs predicts their effect on the beta diversity of the whole community.

\end{abstract}
\newpage
\section{Introduction}\label{sec:introduction}

Natural environments are not static: temperature, pH, or availability of resources change over time. Many studies in microbiology, ecology and evolution have focused on responses to fluctuations in resource abundance in the regime of feast and famine periods \citep{Hengge-Aronis1993,Vasi1994,Srinivasan1998,Xavier2005, Merritt2018, Himeoka2019}. These models capture the dynamics within many natural ecosystems. For example, the gut microbiome of a host is exposed to fluctuating resources that depend on its host's feeding rhythm, which may affect microbiota diversity \citep{Cignarella2018, Li2017a, Thaiss2014a}. In addition to their magnitude, environmental fluctuations can also differ in their time scales -- for the gut microbiota, a host's feeding rhythm may vary from hourly to daily, or even monthly if feeding depends on seasonal changes \citep{Davenport2014,Smits2017} -- or in the type of substrates taken up, which can be nutritious or harmful for the microbiota. 
\par
How environmental fluctuations (EFs) affect species diversity has been a highly contested topic in ecology. Here, EFs refer to changes that are not caused by the organisms themselves, but nevertheless affect their dynamics (e.g., abiotic resource supplies). The intermediate disturbance hypothesis argues that intermediate intensity and frequency of disturbances  maximize species diversity \citep{Connell1978,  Grime1973a} because highly competitive species dominate at a low level of disturbance, while only species that have adapted to the disturbance dominate at high disturbance \citep{Grime1977}. This hypothesis is controversial \citep{Fox2013} and other relationships between disturbance and species diversity have been reported both empirically and theoretically \citep{Mackey2001, Miller2011}. 
\par
Another framework that is used to predict species diversity under EFs is the modern coexistence theory \citep{Chesson1994}, which explains the maintenance of diversity through species' differing responses to and preferences for environmental conditions, which can vary over spatial and/or temporal scales through fluctuations \citep{Amarasekare2019, Chesson2000a, Chesson2000b, Letten2018b,Barabas2018,Ellner2019}. 
The modern coexistence theory divides environmental factors into those that are independent of species abundances (e.g., temperature) and those that depend on them (e.g., amounts of resources). The latter environmental factors  mediate the sign and/or magnitude of interspecific interactions \citep{Hoek2016, Piccardi2019, Zuniga2019}, and whether species tend to cooperate or compete can, in turn, drive community diversity and stability \citep{Mougi2012,Coyte2015, Marsland2019, Butler2018, Butler2019b}. Microbial communities often experience extreme environmental fluctuations that can alter interactions between species and hence affect species diversity \citep{Rodriguez-Verdugo2019, Nguyen2020b}. 
\par
Another important factor potentially influencing the outcome of interactions between species is demographic noise (DN) arising from randomness in birth and death events in finite populations. DN is negligible in large populations but strong in small ones, where it can lead to species extinction or fixation, and affect community composition \citep{Roughgarden1979, Ewens2004mpg}. As EFs affect population sizes, they modulate the strength of demographic noise, leading to a coupling of EFs and DN. This interdependence has been understudied until recently, despite its potentially important consequences on eco-evolutionary dynamics \citep{Wienand2017a, Wienand2018, West2019a, Taitelbaum2020}. 

To understand how the interplay between EFs and DN affect community diversity, we set up a stochastic model of multiple species subject to a varying supply of nutrients and/or toxins. Our model then allows us to ask: How do EFs, coupled to DN, affect species interactions and diversity? 
\par
We include toxins in our model as they are important in natural communities, but often missed in similar models. Toxins that typically come to mind are pesticides or antibiotics \citep{Perez2005,Xu2011}, but in principle they can be anything that inhibits microbial growth compared to some optimal condition. 
 For example, oxygen is harmful to anaerobic microbes \citep{Guittar2021} and bile acids are toxic to microbes in the human gastrointestinal tract \citep{Ruiz2013, Molinero2019}. These toxins can, however, be degraded by microbes. In the example of bile acids, \textit{Lactobacillus} and \textit{Bifidobacterium} strains degrade them by producing bile-salt hydolases and other extracellular enzymes \citep{Ruiz2013,Molinero2019}.
Such degradation of toxic compound by microbes is expected to be quite common, as microbes will be selected to become resistant to toxins by diminishing environmental toxicity. 
In our model then, toxins contribute to environmental harshness and reduce population sizes, which can modulate the strength of DN. We also expect toxins to affect inter-species interactions, as species that absorb or degrade them can potentially facilitate the growth of others \citep{Hoek2016, Piccardi2019, Zuniga2019}. 
\par
The next section describes our theoretical model, explaining how DN and EFs are implemented and how species interactions and diversity are measured (section \ref{sec:model}). We first use it to explore how often one single species goes extinct due to environmental switching coupled with DN (section \ref{sec:mono_extinction}). We then add a second slower-growing species, still focusing on the behavior of the first, and ask how its extinction probability is affected by the slower-grower, and how different properties of the two species change the effects of the switching rate (section \ref{sec:species-interaction}). We find that sensitivity to toxins increases the strength of DN because it decreases species abundances. 
At toxin sensitivities that are high enough to increase DN but low enough that either species is likely to survive, the slower-growing species can outcompete the fast-growing
 one (we call this phenomenon \textit{exclusion of the fittest}, sections \ref{sec:CE-change} and \ref{sec:scenarios}), a result that is a direct consequence of coupling DN and EFs. Finally, we expand our model to larger communities and show that our first analysis of the exclusion of the fittest species predicts beta diversity patterns in larger communities (section \ref{sec:beta-div}): community composition is the most diverse when strong species are most likely to be excluded. Section \ref{sec:discussion} discusses the importance of coupling EFs and DN on species interactions and diversity. Finally, we compare our results with the intermediate disturbance hypothesis and the modern coexistence theory. Technical and computational details are given in a series of Appendices.
\section{Model \& Methods}\label{sec:model}
In order to investigate the interdependence of EFs and DN on species interactions and diversity, we study an idealized chemostat model that 
combines DN and environmental switching (Fig. \ref{fig:cartoon}). The chemostat is meant to represent natural ecosystems subject to in- and outflow, such as a gut or a river. Within the chemostat, we consider  a well-mixed population consisting of $N$ species of bacteria, and $N/2$ types of resources  as well as $N/2$ types of toxins (without loss of generality, $N$ is assumed to be even). Our system can exhibit different levels of species richness at equilibrium depending on parameter values: only one species may persist as species interact with all resources and toxins, or all species may coexist as the environment has as many limiting factors (resources and toxins) as the number of species. The special case of a model with a single resource
and a single toxin in a mono-culture is considered for completeness in Section \ref{sec:mono_extinction}.  Although many models ignore the role of toxins, they play two important roles in our model: first, sensitivities to toxins changes the strength of DN because population sizes of sensitive species are low. Second, toxins can enable a pair of species to facilitate each other \citep{Piccardi2019} because toxin absorbance or degradation decreases the death rate of other species. In this manuscript, we consider toxins that are not released from cells when they die, e.g., antibiotics that inhibit DNA or RNA synthesis \citep{Cangelosi2014}, although the release of toxins from dead cells could be introduced with a small modification of our model (see \citet{Huang2013} for example). The absence of toxins is similar to having a low toxin sensitivity (Fig. \ref{fig:notoxin}).
\par
In our model, communities follow a continuous-time multi-variate birth-and-death process (see,  e.g. \citet{Novozhilov2006a,Allen2010}) in  a time-fluctuating binary environment (see, {\it e.g.}, \citet{Wienand2017a,Wienand2018,West2019a,Taitelbaum2020}).
More specifically, we consider that at time $\sigma$ the community consists of an amount $r_i(\sigma)$ of resources of type $i$ ($i=1,\ldots, N/2$), an amount $t_j(\sigma)$ of toxin of type $j$ ($j=1,\ldots, N/2$), and an abundance $s_k(\sigma)$ of microbial species $k$ ($k=1,\ldots, N$).  Here, resources are assumed to be nutrients for all species, allowing them to grow at different rates, while the toxins kill all species at different rates when species have different toxin sensitivities. 
 hence the species extinction probability.
\par
In a static environment (no EFs), this chemostat model evolves according to the   birth-and-death process defined by the following ``birth'' and ``death'' reactions:
\begin{subequations}
\label{eq:reactions}
\begin{align}
r_i &\xrightarrow{\tau_{r_i}^+} r_i+1,    \label{eq:transition-x+}\\
r_i &\xrightarrow{\tau_{r_i}^-} r_i-1,    \label{eq:transition-x-} \\
t_j &\xrightarrow{\tau_{t_j}^+} t_j+1,     \label{eq:transition-z+}\\
t_j &\xrightarrow{\tau_{t_j}^-} t_j-1,   \label{eq:transition-z-}\\
s_k &\xrightarrow{\tau_{s_k}^+} s_k+1,    \label{eq:transition-y+} \\
s_k &\xrightarrow{\tau_{s_k}^-} s_k-1,     \label{eq:transition-y-}
\end{align}
\end{subequations}
occurring with transition rates:
\begin{subequations}

\begin{align}
    \tau_{r_i}^+&=\alpha R_{i},\label{eq:transitionA} \\
    \tau_{r_i}^-&=\sum_{k=1}^N\frac{\mu_{ik}}{Y^r_{ik}}\frac{r_i}{r_i+K^r_{ik}}s_k+\alpha r_i,\\
    \tau_{t_j}^+&=\alpha T_j,\\
    \tau_{t_j}^-&=\sum_{k=1}^N\frac{\delta_{jk}}{Y^t_{jk}}\frac{t_j}{t_j+K^t_{jk}}s_k+\alpha t_j,\label{eq:transitionD}\\
    \tau_{s_k}^+&=\sum_{i=1}^{N/2}\mu_{ik}\frac{r_i}{r_i+K^r_{ik}}s_k,\label{eq:transitionE}\\
    \tau_{s_k}^-&=\sum_{j=1}^{N/2}\delta_{jk}\frac{t_j}{t_j+K^t_{jk}}s_k+\alpha s_k,\label{eq:transitionF}
\end{align}
\end{subequations}
where $\alpha$ is the dilution rate of the chemostat, $\xi=\pm1$ (see below) represents changing environmental conditions in terms of in-flowing resource and/or toxin amount. $R_i(\xi)$ ($T_j(\xi)$) is resource $i$'s (toxin $j$'s) supply under the environmental condition $\xi$, $Y^r_{ik}$ ($Y^t_{jk}$) is species $k$'s biomass yield for resource $i$ (toxin $j$), $\mu_{ik}$ is the maximum growth rate of species $k$ by resource $i$, $\delta_{jk}$ is the maximum death rate of species $k$ by toxin $j$ (species $k$'s sensitivity to toxin $j$), and $K^r_{ik}$ ($K^t_{jk}$) is the amount of resource $i$ (toxin $j$) that gives the half-maximum growth (death) rate for species $k$ (see also Table \ref{tab:prameter}). These transition rates 
hence reflect that (i) the amounts of resources and toxins increase depending on the product of their in-flow concentrations and  the dilution rate, (ii) the amounts of resources and toxins decrease with the dilution rate and with the consumption/absorption by species, (iii) the  growth and death rates depend in Monod functional forms on the amounts of resources and toxins, respectively, (iv) the dilution rate $\alpha$ sets the time scale at which the environment attains the state of abundance or scarcity (after a time $\sim 1/\alpha$), see below and \ref{sec:OnlyEF} and (v) all effects are additive. 
When $R_i$ and $T_j$ are constant, the environment is static and the birth-and-death process defined by Eqs \eqref{eq:transition-x+} -- \eqref{eq:transitionF} naturally accounts for the DN arising in the population, which is the sole source of noise.
 \par
We model EFs by considering a
time-fluctuating binary environment resulting in $R_i$ and/or $T_j$
to be {\it dichotomous random variables}, i.e., $R_i=R_i(\xi(\sigma))$ and/or $T_j(\xi(\sigma))$, where $\xi(\sigma)=\pm 1$ represents the binary state of the environment ($\xi=1$ represents a mild environment while $\xi=-1$ represent a harsh environment).
We hence assume that $R_i$ and/or $T_j$ switch between two values at a rate $\nu$, according to a time-continuous colored dichotomous Markov noise (random telegraph noise)  \citep{Bena2006,Horsthemke2006}
\begin{align}
    \xi &\xrightarrow{\nu} -\xi. \label{eq:transition-switch}
\end{align}

We call $\nu$ a ``switching rate'' because we implement a symmetrically switching environment as a simple example of EFs (but see \ref{sec:OnlyEF} for a further discussion of this choice and \ref{sec:variousflucutuation} for other forms of EFs). Although we use arbitrary units in the model, the unit of the switching rate applies to all events. For example, if the unit for the dilution rate is per hour, $\nu=10^0$ means that the environment switches every hour on average. We investigate three environmental switching scenarios, where either or both resource and toxin supplies fluctuate over time, see Table \ref{tab:resoure-toxin-switch}. In the main text, we focus on the scenario where only resource supplies switch between abundant and scarce supplies ($R^+_i$ and $R^-_i$, respectively, such that $R^+_i>R^-_i>0$) while the amounts of toxin supplies remain constant over time ($\langle T_j\rangle \equiv (T_j^+ +T_j^-)/2$); see \ref{sec:alternative} for the remaining scenarios. The initial resource supply in the main text is $R_i\left(\xi\left(0\right)\right) = R_i^+$ with probability $0.5$ and otherwise $R_i\left(\xi\left(0\right)\right) = R_i^-$.
\par
We assume that $\xi$ switches symmetrically between the states $\pm 1$ (see \citet{Taitelbaum2020} and \ref{sec:variousflucutuation} for the cases of asymmetric switching). In all our simulations, $\xi$ is stationary and thus $\xi$ has zero mean, $\langle \xi(\sigma)\rangle=0$, autocorrelation $\langle \xi(\sigma)\xi(\sigma')\rangle={\rm exp}\left(-2\nu|\sigma-\sigma'|\right)$, where $\langle \cdot\rangle$ denotes the ensemble average over the  environmental noise, and finite correlation time $1/(2\nu)$. A great advantage of dichotomous noise  is its simplicity: $\xi$ is bounded ($R_i$ and $T_j$ are always well defined) and straightforward to simulate. This choice allows us to model suddenly changing environmental conditions, which reflect situations in microbial life such as exposure to resource or toxin oscillations that can be reproduced in lab-controlled experiments \citep{Sunya2013}. Other forms of EFs are also possible, e.g. $\xi$ could be a Gaussian random variable, but then $R_i$ and $T_j$ would  be unbounded and vary continuously and could take unrealistic values. Modeling EFs with Eq \eqref{eq:transition-switch} is arguably the simplest biologically-motivated choice  to couple fluctuations in resource/toxin supplies with demographic noise, and allows us to investigate questions that are not specific to dichotomous noise, see \ref{sec:OnlyEF}. Interestingly, the analysis of the long-term dynamics of the  
 two-species, one-resource-one-toxin model under symmetric dichotomous noise in \ref{sec:OnlyEF} reveals that this simple form of
 environmental variability leads to distributions of the total population size $n$ (i.e., the sum of species, resources and toxins abundances $n\equiv \sum_k s_k +\sum_i r_i+\sum_j t_j$) that varies greatly with the rate of switching relative to the dilution rate: when $\nu/\alpha \gg 1$ (fast switching),
the total population size $n$ is unimodal;  when  $\nu/\alpha \ll 1$ (slow switching), 
$n$ is 
 bimodal and fluctuates between two very different values; intermediate scenarios 
 interpolating between unimodal and bimodal distribution of $n$
 arise when  $\nu/\alpha \sim 1$ (intermediate switching), see \citet{Wienand2017a,Wienand2018,West2019a,Taitelbaum2020} and Fig. \ref{fig:PDMP}. This results in an explicit coupling of DN to EFs in multi-species communities, via the modulation of the DN intensity by $\nu/\alpha$. This is a distinctive feature of our model comparing previous studies. For example, some models \citep{GilesLeigh1981, Engen1996, Kalyuzhny2015} do not couple DN and EFs, while other models \citep{Engen1996,Kamenev2008, Chisholm2014,Fung2015} study the coupling of DN and EFs in single species scenarios. See \ref{sec:OnlyEF} for more detailed discussion and analysis.
\par
The master equation for this model is defined by combining the dynamics of the amounts of resources and toxins, abundances of species, with the environmental switching (Eqs \eqref{eq:transition-x+} - \eqref{eq:transition-switch}):
\begin{align}
    \frac{d}{d\sigma}P\left(\vec{r}, \vec{t}, \vec{s}, \xi, \sigma \right)=
    &\sum_{i=1}^{N/2}\left(\mathbb{E}^-_{r_i}-1\right)\left\{\tau^+_{r_i} P\left(\vec{r}, \vec{t},\vec{s}, \xi, \sigma\right)\right\} \nonumber \\
    &+\sum_{i=1}^{N/2}\left(\mathbb{E}^+_{r_i}-1\right)\left\{\tau^-_{r_i} P\left(\vec{r}, \vec{t}, \vec{s}, \xi, \sigma\right)\right\} \nonumber \\
    &+\sum_{j=1}^{N/2}\left(\mathbb{E}^-_{t_j}-1\right)\left\{\tau^+_{t_j} P\left(\vec{r}, \vec{t}, \vec{s}, \xi, \sigma\right)\right\} \nonumber \\
    &+\sum_{j=1}^{N/2}\left(\mathbb{E}^+_{t_j}-1\right)\left\{\tau^-_{t_j} P\left(\vec{r}, \vec{t}, \vec{s}, \xi, \sigma\right)\right\} \nonumber \\
    &+\sum_{k=1}^N\left(\mathbb{E}^-_{s_k}-1\right)\left\{\tau^+_{s_k} P\left(\vec{r}, \vec{t}, \vec{s}, \xi, \sigma\right)\right\} \nonumber \\
    &+\sum_{k=1}^N\left(\mathbb{E}^+_{s_k}-1\right)\left\{\tau^-_{s_k} P\left(\vec{r}, \vec{t}, \vec{s}, \xi, \sigma\right)\right\} \nonumber \\
    &+\nu\left\{P\left(\vec{r}, \vec{t}, \vec{s}, -\xi,\sigma\right) -P\left(\vec{r}, \vec{t}, \vec{s}, \xi, \sigma\right)\right\}
    \label{eq:master_equation}
\end{align}
where $P\left(\vec{r}, \vec{t}, \vec{s}, \xi, \sigma \right)$ 
gives the probability to find the population in state $\left(\vec{r}, \vec{t}, \vec{s}, \xi\right)$ at time $\sigma$, with 
 $\vec{r}=\left(r_i\right)$, $\vec{t}=\left(t_j\right)$, $\vec{s}=\left(s_k\right)$. Here,  $\mathbb{E}_{r_i}^\pm$ is a shift operator such that 
\begin{align}
\mathbb{E}_{r_i}^{\pm}P\left(\vec{r}, \vec{t}, \vec{s},\xi,\sigma\right)=P\left(r_1, \ldots, r_i \pm1, \ldots, r_{N/2},\vec{t}, \vec{s},\xi,\sigma\right),
\label{eq:shift_operation}
\end{align}
and $\mathbb{E}_{t_j}^\pm$ and $\mathbb{E}_{s_k}^\pm$ are the equivalent shift operators for $t_j$ and $s_k$, respectively. Note that $P\left(\vec{r}, \vec{t}, \vec{s}, \xi, \sigma \right)=0$ whenever any of $r_i, t_j, s_k <0$.
The first to sixth lines on the right-hand-side of Eq  \eqref{eq:master_equation} represent the birth-and-death processes 
\eqref{eq:reactions}, while the last line accounts for environmental switching. 
\par
The master equation \eqref{eq:master_equation}
fully describes the model dynamics and can be simulated exactly
with the Gillespie algorithm \citep{Gillespie1977}.
Owing to the stochastic nature of the model, 
after a time that diverges exponentially with the community size, DN will cause the eventual collapse of the population \citep{Spalding2017, Assaf2017}.
This phenomenon is practically unobservable
when resource levels remain sufficiently large,  i.e. if $R_i^-\gg 1$, and the population settles in a long-lived quasi-stationary distribution. Here we focus on the quasi-stationary regime that is attained when distributions of species abundances appear to be stationary for a long time
(see \href{https://github.com/ShotaSHIBASAKI/Switching_Environment/blob/master/animation/TwoSpecies_hist.gif}{ supplementary video}): the distributions of two species' abundances change little from time $\sigma=130$ onward when the parameter values are as shown in Table \ref{tab:prameter} with $\nu=10^{-1}$ and $\delta=0.2$, which are typical parameter values used in this study. 
It would be reasonable then to set $\sigma_{end}=200$ expecting that species' abundances reach a quasi-stationary state in many of our chosen parameter values.
 
\subsection{Evaluating species interactions}\label{sec:model-interaction}
First, we analyze how the net effect of species interactions (i.e., resource competition and facilitation via detoxification) can change under DN and EFs by measuring the extinction probabilities in the presence or absence of another species at time $\sigma_{end}$ (see \ref{sec:detail-simulation1} for details). We begin with two species ($N=2$) where the sign and magnitude of species interactions can change because the amounts of resources and toxins can change the net effect of resource competition and facilitation via detoxification \citep{Piccardi2019}.  We used parameter values such that species 1, if it persists, always outcompetes species 2 in the absence of DN (i.e., species 1 grows faster than species 2, see \ref{sec:analysisODE} for analysis and Table \ref{tab:prameter} for parameter values) to identify the effects of DN. Importantly, EFs alone do not change species 1's extinction probabilities in mono- versus co-culture in the absence of DN. Under DN coupled with EFs in our chosen parameter range, either of the two species or both species tend to go extinct. As a proxy for interactions, we focus on the net effect of species 2 on species 1, which is defined by the extinction probability of species 1 in mono-culture minus that in co-culture with species 2 (the so-called \textit{difference in extinction probability}, see also Eq \eqref{eq:diffProb}):  species 2 has a negative (positive) effect on species 1 if 
species 1 more (less) frequently goes extinct 
in co-culture with species 2 than in mono-culture. When species 2 has a negative effect on species 1, one can consider two possibilities in co-culture: (i) species 2 outcompetes species 1 (see Fig. \ref{fig:ex_exclusion}) or (ii) both species 1 and 2 go extinct. As explained in section \ref{sec:CE-change}, we focus on the former probability, the so-called {\it probability of exclusion of the fittest} to understand the net effect of species 2 on species 1. We performed $10^5$ simulations for each switching rate and toxin sensitivity in mono- and co-cultures. We calculated  95\% of highest posterior density intervals (HPDI) to measure the uncertainty of species 1's extinction probabilities (see \ref{sec:detail-simulation1}) but these intervals are too small to be visible on our plots due to the large number of simulations we ran.

\subsection{Evaluating species diversity}\label{sec:model-diversity}
To explore how species diversity changes with the switching rate, we ran simulations at different numbers of species ranging from $N=2$ species to $N=10$ species and different mean toxin sensitivities (see \ref{sec:detail-simulation2} for details). For each condition (one number of species and one mean toxin sensitivity), we sampled 100 sets of parameters from certain probability distributions. These sets of parameters represented 100 communities composed of $N$ species. In this analysis (section \ref{sec:beta-div}) we consider a more general scenario and all previous simplifying assumptions on the parameter values are relaxed: all species have randomly drawn (see \ref{sec:detail-simulation2}), and hence typically different, growth and death rates.
\par
The dynamics of each community were independently simulated 100 times to see whether the species composition was robust against DN and EFs. These 100 replicate runs can be seen as 100 independent ``patches''that initially consist of the same species set, but no species migrate from one patch to another.  We measured the beta diversity of each community \citep{Jost2007,Chao2012} and species richness (number of surviving species) as functions of the environmental switching rate
in the quasi-stationary distribution (at time $\sigma_{end}$). Beta diversity accounts for the heterogeneity of each community across 100 replicates 
For example, if beta diversity is larger than one but species richness is one in all replicates,  different species fixate in each replicate. In contrast, beta diversity is one if all communities show identical species compositions; for example, in the  two-species scenario with parameter values shown by Table \ref{tab:prameter} and in the absence of DN, species 1 always outcompetes species 2 and thus beta diversity is one. 
This baseline corresponds to a perfectly deterministic scenario. One could instead consider using another baseline corresponding to a perfectly stochastic or neutral scenario where all species' parameter values are identical (e.g., Fig. \ref{fig:neutral}), in which case species compositions are determined only by demographic noise coupled with environmental fluctuations. We choose to focus on the deterministic baseline, as beta diversity is always $=1$ regardless of the number of species, their parameter values, and the environmental switching rate. In contrast, beta diversity in the neutral scenario changes with these parameter values, making it more difficult to compare across conditions. 

\par
We compared the patterns of beta diversity of two- or ten-species communities with the probability of exclusion of the fittest in species pairs sampled from these communities (see also \ref{sec:qunatify_similarity}). The sampled species pairs may stably coexist, in which case the fittest species in a pair is the one that is more abundant in the absence of any noise. If both species go extinct in the absence of noise, either of the two species is randomly labeled as the fittest. This labeling generalizes what is used in the species interaction analysis above where species 1 (the fittest species) always grows faster than species 2, such that their long term coexistence is impossible in the absence of DN. 

\subsection{Statistical analysis}\label{sec:statistics}
Statistical analysis was performed with Python 3.7.6 incorporating Scipy 1.4.1. and pymc3 3.10.0. For statistical tests of Pearson's correlation and  Spearman's rank-order correlation, scipy.stats.pearsonr and scipy.stats.spearmanr were used, respectively. For calculation of HPDIs, pymc3.stats.hpd was used.
\section{Results}\label{sec:results}
\subsection{Toxin sensitivity determines single species' response to coupled DN and EFs}\label{sec:mono_extinction}
\begin{figure}[tbh]
    \centering
    \includegraphics[scale=0.5]{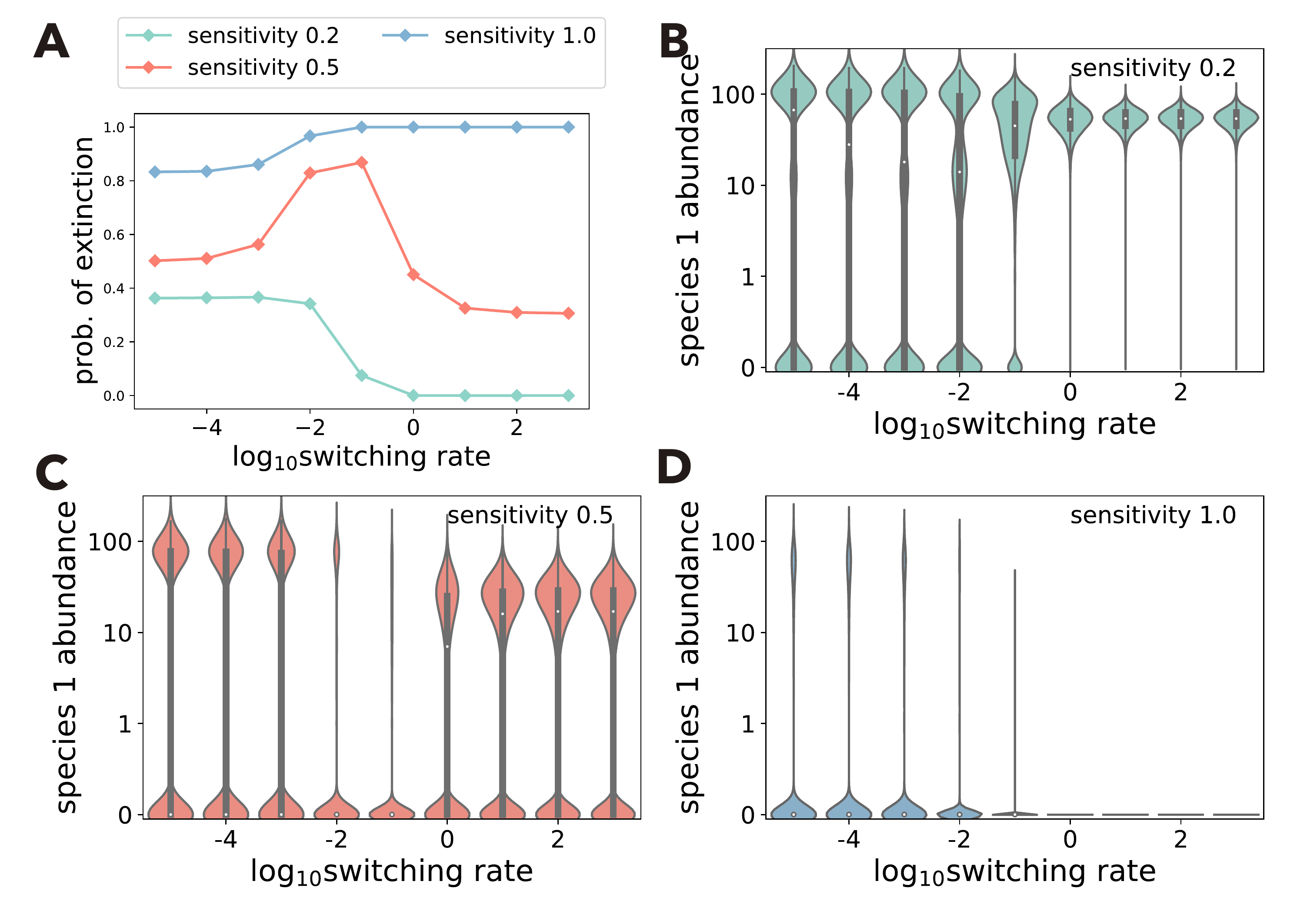}
    \caption{Extinction probability and species abundance in mono-culture}
    \label{fig:mono}
    \begin{flushleft}
    \small{A: Extinction probability of species 1 in mono-culture when the toxin sensitivity is low (green), moderate (orange), or high (blue). 95\% HDPIs are too small to see. B--D: violin plots of species 1's abundance at the end of $10^5$ simulations with a low (B), moderate (C), or high (D) toxin sensitivity. White dots and black bars represent median values of the abundances and their interquartile ranges, respectively.}
    \end{flushleft}
\end{figure}

To establish our intuition on how the coupling of EFs and DN affects the dynamics, we first analyse the extinction probabilities of a single species (species 1) in mono-culture with one type of resource and toxin (Fig. \ref{fig:mono}A). As the switching rate decreases, the duration of the harsh or mild environments become longer. In the presence of DN, this duration determines whether or not the species goes extinct, together with its sensitivity to toxins in the environment, which can be seen to modulate environmental harshness. When the switching rate is very low ($\nu \to 0$), the species is exposed to the static environment with either abundant or scarce resources (with probability $0.5$, respectively) depending on the initial environmental condition $\xi(0)$: it mostly goes extinct under scarce resource supplies even when their sensitivity to toxins is low (Fig. \ref{fig:mono}B). On the other hand, abundant resource supplies maintain species 1 with some probability even if it is very sensitive to the toxin (Fig. \ref{fig:mono}D). Over many simulations, low fluctuation rates therefore result in a bimodal distribution of the species' abundance (e.g., Fig. \ref{fig:mono}B). At the other extreme, very high switching rates ($\nu \to \infty$) expose the bacteria to an environment with mean abundance of resources \citep{Wienand2017a,Wienand2018,West2019a,Taitelbaum2020}. This is enough to rescue the species  with a low toxin sensitivity (Fig. \ref{fig:mono}B) but not with a high sensitivity (Fig. \ref{fig:mono}D). At an intermediate toxin sensitivity (Fig. \ref{fig:mono}C), the worst situation lies in the intermediate fluctuation rate: the duration of  the harsh environment is long enough to drive them extinct, but the time with abundant resource supply is  not long enough to rescue them fully. In sum, even when only a single species is present, we see non-trivial patterns in its response to EFs coupled with DN, which depends on its sensitivity to toxins.
\subsection{Toxin sensitivity changes how switching rate affects two-species competition} \label{sec:species-interaction}
Next, we add another species (species 2) that grows slower than species 1 into the environment and ask how it interacts with species 1 in our model.
Rather than measuring interactions through the effect of each species on the other's abundances, we focus on species 1 and analyze how its extinction probability is affected by the presence of species 2, compared to  mono-culture (Fig. \ref{fig:mono}). Our reasoning is that (i) species 1 should always out-compete species 2 in the absence of  DN, and that (ii) we already know the extinction probability of species 1 under EFs and DN in mono-culture; measuring any deviation from the mono-culture outcome allows us to quantify how likely it is for the fitter species to be lost in a given community. Such species loss events can be seen as ecological drift. 
We again explore changes in the species' toxin sensitivity, as we learned above that it affects species abundances via DN (Fig. \ref{fig:mono}), but also because we expect it to affect species interactions \citep{Piccardi2019}. For now, we varied sensitivity to toxins simultaneously for both species, an assumption that we relax later. 
\par
When both species were highly sensitive to the toxin, species 2 had a positive effect on species 1, reducing its extinction probability. This occurs because in the simulations, toxic compounds are degraded more quickly in co-culture than in mono-culture due to the larger initial number of individuals ($s_1(0)+s_2(0) > s_1(0)$). This effect can be recapitulated by a mono-culture with larger initial abundance (Fig. \ref{fig:init_size}). A larger total initial species abundance in co-culture decreases death rates, which outweighs competition for nutrients in toxic environments \citep{Piccardi2019}. However, for most parameter values in Fig. \ref{fig:interaction}A, species 2 has a negative effect on species 1 by increasing its extinction probability.  We therefore focus on competitive interactions in the main text and consider positive interactions in  \ref{sec:ChangeResourceSupply}.
\par
As in the single species case (Fig. \ref{fig:mono}), the extinction of species 1 was highly dependent on the toxin sensitivity of the two species, as we varied the fluctuation rate: monotonically increasing, monotonically decreasing, or non-monotonically changing with a minimum or maximum value at an intermediate switching rate (Fig. \ref{fig:interaction}B). Interestingly, this pattern does not match the single-species behavior (compare Fig. \ref{fig:mono}A and \ref{fig:interaction}\Add{A}\Erase{B}). 

\subsection{Behavior at extreme switching rates explains non-monotonic changes in exclusion of the fittest}\label{sec:CE-change}
To better understand why the faster-grower goes extinct in the observed parameter ranges (Fig. \ref{fig:interaction}B), we decompose species 2's effect on species 1 into the probability that species 2 persists but species 1 goes extinct (hereafter, called \textit{probability of exclusion of the fittest}) and the probability of any other outcome (e.g., extinction of both species, see Eq \eqref{eq:decomposedDiff}). We focus on the probability of exclusion of the fittest, as it explains the variation in species interaction strength in most cases (compare Figs. \ref{fig:interaction}A and C) and investigate how it changes over the switching rate and depends on toxin sensitivity. 
\par
\begin{figure}
    \centering
    \includegraphics[scale=0.35]{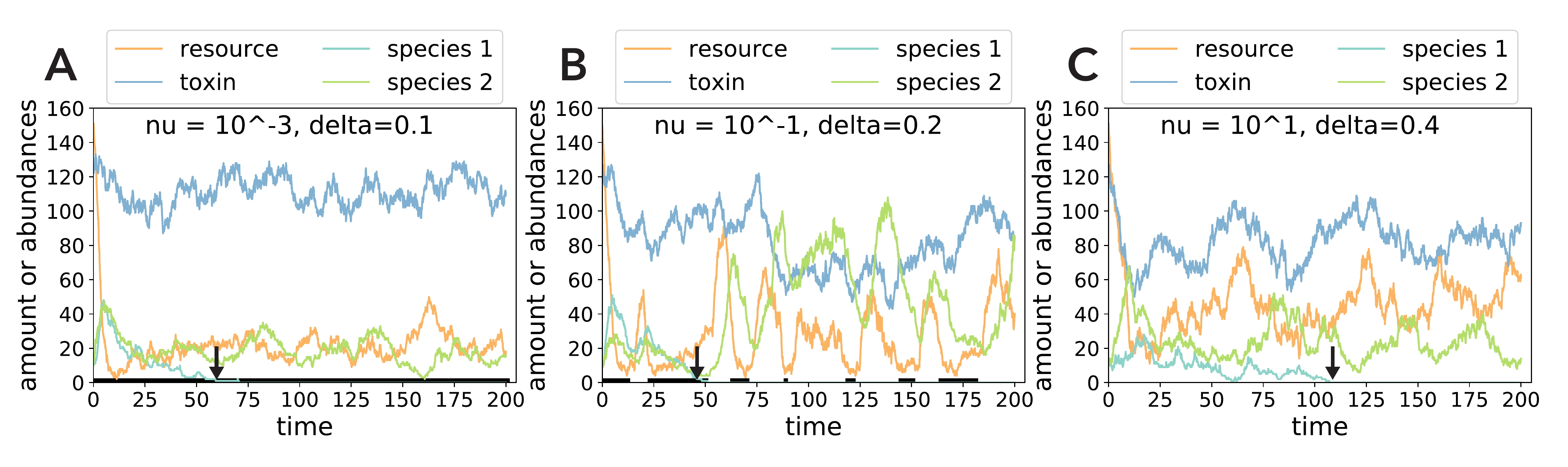}
    \caption{Examples of the dynamics with exclusion of the fittest}
    \label{fig:ex_exclusion}
    \begin{flushleft}
    {\small
    In these examples (A: $\nu=10^{-3}$, and $\delta=0.1$, A: $\nu=10^{-1}$, and $\delta=0.2$, C: $\nu=10^{1}$, and $\delta=0.4$), species 1 goes extinct but species 2 survives at the end of simulation $\sigma_{end}=200$ due to the DN-EFs coupling. Black lines on  the $x$-axis represents times when the resource supply is scarce ($\xi(t)=-1$) while white lines represent times when the resource supply is abundant ($\xi(t)=1$). In panels A and B,  species 1 decreases its abundance and goes extinct during the scarce resource supply condition (pointed by arrows). In panel C, the environmental conditions are not shown because they are not visible due to the fast switching.}
    \end{flushleft}
\end{figure}

We again let the two extreme switching rates guide our intuition (see also Fig. \ref{fig:diff_constant}). At a very slow rate ($\nu \to 0$), the resource supply remains either scarce or abundant, while a very fast environmental switching rate ($\nu\rightarrow\infty$) drives resource supply to the mean concentration. 
We explore system behavior at these three constant resource supplies (scarce, abundant or mean) and different toxin sensitivities, which together represent how harsh the environment is. 
When the environment is harsh  -- due to resource scarcity, high toxin supplies, or high toxin sensitivity --, both species are most likely to go extinct rather than to outcompete each other (Figs. \ref{fig:CompExcl}A -- C). As toxin sensitivity goes down and species survival becomes more likely, DN becomes more important and we see a higher probability that even the fitter species (species 1) will be excluded. The more resources are available, the more likely it is that species survive -- in particular, that species 1 out-competes species 2, and the peak of competitive exclusion moves to a higher toxin sensitivity (see arrows in Figs. \ref{fig:CompExcl}A -- C). When it is easy for both species to survive (toxin sensitivity is low and/or resources are abundant), DN no longer plays an important role and the faster-growing species 1 is unlikely to be excluded. ntuitively then, the exclusion of the fittest is caused by the coupling of DN with EFs: Harsh environments, where both species' abundances are positive but low (see \ref{sec:OnlyEF}), lead to stronger DN and a higher probability
of extinction of the fittest species compared to a static or mild environment (Fig. \ref{fig:ex_exclusion}). 
It is important to stress that the exclusion of the fittest never occurs without DN, regardless of EFs (see \ref{sec:appendix2}).
\par
Hereafter, we refer to the toxin sensitivity that maximizes the probability of exclusion of the fittest in the absence of environmental switching as the ``critical toxin sensitivity'' (arrows in Figs. \ref{fig:CompExcl}A -- C).
We see two critical toxin sensitivities (at $0.1$ and $0.8$ in Figs. \ref{fig:interaction}C and D, Figs. \ref{fig:CompExcl}A and C) at $\nu \rightarrow 0$ that correspond to the long time spent with either scarce or abundant resources. Instead, at $\nu \rightarrow \infty$, where resources remain at mean abundance, there is a single critical toxin sensitivity (at $0.4$ in Figs. \ref{fig:interaction}C and D) where exclusion of the fittest is most likely (Fig. \ref{fig:CompExcl}B). Toxin sensitivities between these critical values can show a maximum or minimum probability of exclusion of the fittest at an intermediate switching rate, resulting in the rugged landscape of Fig. \ref{fig:interaction}C (see \ref{sec:landscape} for more detail).

\subsection{Competition strength changes non-monotonically under different scenarios} \label{sec:scenarios}
We have shown that using a given set of parameters, the rugged landscape shown in Fig. \ref{fig:interaction}C causes the competitive exclusion of a faster-growing species to either increase, decrease or vary non-monotonically across switching rates, depending on toxin sensitivity. We next explore the generality of this finding. In the appendix, we explore scenarios where (i) switching occurs in toxin rather than resource supplies, where (ii) both resource and toxin supplies switch (Table \ref{tab:resoure-toxin-switch}, see \ref{sec:alternative}), or where (iii) we change the amounts of scarce and abundant resource supplies (\ref{sec:ChangeResourceSupply}). 
\par
In all these scenarios, the landscapes of species 1's difference in extinction probability and  probability of exclusion of the fittest are qualitatively similar (Figs.\ref{fig:interaction} C and D, \ref{fig:alternative_scenario}, and  \ref{fig:ChangeResourceSupply}). However, each scenario differs in the three critical toxin sensitivities and likelihood that the difference in extinction probability non-monotonically changes over the switching rate. Accordingly, we asked whether the distances between critical toxin sensitivities  might predict the probability of observing non-monotonic behavior. In Table \ref{tab:sum_criticall} and Fig. \ref{fig:monotonic_toxin_crti},  we show that the distance between critical sensitivities under harsh and mean environments (i.e., very fast environmental switching) correlates positively with the likelihood of observing non-monotonic effects of the switching rate on competition (Fig. \ref{fig:monotonic_toxin_crti}, black circles; Spearman's $\rho=0.77$, P-value: $0.043$), but no significant correlation was found with the distance between the critical toxin sensitivities under the mean and mild, or the harsh and mild environments (Fig. \ref{fig:monotonic_toxin_crti}, grey diamonds and cross marks; Spearman's $\rho = -0.22$, P-value: $0.64$, and $\rho = 0.42$, P-value: $0.35$, respectively). Therefore, the non-monotonic change of the difference in extinction probability is likely when there is a large difference in the critical toxin sensitivities under harsh and mean conditions.

\subsection{Beta diversity changes similarly to exclusion of the fittest}\label{sec:beta-div}
In the previous sections, we focused on interactions between two species and the conditions under which one may drive the other extinct. Ultimately, however, our interest is to predict how whole communities comprised of tens, hundreds or even thousands of species are affected by fluctuations in the environment. 
\par
We set up a model of 100 communities composed of between 2 and 10 species each. Species within each community were defined by parameter values that were randomly sampled from the same distributions with the exception of toxin sensitivity $\delta$, which was sampled from beta distributions with different means, ranging from $\bar{\delta}=0.1$ to $1$. We generated a new set of 100 communities with different numbers of species and different fluctuation rates as above, and ran 100 replicate simulations for each of the 100 communities in each set. 
We then measured beta diversity across the 100 replicate simulations per community and final species richness (number of surviving species) over all 100 runs for the 100 communities (total: 10'000). In this model design, the 100 replicate runs represent independent  ``patches'' without migration. Their beta diversity then indicates how different the species compositions were across all patches in a given environment (e.g. a given fluctuation rate), and we repeat the exercise 100 times with different species sampled from the same distributions to see generality of the results. 
A high beta diversity would then indicate that we have different community compositions in each patch, while a beta diversity of 1 would tell us that all patches have the same species composition.

\par
In two-species communities, we obtained qualitatively similar patterns of exclusion of the fittest to those in the species interaction analysis (compare column A in Fig. \ref{fig:diversity} with Fig. \ref{fig:interaction}C), suggesting that our results in Fig. \ref{fig:interaction}C are unlikely to be specific to the choice of species parameter values in Table \ref{tab:prameter}. Beta diversity changes over the environmental switching rate similarly to the probability of exclusion of the fittest for four out of the five tested mean toxin sensitivities (columns A and B in Fig. \ref{fig:diversity}, see also Fig. \ref{fig:corr_BetaExcl2}): both monotonically decrease (mean toxin sensitivity $\bar{\delta} = 0.1$ or $1.0$), or non-monotonically change with maximum ($\bar{\delta} = 0.2$) or minimum ($\bar{\delta} = 0.6$) values at intermediate switching rates. This similarity can be explained as follows: ignoring extinction of both species, a small probability of exclusion of the fittest indicates that the stronger species 1 fixates in most simulations, a homogeneous outcome with small beta diversity. In contrast, when the exclusion of the fittest is more likely, the weaker species 2 is more likely to fixate, leading to significant heterogeneity in the simulation results and large beta diversity (species 2 survives alone in a fraction of the runs, and species 1 in the rest).
At one mean toxin sensitivity $\bar{\delta} = 0.4$, the patterns of the probability of exclusion of the fittest and beta diversity over the switching rate do not match. At this toxin sensitivity, beta diversity remains high at low switching rates ($\nu=10^{-5}, 10^{-4},  10^{-3}$) because both species go extinct in 50\% of the runs but they can coexist in about 7\%, as illustrated by our measure of final species richness (column C in Fig. \ref{fig:diversity}). Beta diversity ignores the cases of both-species extinction but increases in cases of coexistence, see Eq \eqref{eq:beta_div}.
Overall, looking at final species richness (column C in Fig. \ref{fig:diversity}), we see that complete extinction is more likely to occur as sensitivity increases. At the highest toxin sensitivity, the only way a species can survive is if the switching rate is really low and they can benefit from abundant resources for a long time, while at lower toxin sensitivity, complete extinctions only occur at low switching rates, because there is long term exposure to scarce resources.
\par
In communities with ten species (see Figs. \ref{fig:beta-diversity-supp} and \ref{fig:richness-supp} for intermediate community sizes), we observe similar patterns between beta diversity (column D in Fig. \ref{fig:diversity}) and the probability of exclusion of the fittest (column A in Fig. \ref{fig:diversity}) 
Studying interactions between species pairs can therefore predict the behavior of a ten-species community. To explore whether it matters which two species one selects for the interaction analysis, we next repeatedly sub-sampled species pairs from each ten-species community and compared the exclusion of the fittest in the pairs with the beta diversity of the whole community (Fig. \ref{fig:subsample}). Naturally, the more species pairs one samples, the more accurately we can predict the pattern of beta diversity, but this accuracy appears to saturate at around five species pairs (\ref{sec:qunatify_similarity}, Fig. \ref{fig:corr_BetaExcl10}), which is approximately 11\% of all possible species pairs. In addition, large beta diversity does not necessarily reflect a large variation in species richness; large beta diversity with small species richness (see mean toxin sensitivity $\bar{\delta}=0.4$ or $0.6$ at switching rate $\nu \ge 10^0$ in columns D and E of Fig. \ref{fig:diversity}) indicate that different species fixate in each run, which also supports the observed relationship between the exclusion of the fittest and beta diversity.
In sum, estimating the probability of exclusion of the fittest between a few randomly selected species pairs (section \ref{sec:CE-change}) is a good predictor for the beta diversity of larger communities under those same environmental conditions (see \ref{sec:qunatify_similarity} for detailed discussion). This similarity is not coincidental: as for the probability of exclusion of the fittest, beta diversity is also maximized when DN is the strongest, such that different species survive in each patch.
\section{Discussion}\label{sec:discussion}
Understanding how species diversity in microbial communities arises and is maintained is a central question in microbial ecology and evolution. While many theoretical and experimental studies have addressed this question in static environments, community diversity is expected to respond to fluctuations between benign and harsh environmental conditions, which can alter the abundance of different species and the interactions between them (see e.g., \citet{Rodriguez-Verdugo2019}). Strong drops in population sizes caused by harsh conditions can increase the strength of DN, which, coupled with EFs may lead to non-trivial outcomes \citep{Wienand2017a,Wienand2018,West2019a,Taitelbaum2020}. Here we have analysed a mathematical model representing a biological scenario such as a gut or soil microbial community that experiences fluctuations between benign and harsh conditions, such as feast and famine, or drought and rain. The model shows how community diversity -- mediated by inter-specific interactions and DN -- changes with environmental fluctuation rates.
\par
Our study is centred on two main findings. First, we show that the rate at which resource supplies switch changes the ability of a slower-growing species to drive a fast-growing one extinct (Fig. \ref{fig:interaction}). While the fitter species will never be excluded in the absence of DN, in a fluctuating environment with DN, harsh conditions (e.g. scarce resources) strengthen DN to sometimes drive species extinction in spite of their greater fitness. We see this as a form of ecological drift, wherein selection by the environment is not strong enough to maintain the fittest species and thus the fittest species may go extinct due to DN. By changing the length of time spent in the harsh environment, the environmental switching rate affects such competitive exclusion. In addition, the species' ability to withstand environmental harshness can also strengthen DN and lead to stronger ecological drift. 
We confirmed the generality of these results by exploring various forms of EFs (e.g., asymmetric switching environments and cyclically changing environments in \ref{sec:variousflucutuation}).  
When we consider many replicate communities with identical initial species compositions, the increased stochasticity resulting from DN means that which species go extinct under these harsh conditions is less dependent on their relative competitiveness, and species composition will be more random in each replicate, leading to greater beta diversity. Recent studies corroborate this finding: smaller communities show larger varieties in species composition due to DN \citep{Gilbert2017}, while species composition is robust against EFs when species are insensitive to them \citep{Dedrick2021}. 
\par
Although we now understand that beta diversity increases when the probability of exclusion of the fittest is high, precisely when such ecological drift is maximized will be difficult to predict in practice, as it is a function of multiple factors: the form of the EFs (\ref{sec:alternative} and \ref{sec:variousflucutuation}), the magnitude of EFs (\ref{sec:ChangeResourceSupply}), the rate of EFs and the sensitivity of species to environmental harshness. Yet, our simulation approach allows us to investigate these different effects efficiently.
\par
Our second main finding is that a good way to predict how beta diversity will respond to EFs is to measure the probability of exclusion of the fittest across fluctuation rates in few pairs of species randomly sampled from a focal community and use it as an indicator for how the beta diversity of the whole community will behave (Fig. \ref{fig:diversity}\Add{, \ref{sec:qunatify_similarity}}). In two species communities, explaining the similarity between the probability of exclusion of the fittest and beta diversity is straightforward: the latter accounts for the probabilities that both or either of the two species persist. We verified that this similarity also holds in neutral scenarios (i.e., two species are identical except for their labels and thus the fittest species is arbitrarily chosen, Fig. \ref{fig:neutral}). In larger communities, the similarity is not straightforward but our first main finding helps to understand this result. EFs affecting DN strength could exclude species that would persist in a community without noise. If the ``fittest'' species goes extinct with some probability due to DN, community compositions at time $\sigma_{end}$ are heterogeneous and beta diversity increases. Therefore, our two findings emphasize the importance of coupling EFs and DN: competition results can differ from those in static environments and this affects beta diversity. 
\par
This brings us to a hypothesis that has been debated at length in ecology: the intermediate disturbance hypothesis \citep{Connell1978, Grime1973a}, which states that intermediate intensity and frequency of disturbance maximize species diversity. \citet{Fox2013} argues that the intermediate disturbance hypothesis should be abandoned because many examples disagree with it \citep{Mackey2001, Miller2011}. In our model, fluctuations in resource and toxin  can be regarded as disturbances.  In agreement with \citet{Mackey2001} and \citet{Miller2011} then, an intermediate intensity (i.e., toxin sensitivity) or disturbance frequency (environmental switching rate) does not always maximize beta diversity: our analysis shows that intermediate frequencies of disturbance maximize beta diversity only when mean toxin sensitivity is within a certain range. Mean toxin sensitivities at the two thresholds of this range show that beta diversity monotonically decreases or increases over the switching rate. These thresholds depend on scenarios of environmental switching and amounts of resource supplies because these parameters change the probability of competitive exclusion (see \ref{sec:alternative} and \ref{sec:ChangeResourceSupply}). High beta diversity at intermediate disturbances is then a consequence of a change in environmental conditions and not expected to apply generally.
 \par
The relationship between EFs and species diversity is also an important question in the modern coexistence theory, which predicts that fluctuations will affect species coexistence by changing species growth rates when rare \citep{Chesson2000a,Chesson2000b, Barabas2018,Ellner2019,Letten2018a}. Compared to the approach taken in this study -- where we ask how many and which species persist at the end of a long but fixed time frame (i.e., for a quasi-stationary distribution, Fig. \ref{fig:diversity}) --  the modern coexistence theory allows one to analyze whether or for how long a set of species will all coexist \citep{Schreiber2020}. An interesting future direction would be to apply the modern coexistence theory to investigate how environmental fluctuation rates and toxin sensitivities affect the duration of all-species coexistence. This approach would help to propose biological mechanisms behind species coexistence in our setup.

\par
Of course, our model makes some simplifying assumptions and has some limitations. First, we used arbitrary time units, which in practice can be considered to be hours, corresponding to typical bacterial growth rates in relevant experiments \citep{NOVICK1950,Lin2002,Zhao2003}. This implies that species interactions and beta diversity will vary when environmental switching ranges from hourly ($\nu=10^0$) to about once every four days ($\nu=10^{-2}$) on average, which is shorter than in some experimental studies \citep{Benneir1999,Rodriguez-Verdugo2019,Chen2020} but not impractical. That said, under this assumption, changing environmental switching from a daily to an hourly scale, for example, would show different species compositions or diversity. 
\par
Second, our model focuses on competitive exclusion but other types of interactions can also affect diversity \citep{Rodriguez-Verdugo2019}. Positive interactions between pairs of species (e.g., cross-feeding), for example, might increase alpha and gamma diversities, because such interactions enable species to coexist \citep{Sun2019}. This could result in an increase in beta diversity because the extinction of one species increases its partner species' extinction probability \citep{Dunn2009,Goldberg2020}. 
\par
Third, one can consider more complex species-resource (and species-toxin) interaction  functions. For example, species' growth rates can be limited by the resource with the smallest amount when the resources are complementary \citep{Leon1975}. In addition, absorption rates of toxin might correlate with species' growth rate if the toxin targets cell metabolism. These functional forms may enable more species to coexist, collapsing the similarity between beta diversity and the probability of exclusion of the fittest. Our study would also be more general if species could change their growth rates depending on resource concentrations \citep{Nguyen2020a} or types \citep{Balakrishnan2021}. Introducing such plasticity would affect species' extinction probability and diversity. One could also consider building species-compounds interaction networks. In the current manuscript, we assume that each resource (toxin) has a positive (negative) effect on each species. However, we know that some compounds, (e.g., pesticides) can be resources for some species but toxic to others \citep{Muturi2017}, and other compounds (e.g., that affect pH \citep{Ratzke2020} and osmolarity \citep{Larsen1986,Oren2008}) can have either positive or negative effects on growth depending on their concentrations. The way in which species and compounds interact could affect exclusion of the fittest and species diversity. 
\par
Although we focused on beta diversity to measure the heterogeneity of communities in this manuscript, other metrics of species diversity could be considered. For example, \cite{Fung2015} and \cite{Kalyuzhny2015} analyze how demographic noise and environmental fluctuations affect species abundance distributions (SAD). \cite{Grilli2020} instead calculate the mean abundance distribution (MAD), which is defined as the distribution of mean species abundances over communities and follows a log-normal distribution. 
While SAD characterizes species diversity within a single community, it does not explain the heterogeneity of species compositions across communities, which is what we can capture with beta diversity. Combining multiple SADs or probability distributions of SADs is possible, but would not be as easy to analyze. Similarly, MAD ignores the variation of species' abundances across communities and may therefore not capture the heterogeneity of communities caused by demographic noise and environmental fluctuations.
\par
Finally, our community analysis considers up to ten microbial species, which is orders of magnitude below the size of natural microbial communities, according to genomic sampling \citep{Gans2005,Roesch2007}. However, it may also be reasonable to assume that species live in structured environments where they cannot possibly interact with more than a handful of other genotypes \citep{Tecon2019}. This suggests that a 10-species community may already be biologically meaningful.

\par 
In conclusion, the time scale of environmental fluctuations changes the importance of species fitness for survival and thus community beta diversity. This occurs in our model when EFs affect the strength of DN, leading to the occasional exclusion of strong species. Predicting how the strength of DN changes is not simple because it is affected by both environmental and species' parameters (resource and/or toxin supplies and toxin sensitivities in our model). This may be one explanation as to why the intermediate disturbance hypothesis does not always hold, but rather there are many relations between diversity and disturbance \citep{Mackey2001, Miller2011}. Nevertheless, we found similarities between how competitive exclusion plays out between species pairs and beta diversity at the community level. In the event that we would like to predict how the diversity of a given ecosystem, such as a soil community or a bioremediation ecosystem, responds to environmental fluctuations, it may be sufficient to isolate a few culturable species and analyze their interactions over different fluctuation rates. This approach promises to greatly facilitate our ability to study large and complex natural communities and their response to harsh conditions.

\section*{Author Contributions}
SS, MM and SM designed the study, SS performed simulations, analyzed data, and wrote the first draft,  and all authors contributed to revisions.

\section*{Acknowledgement}
We thank four anonymous referees for their helpful comments in the earlier version of the manuscript.

\section*{Data accessibility}
The programming codes and simulation data in csv files for this manuscript are available in \href{https://github.com/ShotaSHIBASAKI/Switching_Environment}{Github}.

\section*{Funding Statement}
 S.S. is funded by the University of Lausanne and Nakajima foundation. S.M. is funded by European Research Council Starting Grant 715097 and the University of Lausanne.  The authors declare no conflict of interest.
 
\clearpage
\bibliography{references.bib, ref3.bib}
\newpage
\begin{figure}[tbh]
    \centering
    \includegraphics[scale=0.5]{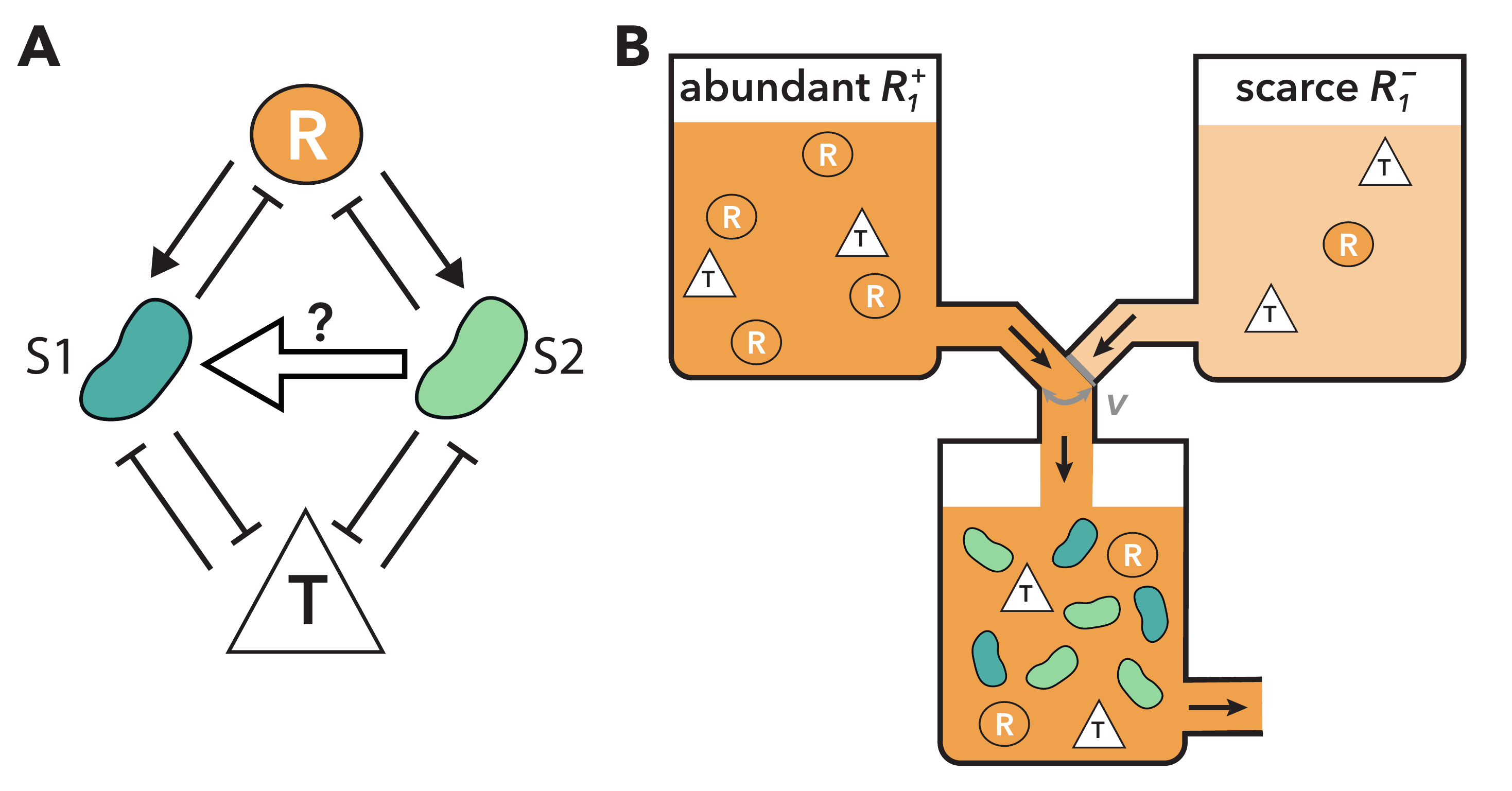}
    \caption{Schematic illustration of the model}
    \label{fig:cartoon}
    \begin{flushleft}
    \small{A chemostat model with environmental switching. A: Interaction network when $N=2$. A $\rightarrow$ B represents that A increases B while A \rotatebox{90}{$\perp$} B represents that A decreases B. Two species compete for the same resource (R in a circle) but are killed by the same toxic compound (T in a triangle). As a proxy for species interactions, we follow the net effect of the slower-growing species 2 on species 1 (large arrow from species 2 to 1).  B: An example of a chemostat model with environmental switching and $N=2$. Environmental switching is realized by changing the media flowing into a chemostat. In this example, the current environmental condition is abundant resource supply $R_1^+$.}
    \end{flushleft}
\end{figure}
\par

\begin{figure}
    \centering
    \includegraphics[scale=0.45]{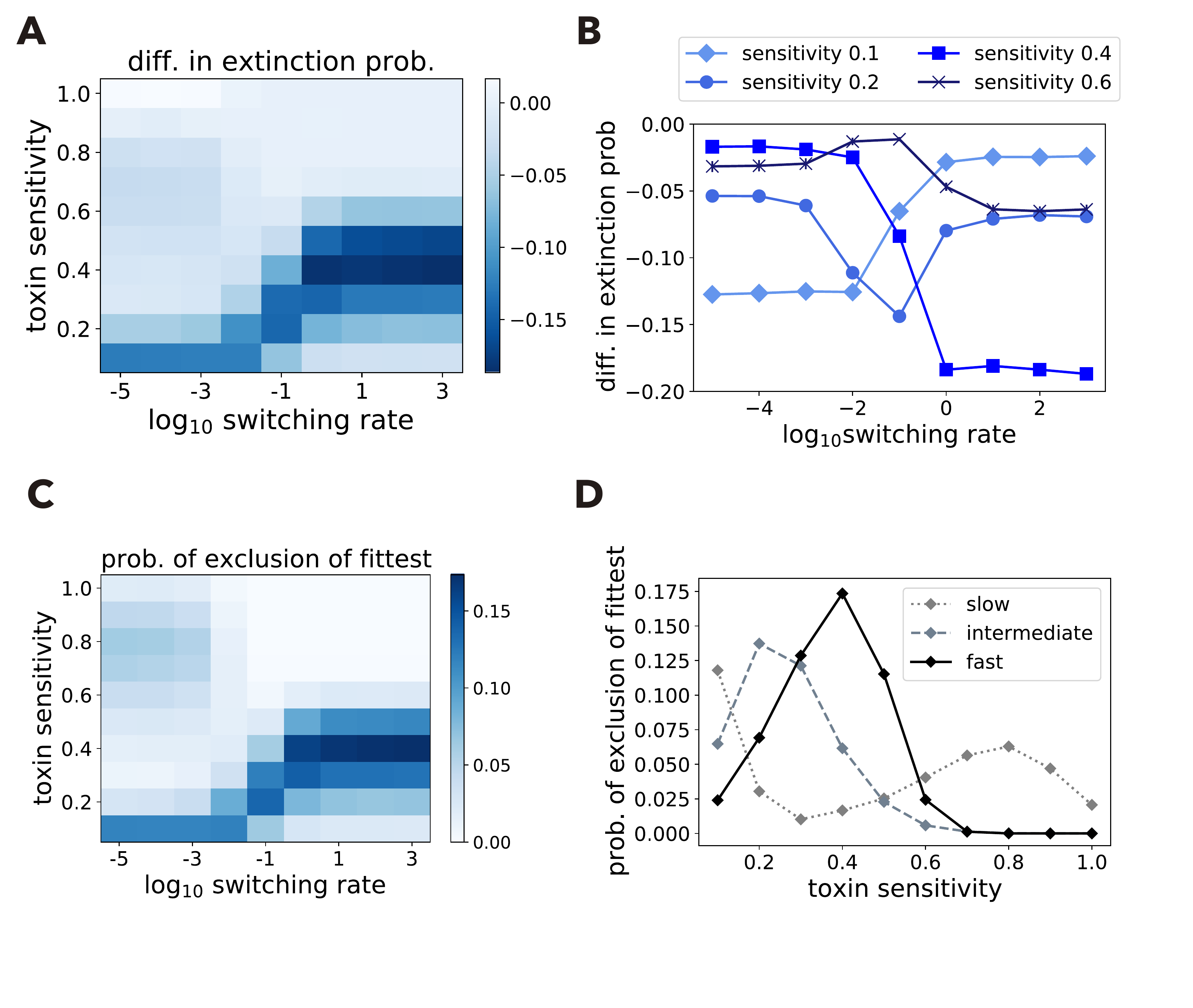}
    \caption{Species interaction strength changes differently over the switching rate}
    \label{fig:interaction}
    \begin{flushleft}
    {\small A:  Difference in species 1's extinction probability in mono-culture minus co-culture with species 2 (Eq \ref{eq:diffProb}) changes over the switching rate $\nu$ and the two species' identical toxin sensitivities. In most of the parameter space, species 2 has a negative effect on species 1 (i.e.,  species 2 increases the extinction probability of species 1). B: Some illustrative examples from panel A plotted differently to show how species 2's effect on species 1 changes over the switching rate at given toxin sensitivities. The difference in extinction probability can monotonically increase (toxin sensitivity $0.1$), monotonically decrease (toxin sensitivity $0.4$), or non-monotonically change with a minimum (toxin sensitivity $0.2$) or a maximum (toxin sensitivity $0.6$) value at an intermediate switching rate. C: Probability that species 2 persists but species 1 goes extinct (i.e., exclusion of the fittest) over the switching rate and the toxin sensitivity. D:  probabilities of exclusion of the fittest over the toxin sensitivity, when the environmental switching rate is slow ($\nu=10^{-5}$), intermediate ($\nu=10^{-1}$), or fast ($\nu=10^3$). In panels B and D, 95\% HDPIs are too small to see.}
    \end{flushleft}
\end{figure}

\begin{figure}
    \centering
    \includegraphics[scale=0.30]{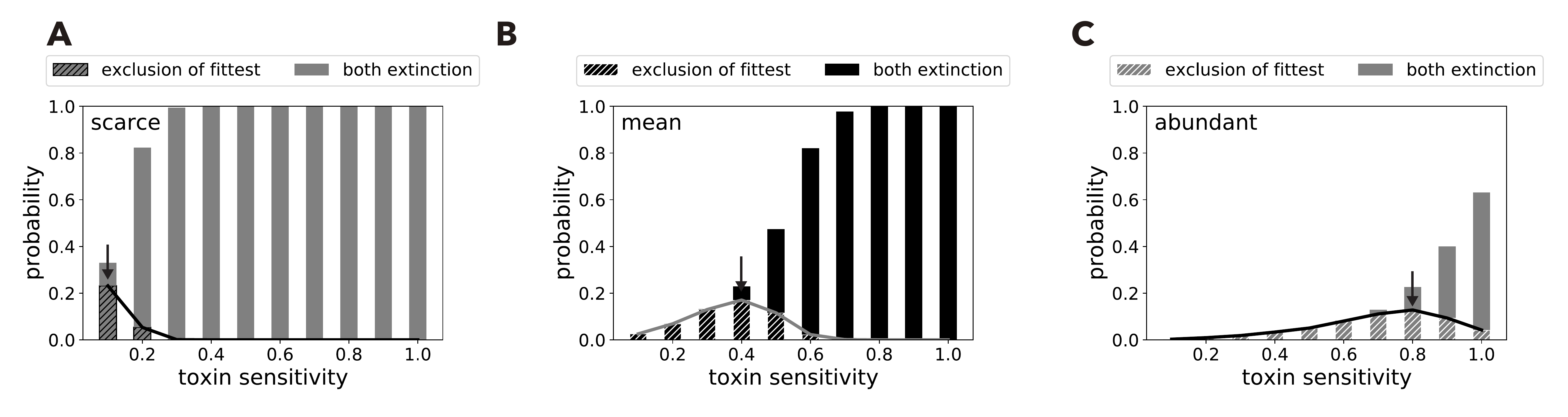}
    \caption{Exclusion of the fittest explains how DN changes with toxin sensitivity and resource supply}
    \label{fig:CompExcl}
    \begin{flushleft}
    {\small Analysis of exclusion of the fittest predicts at which toxin sensitivity DN is strongest. A -- C: In the absence of environmental switching, the probability of exclusion of the fittest (i.e., probability that species 2 excludes species 1 and survives, shown by solid lines and hatched bars) is uni-modal over the toxin sensitivity while the probability of both species extinction (bars) monotonically increases. The toxin sensitivities giving the peak values of probabilities of exclusion of the fittest (critical toxin sensitivities, pointed  by black arrows) depend on the resource supply: scarce $R_1=R_1^-$ (A), mean $R_1=\left<R_1\right>$ (B), or abundant $R_1=R_1^+$ (C). 95\% HDPIs are not observable as they are very small.  }
    \end{flushleft}
\end{figure}
\begin{figure}
\rotatebox{90}{
    \begin{minipage}{1.0\textwidth}
    \centering
    \includegraphics[scale=0.45]{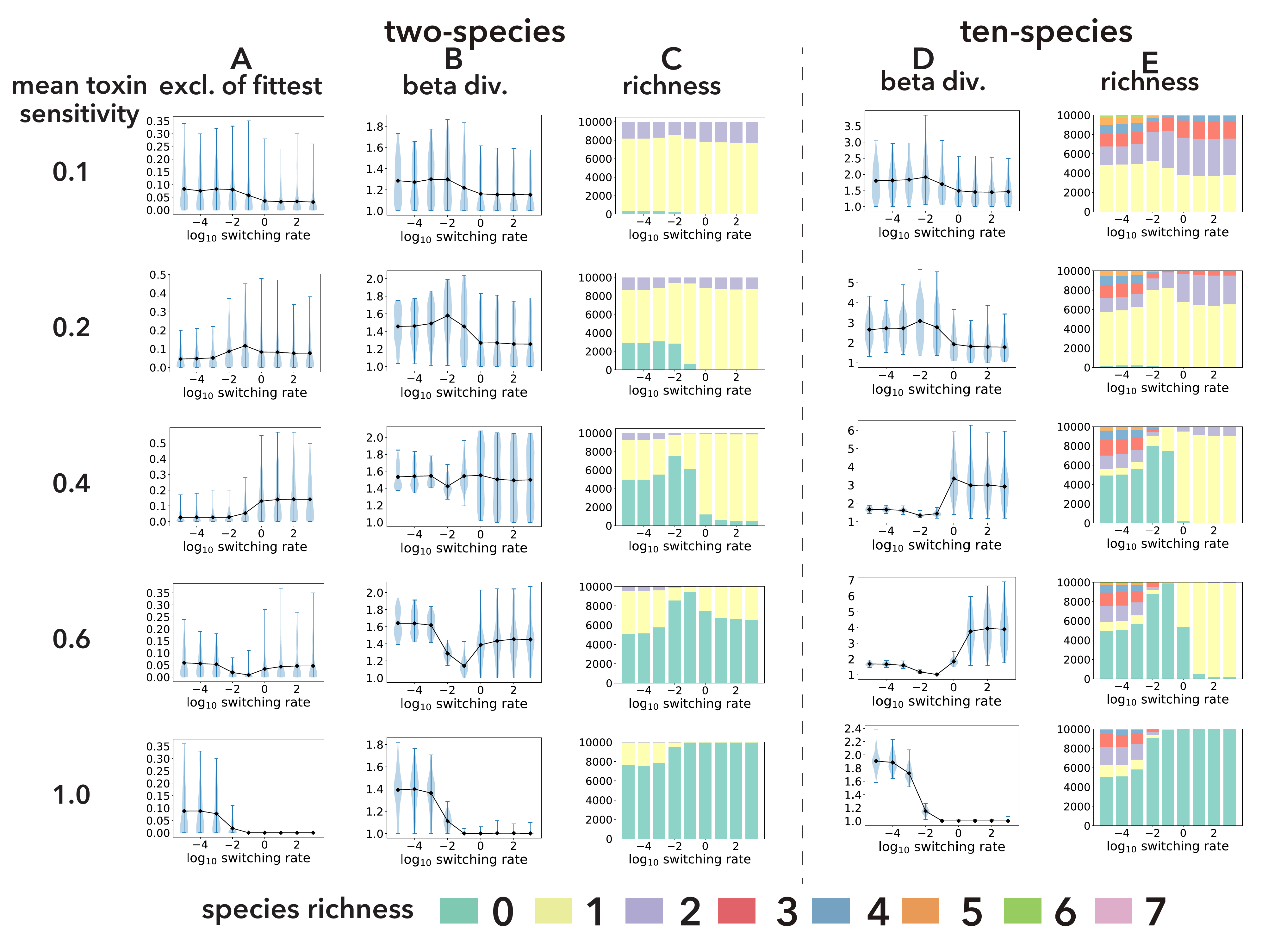}
    \caption{Comparison of  exclusion of the fittest and diversity}
    \label{fig:diversity}
\end{minipage}
}
\begin{flushleft}
    {\small 
     Probabilities of exclusion of the fittest (column A), beta diversities (column B), and species richness (column C) over the switching rate and the mean toxin sensitivities in two-species communities. Here, competitive exclusion refers the event where the slower-growing species excludes the faster-growing species.  Columns D and E show the beta diversity and species richness in ten-species communities, respectively. In the plots of probability of exclusion of the fittest and beta diversity, the black lines show the means and blue areas represent the probability distributions calculated from 10'000 simulations (100 beta diversity and each of them from 100 replicate runs). Each color in the species richness plots represents the proportion of 10'000 runs where at the end of the run there were that number of species surviving. Each row represents the mean toxin sensitivity of communities in those runs.}
    \end{flushleft}
\end{figure}
\clearpage
\begin{table}[tb]
    \centering
    \caption{Different scenarios of environmental switching}
    \begin{tabular}{|c|cccc|}\hline
         Scenario&$R_i\left(\xi=1\right)$&$R_i\left(\xi=-1\right)$&$T_j\left(\xi=1\right)$&$T_j\left(\xi=-1\right)$  \\ \hline
         1 & $R^+_i$ & $R^-_i$ & $\left<T_j\right>$ &$\left<T_j\right>$ \\ \hline
         2 & $\left<R_i\right>$ & $\left<R_i\right>$ & $T^-_j$ &$T^+_j$ \\ \hline
         3 & $R^+_i$ & $R^-_i$ & $T^-_j$ &$T^+_j$ \\ \hline
    \end{tabular}
    \label{tab:resoure-toxin-switch}
    \\
    \begin{flushleft}
    $R_i^+$, $R_i^-$, and $\left<R_i\right>$ represent abundant, scarce, and mean resource supplies, respectively.
    $R_i^+>R_i^-$ and  $\left<R_i\right>=\left(R_i^+ + R_i^-\right)/2$ for $i=1, \ldots,N/2$. Similar representation and relations hold for toxin supply $T_j$. In each condition, $\xi=1$ ($\xi=-1$) means mild (harsh) environment, respectively.
    \end{flushleft}
    
 \end{table}
\clearpage

\newpage
\setcounter{equation}{0}
\setcounter{section}{0}
\setcounter{figure}{0}
\setcounter{table}{0}
\setcounter{page}{1}
\appendix
\renewcommand{\theequation}{A.\arabic{equation}}
\renewcommand{\thefigure}{A.\arabic{figure}}
\renewcommand{\thetable}{A.\arabic{table}}
\renewcommand{\thesection}{Appendix \arabic{section}}
\renewcommand{\bibnumfmt}[1]{[A#1]}
\renewcommand{\citenumfont}[1]{A#1}
\begin{center}
{\Large Supplementary information}
\end{center}
\section{Details of analysis}
\subsection{Species interaction analysis}\label{sec:detail-simulation1}
Here, we summarize the details of the simulations and parameter values used for species interaction analysis in the main text. In the analysis of species interactions, we used the minimal model ($N=2$): two species compete for one resource and absorb and are killed by one toxin. We assumed that species 1 grows faster than species 2 but the other parameter values are identical (Table \ref{tab:prameter}).  In each run, the initial condition is either  $\left(r_1(0), t_1(0), s_1(0), s_2(0)\right) =\left(150, 100, 10, 0\right)$ or $\left(150, 100, 10, 10\right)$, where species 2 is absent or present, respectively. The initial environmental condition is $\xi=1$ with a probability of 0.5; otherwise $\xi=-1$. Each simulation continues until at time $\sigma_{end}=200$ when the distributions of species abundances converge to a quasi-stationary distribution and do not change for a long time. The extinction probability was estimated by running $10^5$ simulations for each condition. 
\par
In the main text, to investigate how species 2's effect on species 1 changes over the switching rate, we began by analyzing the difference in the extinction probability of species 1 in the presence/absence of species 2 as a proxy for species interactions:
\begin{align}
\Delta P\left(s_1(\sigma_{end})=0\right) \equiv P\left(s_1(\sigma_{end})=0;s_2(0)=0\right)- P\left(s_1(\sigma_{end})=0;s_2(0)>0\right).\label{eq:diffProb}
\end{align}
\par
\begin{figure}
    \centering
    \includegraphics[scale=0.6]{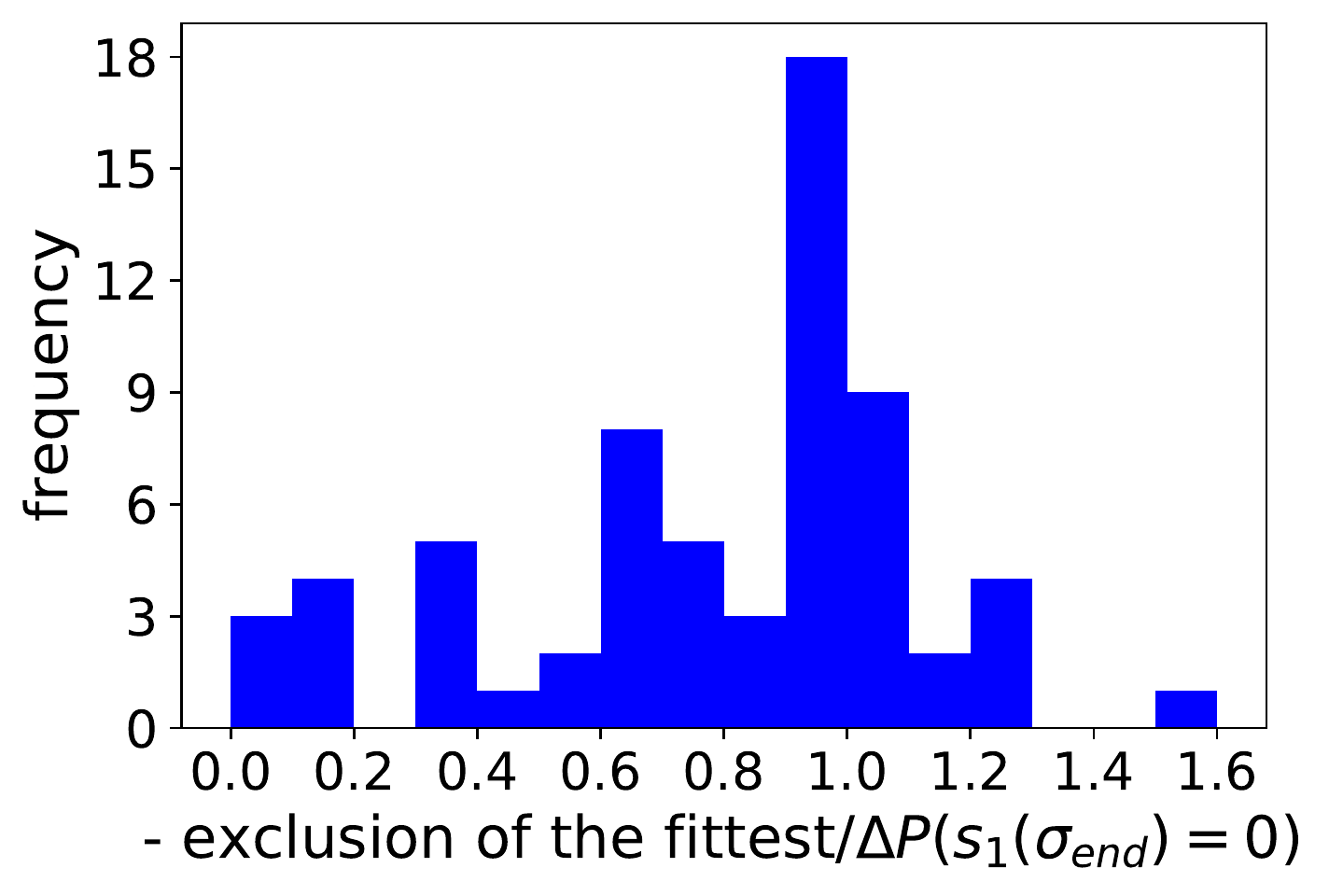}
    \caption{Probability of exclusion of the fittest relative to the difference in extinction probability of species 1}
    \label{fig:rel-comp}
    \begin{flushleft}
    {\small 
    A histogram of the ratio of the probability  of exclusion of the fittest and the difference in species 1's extinction  probabilities alone versus in the presence of species 2, showing the results of 81 sets of the environmental switching rate $\nu$ and the toxin sensitivity $\delta$ (9 values of $\nu=10^{-5}, \ldots, 10^3$ and 9 values of $\delta=0.1, \ldots, 0.9$). For each set of the parameter values, $10^5$ simulations were run to calculate the competitive exclusion probability and the difference in species 1's extinction probabilities in the presence/absence of species 2. In many of these 81 parameter sets, this ratio is close to 1, indicating that both measures yield similar results. As in the manuscript we focus on conditions leading to competition between the two species, we ignore toxin sensitivity $\delta=1.0$ where species 2's effect on species 1 can be positive.
    }
    \end{flushleft}
\end{figure}

Then, we moved to using the probability of exclusion of the fittest (probability that species 2 excludes species 1) instead:
\begin{align}
    \Delta P\left(s_1(\sigma_{end})=0\right)=&-\underbrace{
    P\left(s_1(\sigma_{end})=0, s_2(\sigma_{end})>0;s_2(0)>0\right)}_{\mbox{exclusion of the fittest}} \nonumber\\
    &+\left\{\underbrace{P\left(s_1(\sigma_{end})=0 ;s_2(0)=0\right)}_{\mbox{sp 1 goes extinct in mono-culture}}-\underbrace{P\left(s_1(\sigma_{end})=0, s_2(\sigma_{end})=0;s_2(0)>0\right)}_{\mbox{both species go extinct}}\right\}.
    \label{eq:decomposedDiff}
\end{align}
Although Fig.\ref{fig:interaction}A and C show that these two measures give similar results, Fig. \ref{fig:rel-comp} confirms this conclusion as the ratio of these two measures is around 1. In other words, the second line on the right-hand-side of Eq \eqref{eq:decomposedDiff} is ignorable. Intuitively, this is because the environment is very harsh when both species are likely to go extinct (i.e., small resource supplies and high toxin sensitivity); then species 1 is also likely to go extinct in mono-culture under such harsh environment. For this reason, the competitive exclusion probability (the first line on the right-hand-side of Eq \eqref{eq:decomposedDiff}) changes similarly to difference in extinction probability ($ \Delta P\left(s_1\left(\sigma_{end}\right)=0\right)$) over environmental switching rates and toxin sensitivities (Fig. \ref{fig:interaction}C), although they differ in their signs.
\par
We calculated 95\% of highest posterior density intervals (HDPIs) to see the uncertainty of species '1 extinction probabilities or the probability of exclusion of the fittest. As we were interested in whether species 1 goes extinct (or is excluded by species 2) or not in each simulation run, it is reasonable to assume these probabilities follow beta distributions. As a prior distribution, we assumed the following beta distribution
\begin{align}
    \mbox{Beta}\left(1,1\right),
\end{align}
which is equivalent to a uniform distribution between 0 and 1. After running $10^5$ simulations and observing species 1's extinction (or exclusion) $X$ times, the posterior probability distribution of species 1's extinction probability (or the probability of exclusion of the fittest) is given as follows:
\begin{align}
    \mbox{Beta}\left(X+1, 10^5-X+1\right).
\end{align}
To calculate the 95\% HDPIs, we sampled 10,000 samples from the posterior distributions and used pymc3.stats.hpd function. However, the 95\% HDPIs are very small and not observable in the main text due to the large number of simulations.

\subsection{Landscapes of exclusion of the fittest}\label{sec:landscape}
To understand how the switching rate and toxin sensitivity affect the probability of the exclusion of the fittest, we analyzed three cases in the absence of EFs (Fig. \ref{fig:CompExcl}), deriving three critical toxin sensitivities where the probability of exclusion of the fittest is maximized.
This analysis clarifies what happens at the two extreme switching rates. We see two critical toxin sensitivities (at $0.1$ and $0.8$ in Figs. \ref{fig:interaction}C and D) at $\nu \rightarrow 0$ that correspond to the long time spent with either scarce or abundant resources, while at $\nu \rightarrow \infty$, where resources remain at mean abundance, there is a single critical toxin sensitivity (at $0.4$ in Figs. \ref{fig:interaction}C and D). As the switching rate increases from one extreme to the other, the form of the probability of exclusion of the fittest in Fig. \ref{fig:interaction}C changes from bi-modal to uni-modal.
\par
We now see that the landscape of competitive exclusion in Fig. \ref{fig:interaction}C contains two ``mountain ranges''. The first includes two peaks corresponding to the critical toxin sensitivities under scarce and mean resources ($0.1, 0.4$, respectively). The peak at toxin sensitivity $0.1$ converges to the peak at $0.4$ (Fig.  \ref{fig:interaction}D) with increasing environmental switching. The second mountain range has a single peak corresponding to the critical toxin sensitivity under abundant resources ($0.8$), which vanishes by increasing the switching rate (Fig. \ref{fig:interaction}D).
At toxin sensitivities between the critical values under scarce and mean resource supplies ($\delta=0.2, 0.3$), the probability of exclusion of the fittest changes in a humped shape over the switching rate (species 1's difference in extinction probability changes in a U shape, see Fig. \ref{fig:interaction}B), as it passes over the first mountain range. When the toxin sensitivity is between the critical values under mean and abundant resources (e.g. $\delta=0.6$), the probability of exclusion of the fittest instead has a ``valley'' over the switching rate (the difference in extinction probability changes in a humped shape, see Fig. \ref{fig:interaction}B). At toxin sensitivities larger than the abundant resource critical value ($\delta > 0.8$), exclusion of the fittest is very unlikely because both species frequently go extinct (Fig. \ref{fig:CompExcl}C). 
\par 
In summary, the transition between the two extreme switching rates results in a highly rugged landscape. This means that the stochastic exclusion of the fittest species can be very high or very low at an intermediate switching rate (either in agreement or in contradiction with the intermediate disturbance hypothesis), depending on species' toxin sensitivities, our proxy for environmental harshness. This makes predicting the outcome at a given switching rate very difficult, as it is dependent on precise parameter values.
\subsection{Species diversity analysis}\label{sec:detail-simulation2}
In the community diversity analysis, we changed the number of species from $N=2$ to $N=10$. Some parameter values were not fixed in this analysis, and we sampled them from the following probability density functions:
\begin{subequations}
\begin{align}
    \mu_{ik} &\sim \mathcal{N}\left(1, 0.1^2\right), \\
    \delta_{jk} &\sim \mbox{Beta}\left(100\bar{\delta}, 100\left(1-\bar{\delta}\right)\right),\\
    K^r_{ik}, K^{t}_{jk} &\sim \mathcal{N}\left(100, 10^2\right).
 \end{align}
\end{subequations}
Here, each function is uni-modal with a mean of $1.0$ for $\mu_{ik}$, $\bar{\delta}$ for $\delta_{jk}$, and $100$ for $K^r_{ik}, K^{t}_{jk}$. For $\mu_{ik},K^r_{ik}$, and $K^{t}_{jk}$, the mean values are the same as in Table \ref{tab:prameter} and they are rarely negative due to the small variances. We set the mean of $\mu_{ik}$ as 1 so that $\mu_{ik}$ is likely to be larger than $\delta_{jk}$ unless $\bar{\delta}=0.99$: species would easily go extinct when $\mu_{ik}< \delta_{jk}$. The means of $K^r_{ik}$ and $K^{t}_{jk}$ are chosen so that amounts of resource and/or toxin inflows are larger (smaller) than these values under the abundant (scarce) supply condition. We expected that the growth and/or death rates of species would largely change depending on the environmental conditions with this setting. We sampled $\delta_{jk}$ from a beta distribution so that $0\leq \delta_{jk} \leq1$. $\delta_{jk}$ should be non-negative by definition and not be larger than $1$ because a large $\delta_{jk}$ is likely to drive species $k$ extinct.
Beta distribution satisfies these requirements regardless value of mean $\bar{\delta}$.
We systematically vary the value of the mean toxin sensitivity $\bar{\delta}$ to be $\bar{\delta}=0.1,0.2, 0.4, 0.6$ or $0.99$ in each set of simulations. We used $\bar{\delta}=0.99$ instead of $\bar{\delta}=1.0$ because the beta distribution did not generate different values of $\delta_{jk}$ with $\bar{\delta}=1.0$. For each number of species $N$ (2, 4, 6, 8 or 10) and $\bar{\delta}$, 100 sets of parameter values are sampled. With each parameter set,  we performed the simulation $100$ times until $\sigma_{end}=200$ at each value of the switching rate $\nu$. Then, we calculated the species richness (the number of surviving species) and beta diversity.
\par
In each run, initial resource amounts, species abundances, and toxin amounts are given by $\left(r_i(0), t_j(0), s_k(0)\right)=\left(150, 150, 10\right)$ for any $i,j,k$. As an  environmental switching scenario, we chose scenario 1 (Table \ref{tab:resoure-toxin-switch}) with $R_i^+=200$, $R_i^-=50$, and $\left<T_k\right>=125$ for any $i$ and $k$. The initial environmental condition is $\xi=1$ with probability of 0.5; otherwise $\xi=-1$.
\par
\begin{figure}
    \centering
    \rotatebox{90}{
    \includegraphics[scale=0.4]{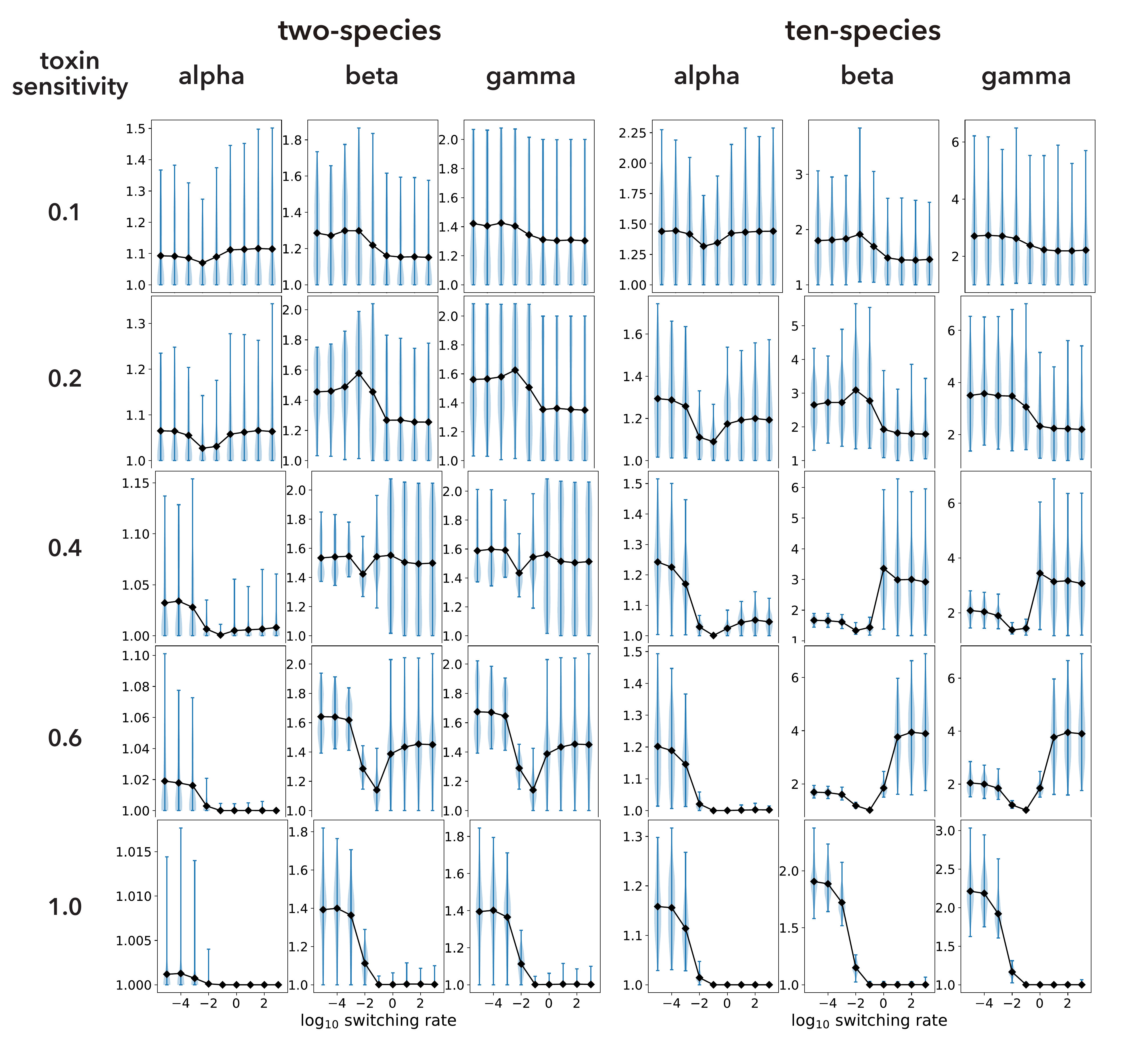}
    }
    \caption{Changes of alpha, beta, and gamma diversities}
    \label{fig:my_ABG}
    \begin{flushleft}
    {\small Alpha, beta, and gamma diversities over the switching rate and mean toxin sensitivity in two- and ten-species communities at the end of simulations. As alpha diversity is always closed to one, beta diversity and gamma diversity show similar trend. The black lines and blue areas represent the mean values and the probability distributions of the diversities calculated from 10'000 simulations. See \ref{sec:detail-simulation2} for more detail.
    }
    \end{flushleft}
\end{figure}

Then, at a quasi-stationary distribution ($\sigma_{end}$), we evaluated beta diversity and species richness. Beta diversity is calculated as follows:
\begin{align}
      ^1 D_{\beta} \left(\sigma_{end}\right)&\equiv  \frac{^1 D_{\gamma} \left(\sigma_{end}\right)}{ ^1 D_{\alpha} \left(\sigma_{end}\right)},
      \label{eq:beta_div}
\end{align}
with alpha and gamma diversities defined as below:
\begin{align}
     ^1 D_{\alpha} \left(\sigma_{end}\right) &\equiv  \exp\left(-\sum^{100}_{l=1}\sum^{N}_{k=1}w_lp_{lk} \left(\sigma_{end}\right)\ln p_{ln} \left(\sigma_{end}\right)\right)\label{eq:alpha-div}, \\
     ^1 D_{\gamma} \left(\sigma_{end}\right) &\equiv  \exp\left(-\sum^{N}_{k=1}\bar{p}_{k}\ln \bar{p}_{k} \left(\sigma_{end}\right)\right). \label{eq:gamma-div} 
\end{align}
 $w_l$ is a weight for community $l$ calculated by size of community $l$ (sum of species abundances in community $l$ relative to the sum of community sizes over $l)$, $p_{lk}$ is the relative abundance of species $k$ in community $l$ (i.e., in community $l$, $p_{lk} \left(\sigma_{end}\right)=s_k \left(\sigma_{end}\right)/\sum_k s_k \left(\sigma_{end}\right)$), and $\bar{p}_{k}=\sum_lw_l p_{lk}$ is  the mean relative abundance of species $k$ among communities $l=1, \ldots, 100$.  If all species go extinct in community $l$, it does not affect alpha, beta and gamma diversities as $w_l=0$. If all species go extinct in all communities, beta diversity becomes $^1D_\beta  \left(\sigma_{end}\right)=1$. Fig. \ref{fig:my_ABG} represents how alpha, beta, and gamma diversities change over the switching rate and mean toxin sensitivities.  
 
\section{Analysis excluding noise}\label{sec:appendix2}
\subsection{Deterministic scenarios (absence of DN and EFs)}\label{sec:analysisODE}
In this section, we analyse the equilibrium state of a two-species, one-resource, and one-toxin model, in the absence of environmental switching and demographic noise. Although an equilibrium state cannot be analytically obtained, we shall see that there exists at most one stable and feasible equilibrium state.
By removing demographic noise and environmental switching from Eq \eqref{eq:master_equation} in the main text, the dynamics are governed by the following ordinary differential equations:
\begin{subequations}
\begin{align}
    \dot{r_1}&= \alpha\left(R_1-r_1\right)-\sum_{k=1,2}\frac{\mu_{1k}}{Y^r_{1k}}\frac{r_1}{r_1+K^r_{1k}}s_k, \label{eq:resource}\\
    \dot{t_1}&= \alpha\left(T_1-t_1\right)-\sum_{k=1,2}\frac{\delta_{1k}}{Y^t_{1k}}\frac{t_1}{t_1+K^t_{1k}}s_k, \label{eq:toxin} \\
    \dot{s_k}&= \left(\mu_{1k}\frac{r_1}{r_1+K^r_{1k}}-\delta_{1k}\frac{t_1}{t_1+K^t_{1k}}-\alpha\right)s_k, \label{eq:species}
\end{align}
\end{subequations}

where the dot denotes the time derivative.
In spite of the simplicity of the model, we learn from \eqref{eq:resource} -\eqref{eq:species}
that, in the absence of consumption by the species,  the resources and toxins relax towards the in-flowing values $R_1$ and $T_1$ on a timescale of order $1/\alpha$. 
Furthermore, we infer from \eqref{eq:resource} -\eqref{eq:species} that there are feedback loops between species abundances and resource and toxin concentrations (see Fig. \ref{fig:cartoon}A): as $s_k$ increases, both $r_1$ and $t_1$
decrease, see \eqref{eq:resource} and \eqref{eq:toxin}, which in turn can result either in a decrease (negative feedback loop) or in an increase of $s_k$ (positive feedback loop), depending on the sign of the parenthesis on the right-hand-side of 
\eqref{eq:species}.

\par
When only species $k$ persists in the system, a feasible\footnote{Here, feasibility means $r_{1k}^*, t^*_{1k}, s_k^*>0$} equilibrium state of Eqs \eqref{eq:resource} -\eqref{eq:species}, $\left(r_1,t_1,s_k\right)=\left(r^*_{1k},t^*_{1k}, s^*_k\right)$, should satisfy below:
\begin{subequations}
\begin{align}
    \alpha\left(R_1-r^*_{1k}\right) &= \frac{\mu_{1k}}{Y^r_{1k}}\frac{r_{1k}^*}{r_{1k}^*+K^r_{1k}}s_k^*, \\
        \alpha\left(T_1-t_{1k}^*\right) &= \delta_{1k}Y^t_{1k}\frac{t_{1k}^*}{t_{1k}^*+K^t_{1k}}t_k^*, \\
    \mu_{1j}\frac{r_{1k}^*}{r_{1k}^*+K^r_{1k}} &= \delta_{1k}\frac{t_{1k}^*}{t_{1k}^*+K^t_{1k}} + \alpha \label{eq:obtaining_species}.
\end{align}
\end{subequations}
By rearranging the first and second equations, we see that they represent quadratic functions of $r^*_{1k}$ and $t^*_{1k}$:
\begin{align}
    -r_{1k}^{*2} +\left\{R_1-K^r_{1k}-\mu_{1k} s_k^*/(\alpha Y^r_{1k})\right\}r_{1k}^*+K^r_{1k}R_1&=0,\label{eq:root_resource}\\
     -t_{1k}^{*2} +\left\{T_1-K^t_{1k}-\delta_{1k} s_k^*/(\alpha Y^t_{1k})\right\}t^*_{1k}+K^t_{1k}T_1&=0. \label{eq:root_toxin}
\end{align}
Notice that $K^r_{1k}R_1$ and $K^t_{1k}T_1$ are positive. This implies that the above equations always have a unique positive root. In other words, we can obtain unique solutions for $r_{1k}^*$ and $t_{1k}^*$ once we obtain $s_k^*$. 
\par
By substituting the positive roots of Eqs \eqref{eq:root_resource} and \eqref{eq:root_toxin} into Eq \eqref{eq:obtaining_species}, we obtain the following equation whose positive root is $s^*_k$:
\begin{align}
f(s) &= \frac{1}{2\alpha}\left\{\left(-\mu_{1k}+2\alpha+\delta_{1k}\right)s+\sqrt{Q_1\left(s\right)}-\sqrt{Q_2\left(s\right)}+\alpha\left(-K^r_{1k}Y^r_{1k}+ K^t_{1k}Y^t_{1k}- Y^r_{1k}R_1 + Y^t_{1k}T_1\right)\right\}\label{eq:funcy}
\end{align}
where $Q_1(s)$ and $Q_2(s)$ are quadratic functions of $s$:
\begin{subequations}
\begin{align}
    Q_1\left(s\right)&= \left\{\mu_{1k} s + \alpha Y^r_{1k}\left(K^r_{1k}-R_1\right)\right\}^2+4\alpha Y^r_{1k}K^r_{1k}R_1>0\\
    Q_2\left(s\right)&= \left\{\delta_{1k} s + \alpha Y^t_{1k}\left(K^t_{1k}-T_1\right)\right\}^2+4\alpha Y^t_{1k}K^t_{1k}T_1>0.
\end{align}
\end{subequations}
Notice that $f(s)$ always has a root $s=0$ because
\begin{align}
f(0) &= \frac{1}{2\alpha} \left\{\sqrt{Q_1(0)}-\sqrt{Q_2(0)}+\alpha\left(-K^r_{1k}Y^r_{1k}+ K^t_{1k}Y^t_{1k}- Y^r_{1k}R_1 + Y^t_{1k}T_1\right)\right\} \nonumber \\
&=\frac{1}{2\alpha}\left\{\alpha Y^r_{1k}\left(K^r_{1k}+R_1\right)-\alpha Y^t_{1k}\left(K^t_{1k}+T_1\right)+\alpha\left(-K^r_{1k}Y^r_{1k}+ K^t_{1k}Y^t_{1k}- Y^r_{1k}T_1 + Y^t_{1k}T_1\right)\right\}\nonumber\\
&=0.
\end{align}
Although Newton's method numerically provides a root of $f(s)$, this root depends on the initial value used in Newton's method. In addition, as $f(s)$ has root $s=0$, Newton's method may provide this root with various initial values, which does not always mean that $f(s)$ does not have positive roots (i.e., $s^*_k$). In other words, it is recommended to investigate how many positive roots $f(s)$ has before using Newton's method.
\par
To investigate the number of $f(s)$'s positive roots, it is useful to obtain $df/ds$:
\begin{align}
    \frac{df}{ds}=\frac{1}{2\alpha}\left\{\left(-\mu_{1k}+2\alpha+\delta_{1k}\right)+\frac{dQ_1/ds}{2\sqrt{Q_1\left(s\right)}}-\frac{dQ_2/ds}{2\sqrt{Q_2\left(s\right)}}\right\}.\label{eq:dfdy}
\end{align}
Although it is analytically difficult to obtain the solution(s) of $df/ds$, we can obtain the maximum number of positive roots of $f(s)$ by analyzing the number of $df/ds$'s sign changes. Notice that $dQ_1/ds$ and $dQ_2/ds$ are linear functions of $s$:
\begin{subequations}
\begin{align}
    \frac{dQ_1}{ds}&=2\mu_{1k}\left\{\mu_{1k}s+\alpha Y^r_{1k}\left(K^r_{1k}-R_1\right)\right\}, \\
    \frac{dQ_2}{ds}&=2\delta_{1k}\left\{\delta_{1k}s+\alpha Y^t_{1k}\left(K^t_{1k}-T_1\right)\right\}.
\end{align}    
\end{subequations}
In addition, $Q_1(s)$ and $Q_2(s)$ are always positive. Then, the second and third terms of Eq \eqref{eq:dfdy} change their sign at most once by increasing $s$. The maximum number of $df/ds$'s sign change is, therefore, two. This implies that the maximum number of positive roots of $f(s)$ is also two. To obtain the exact number of positive roots of $f(s)$, it is necessary to numerically calculate the root(s)  of $df/ds$. Substituting the root(s) into Eq \eqref{eq:funcy} and calculating the sign of $f(s)$, it is possible to obtain the exact value of $s_k^*$. 
\par
\begin{figure}[p]
    \centering
    \includegraphics[scale=0.5]{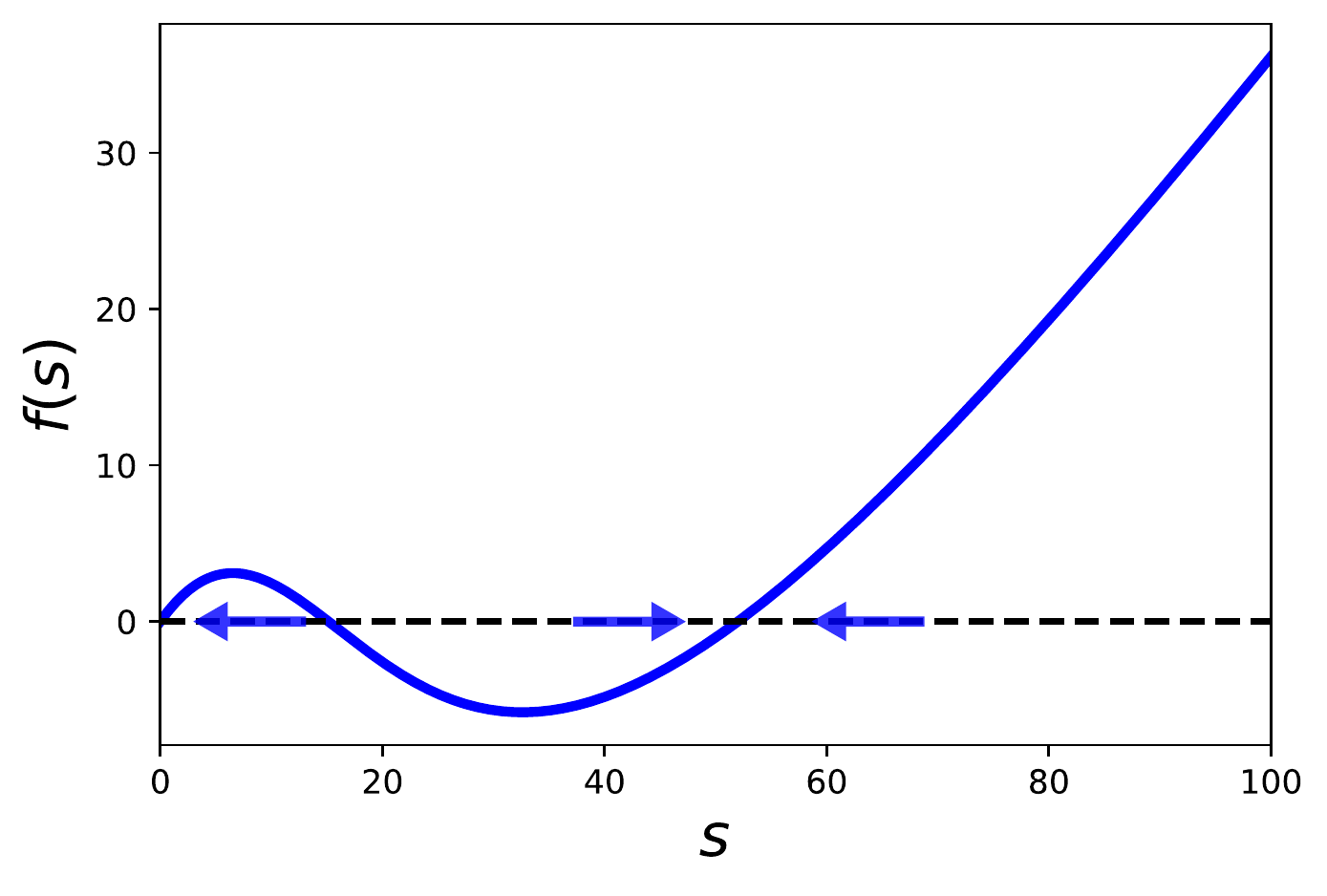}
    \caption{Feasible and stable equilibrium state}
    \label{fig:FeasbleStable}
    \begin{flushleft}
    {\small An example of $f(s)$ in Eq \eqref{eq:funcy} and equilibrium states. In this example, $f(s)$ has three roots: $s=0$, and two feasible equilibria. The blue arrows indicate that $s$ decreases or increases when $f(s)$ is positive or negative, respectively. As $df/ds$ is negative at the left equilibrium state, this equilibrium state is unstable. On the other hand, the right feasible equilibrium has a positive $df/ds$ and thus this equilibrium can be stable. Note that the equilibrium state corresponding to $s=0$ can be also stable. Parameter values are $\delta_{1k}=1.2$, $R_1=200$, $T_1=125$ and as in Table \ref{tab:prameter} otherwise.}
    \end{flushleft}
\end{figure}

Once we have obtained a feasible equilibrium state $\left(r_{1k}^*,s_{k}^*, t^*_{1k}\right)$, it is necessary to analyze the stability of this equilibrium state. Although the stability analysis requires the evaluation of the Jacobian matrix at each equilibrium state, we can see that there exists at most one feasible and stable equilibrium state without such a stability analysis. Notice that:
\begin{align}
    \dot{s_j}&\lessgtr 0 \nonumber \\
    \Leftrightarrow \mu_{1k}\frac{r_{1k}^*}{r_{1k}^*+K^r_{1k}} &\lessgtr \delta_{1k}\frac{t_{1k}^*}{t_{1k}^*+K^t_{1k}} + \alpha\nonumber\\
     \Leftrightarrow f\left(s\right)&\gtrless0.
\end{align}
These inequalities imply that a stable equilibrium state satisfies the following inequality:
\begin{align}
    \left.\frac{df}{ds}\right|_{s=s^*_k}>0.
\end{align}
Although there can exist at most two feasible equilibria, only one of them satisfies the above inequality (Fig. \ref{fig:FeasbleStable}). The number of feasible and stable equilibria is, therefore, one at most.
\par
\begin{figure}
    \centering
    \includegraphics[scale=0.4]{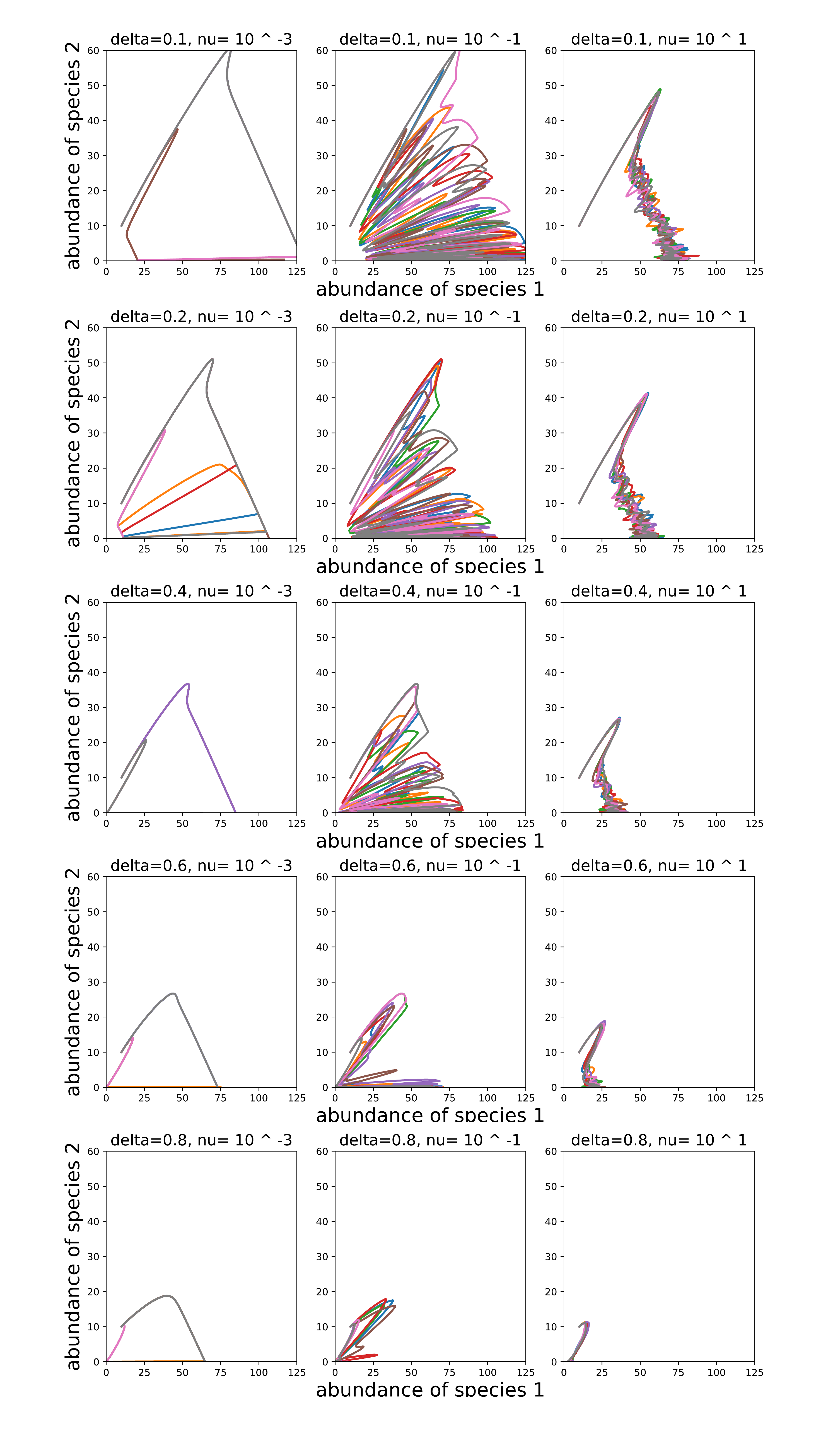}
    \caption{Examples of the dynamics with only the environmental fluctuations}
    \label{fig:ex_dynamics_onlyEF}
    \begin{flushleft}
    {\small
    State transition of two species abundances $(s_1, s_2)$ in the absence of DN but the presence of the EFs are shown. Here, EFs  switch resource supply (scenario 1 in Table \ref{tab:resoure-toxin-switch}) and the initial population abundances are $s_1(0)=s_2(0)=10$. In this setting, species 2 always goes extinct ($\lim_{\sigma \rightarrow \infty} s_2(\sigma)=0$) regardless of the values of $\delta$ and $\nu$. On the other hand, species 1 survives ($\lim_{\sigma \rightarrow \infty} s_1(\sigma)>0$) if the environment is not too harsh. In each panel, different colors represent different samples of the dynamics with EFs alone. The values of $\nu$ and $\delta$ are shown on the top of each panel and the rest parameter values are shown in Table \ref{tab:prameter}.}
    \end{flushleft}
\end{figure}

For the sake of simplicity but without loss of generality, in the main text, we assumed that the maximum growth rate of species 1 is larger than species 2 but the remaining parameter values are identical, resulting in a  \textit{per-capita} growth rate of species 1 always larger than species 2. In this setting, in the absence of DN,  species 2 always goes extinct after a finite time while species 1 persists if it has a feasible and stable equilibrium state. It is worth noting that this feature also characterizes the two-species, one-resource and one-toxin model in the presence of environmental switching without DN: in this case species 2 always goes extinct in a finite time, and at equilibrium one has either $s_1^*=s_2^*=0$ (extinction of both species) or $s_1^*>0, s_2^*=0$ (survival of species 1, extinction of species 2), see Fig. \ref{fig:ex_dynamics_onlyEF}.

\subsection{With environmental fluctuations alone}\label{sec:OnlyEF}

It is also instructive to analyze the long-time dynamics of the  
 two-species, one-resource, one-toxin model in the presence of EFs and without DN. This describes the dynamics of a sufficiently large community in which 
 DN is always negligible and whose time evolution is described by Eqs \eqref{eq:resource}-\eqref{eq:species}, but now with $(R_1,T_1)=(R_1(\xi),T_1(\xi))$  switching randomly with rate $\nu$ between two states $\xi =\pm 1$, which is applicable to all three scenarios in Table \ref{tab:resoure-toxin-switch}. Here, for the sake of simplicity and concreteness, we consider the scenario 1 of Table \ref{tab:resoure-toxin-switch} with $1\ll R_1^-<\left<R_1\right><R_1^+$, while $T_1=\left<T_1\right>$ does not vary with the environment. In this case, Eqs.~\eqref{eq:resource}-\eqref{eq:species} are a system of stochastic differential equations (SDEs).
 In order to appreciate the effect of EFs on DN, it is useful to 
 study the total population size $n \equiv r_1 +t_1+s_1+s_2$ which, according to \eqref{eq:resource}-\eqref{eq:species}, obeys 
 
\begin{align}
\label{eq:totalpop}
\dot{n} = \alpha \left(R_1(\xi) + \langle T_1\rangle-n\right)-2\sum_{k=1,2} \delta_{1,k}~\frac{t_1 s_k}{t_1+K_{1,k}^t},
\end{align}
where, as in the main text, $Y_{ik}^r= Y_{jk}^t=1$. Importantly, $\sqrt{n}$ gives the intensity of DN in a total population of size $n$. Eq \eqref{eq:totalpop} is, however, difficult to analyze, due to the nonlinear coupling of $t_1$ with $s_k$ whose dynamics, according to Eqs \eqref{eq:toxin}-\eqref{eq:species}, also depend on $\xi$. In order to obtain some insight into the stationary probability density $p(n)$ of the population, we have thus we analyzed 
$\hat{n} \equiv r_1-t_1+s_1+s_2$ 
 which is related to the total
population size, since $\hat{n}=n-2t_1$. In this simple example $\hat{n}$, according to Eqs. \eqref{eq:resource}-\eqref{eq:species}, obeys the following linear SDE
\begin{align}
\label{eq:totalpop-eq}
\dot{\hat{n}} = \alpha \left(R_1(\xi) - \left<T_1\right>-\hat{n}\right).
\end{align}

\begin{figure}
    \centering
    \includegraphics[scale=0.3]{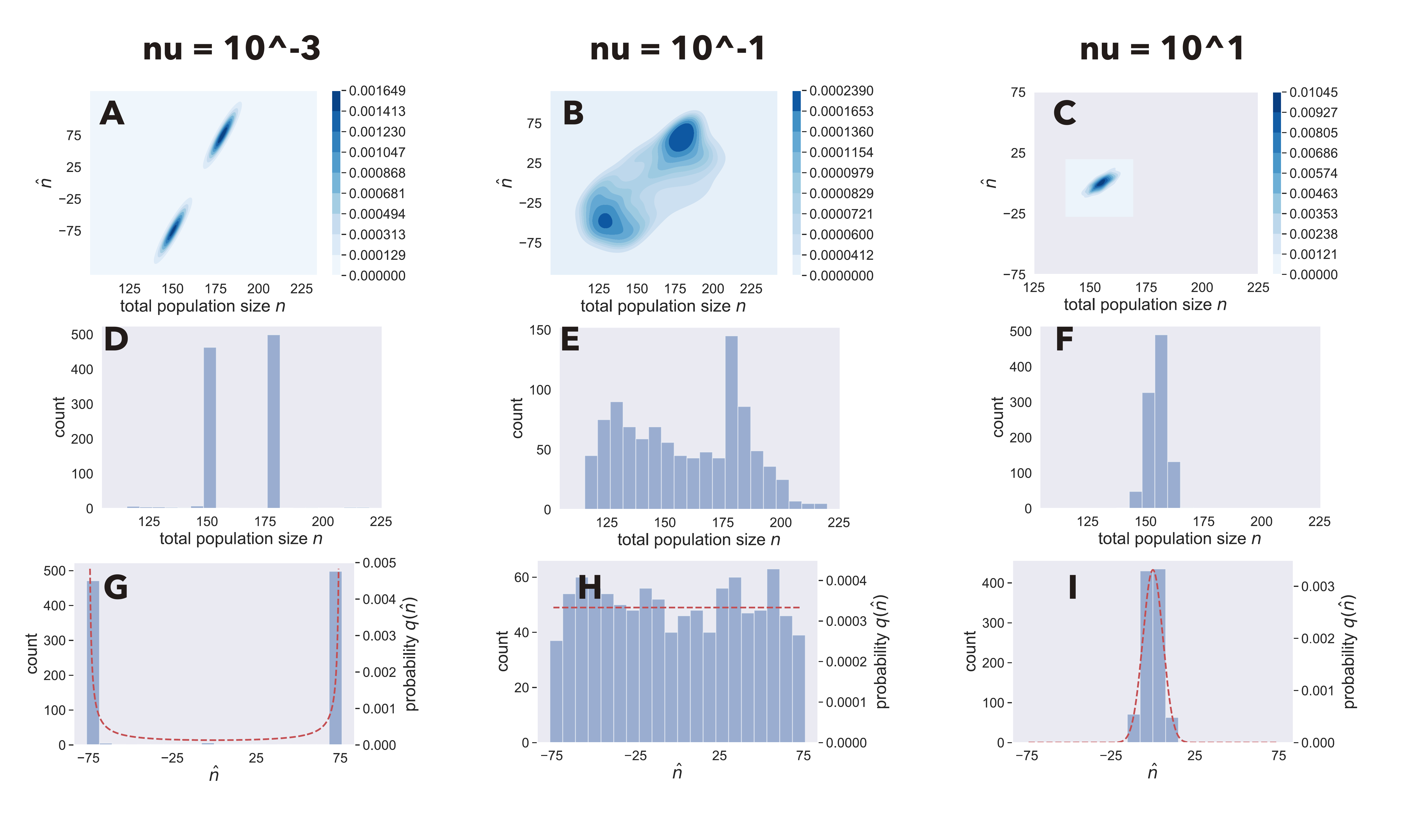}
    \caption{Distributions of total population sizes}
    \label{fig:PDMP}
    \begin{flushleft}
    {\small
    Probability distributions of total population size $n=r_1+t_1+s_1+s_2$ and the  auxiliary quantity  $\hat{n} = r_1-t_1+s_1+s_2$ obtained from 1,000 runs of Eqs \eqref{eq:resource}-\eqref{eq:species} with environmental switching of scenario 1 in Table \ref{tab:resoure-toxin-switch} at the slow $\nu=10^{-3}$ (first column), intermediate $\nu=10^{-1}$ (second column), or fast $\nu=10^1$ (third column) switching rates.  We collected the simulation data at time $\sigma_{end}=200$. At the beginning of each simulation, $\xi=1$ with 50 percents; otherwise $\xi=-1$. (A-C): the contour plots show the joint probability distributions of $n$ and $\hat{n}$: large $n$ corresponds to large $\hat{n}$. (D-F): the histograms show the distributions of the total population size $n$. (G-I): the histograms of $\hat{n}$ (left y-axis) and its theoretical probability distributions $q_{\nu/\alpha}\left(\hat{n}\right)$ (right y-axis), which is given by Eq\eqref{eq:PDMPdens}, are shown in blue bars and red dashed lines, respectively. We used $\delta_{1,1}=\delta_{1,2}=0.2$ and all other  parameter values are shown in Table \ref{tab:prameter}, and thus $\nu/\alpha=0.01$ corresponds to the slow switching rate (first column), $\nu/\alpha=1$ corresponds to the intermediate switching rate (second column), and $\nu/\alpha=100$ corresponds to the fast switching rate (third column), respectively, see \ref{sec:OnlyEF}. Exceptionally, we used $\nu/\alpha =80$ instead of $\nu/\alpha=100$ to show $q_{\nu/\alpha}\left(\hat{n}\right)$ in the third column because $\nu/\alpha=100$ causes overflow during the calculation of $q_{\nu/\alpha}\left(\hat{n}\right)$, but this modification does not change the qualitative feature of $q_{\nu/\alpha}\left(\hat{n}\right)$.}
    \end{flushleft}
\end{figure}

As seen above, the system reaches its equilibrium and $(s_1,s_2,t_1)\to (s_1^*,s_2^*=0,t_1^*)$ after a finite time while the environment varies endlessly .The SDE \eqref{eq:totalpop-eq} defines a simple piecewise deterministic Markov process (PDMP) \citep{Davis1984a}
whose (marginal) probability density $p_{\nu/\alpha}(\hat{n})$ can be obtained analytically, see \citet{Bena2006} and \citet{Horsthemke2006}:

\begin{align}
 \label{eq:PDMPdens}
q_{\nu/\alpha}(\hat{n})={\cal Z}~\left[\left\{R_1^+ -\left<T_1\right>-\hat{n}\right\}\left\{\hat{n}- R_1^- + \left<T_1\right>\right\}\right]^{\frac{\nu}{\alpha}-1},
\end{align}

where ${\cal Z}$ is the normalization constant. Hence,
in the absence of DN, with environmental switching as the sole source of randomness, the dynamics leads to a population consisting of only individuals of species 1 (species 2 is wiped out) and toxin and resources, with respective abundances $s_1,  r_1,  t_1$, in a fluctuating population whose scaled size 
$\hat{n}=s_1+ r_1- t_1$
is distributed according to   $q_{\nu/\alpha}(\hat{n})$ of finite support $\left[R_1^- - \left<T_1\right>,R_1^+ -\left<T_1\right>\right]$ (i.e. $R_1^- - \left<T_1\right>\leq n \leq R_1^+ - \left<T_1\right>$).
While this PDMP probability density 
ignores DN, it is known to generally provide a useful approximate description of how the quasi-stationary (at finite time $\gtrsim \sigma_{end}$) distribution varies with the environment in a large yet finite system, see, {\it e.g.}, 
\citep{Wienand2017a,Wienand2018,West2019a,Taitelbaum2020}. Here, Eq. \eqref{eq:PDMPdens}
readily sheds light on the effect of $\nu$ and $\alpha$ on 
$\hat{n}=n-2t_1$. It is clear from Eqs.~\eqref{eq:resource}-\eqref{eq:species}, that $t_1$  varies with $\xi$
and thus the marginal probability density of population size, $p(n)$, cannot be immediately obtained
from \eqref{eq:PDMPdens}. Yet,
we can obtain useful information about how
the population varies with  $\nu$ and $\alpha$
by focusing on the fast and slow switching regimes
(see Fig. \ref{fig:PDMP}):

\begin{itemize}
\item when $\nu/\alpha \ll 1$ (slow varying environment),  
$q_{\nu/\alpha}(\hat{n})$ is bimodal
and $\hat{n}$ is as likely to fluctuate about 
$R_1^+ -\langle T_1\rangle$
or $R_1^- -\langle T_1\rangle$. Hence, when in this slowly switching regime, 
the population size fluctuates with the same probability either 
about $R_1^- -\langle T_1\rangle+ 2t_1^*(\xi=-1)$ or $R_1^+ -\langle T_1 \rangle+ 2t_1^*(\xi=+1)$ (see, the first column of Fig. \ref{fig:PDMP}).
 \item when $\nu/\alpha \gg 1$ (fast changing environment),  $q_{\nu/\alpha}(\hat{n})$ is unimodal
and $\hat{n}$ fluctuates about its average, i.e. 
$\hat{n}\approx \langle R_1 \rangle - \left<T_1\right> $. Hence,
in this fast switching regime, the population
size 
fluctuates about 
$n\approx \hat{n} +2\langle t_1\rangle=\langle R_1 \rangle - \langle T_1\rangle +2\langle t_1\rangle$, where $\langle t_1\rangle=(t_1^*(\xi=1)+t_1^*(\xi=-1))/2$ and $t_1^*(\xi=\pm 1)$ are obtained from the equilibria of Eqs. \eqref{eq:resource}-\eqref{eq:species} with $R_1=R_1(\xi=\pm 1)$ (see, the third column of Fig. \ref{fig:PDMP}).
\end{itemize}
\par
This picture is confirmed by the simulation results reported in Fig. \ref{fig:PDMP}, where we see that the marginal probability densities $p(n)$ and $q_{\nu/\alpha}(\hat{n})$ have qualitatively the same features: both are bimodal and have two well-separated sharp peaks
when $\nu/\alpha \ll 1$, and a single pronounced peak when $\nu/\alpha \gg 1$. At intermediate values of $\nu/\alpha$, the probability densities of $n$ and $\hat{n}$ are much broader  with a generally flat profile (exhibiting one or two ``bumps''), see the second column of Fig. \ref{fig:PDMP}.
This analysis, which can be readily extended to $N>2$ species and to other scenarios of environmental fluctuations,
clearly shows that the population size can greatly vary as the environment changes. In particular, Eq \eqref{eq:PDMPdens} shows that when $\nu$ is low or $\alpha$ is high, half of the simulation runs lead to communities of ``small sizes'' where the effect of DN is expected to be significantly larger than in communities obtained
in the faster switching regime  ($\nu/\alpha \gg 1$) when $R_1^+\gg R_1^-$. 
\par

It is also worth noting from Eqs \eqref{eq:resource}-\eqref{eq:species} and Eq \eqref{eq:totalpop-eq} that if, say, only the maximum growth rates were subject to environmental switching, i.e. $\mu_{ik}=\mu_{ik}(\xi)$
with all other parameters kept constant, 
we always  obtain a large constant population size (if $R_1\gg T_1$) and thus no DN-EFs coupling because the distributions of $n$ and $\hat{n}$ are independent on $\mu_{ik}$. 
On the other hand, if only the maximum death 
 rates were subject to environmental switching, i.e. $\delta_{jk}=\delta_{jk}(\xi)$,
 the DN-EFs coupling would result from a complicated set of coupled stochastic differential equations obtained from \eqref{eq:resource}-\eqref{eq:species} with $\delta_{jk} \to \delta_{jk}(\xi)$. These analyses clarify the significant difference of our model from others. Some previous studies (e.g., \citet{GilesLeigh1981, Kalyuzhny2015} and multi-species model of \citet{Engen1996}) do not include DN-EFs coupling because they assume that EFs affect species' growth rates, but total species abundances or maximum population sizes do not change. Other studies (e.g., \citet{Kamenev2008,Chisholm2014,Fung2015} and the single species model of \citet{Engen1996}) include a form of DN-EF coupling because EFs in their model change both species' growth rates and population sizes. However, these models consist of only one species, and hence do not consider interspecific interactions. Our model includes both DN-EFs and indirect species interactions (resource competition and facilitation via detoxification, see Fig. \ref{fig:cartoon}A), and thus we can analyze how DN-EFs coupling affect species diversity as in Fig. \ref{fig:diversity}. In summary, here we have attempted to make the simplest choice to couple
 DN and EF in a transparent and biologically-relevant way.

\section{Alternative environmental switching scenarios}
\label{sec:alternative}

In the main text, environmental switching affects only the resource supply (scenario 1) while the amount of toxin supply is fixed. Here, the results of other environmental switching scenarios are shown: In scenario 2, environmental switching affects only the toxin supply, while both resource and toxin supplies change and correlate negatively in scenario 3 (see Table \ref{tab:resoure-toxin-switch}). 
\par
\begin{figure}
    \centering
    \includegraphics[scale=0.5]{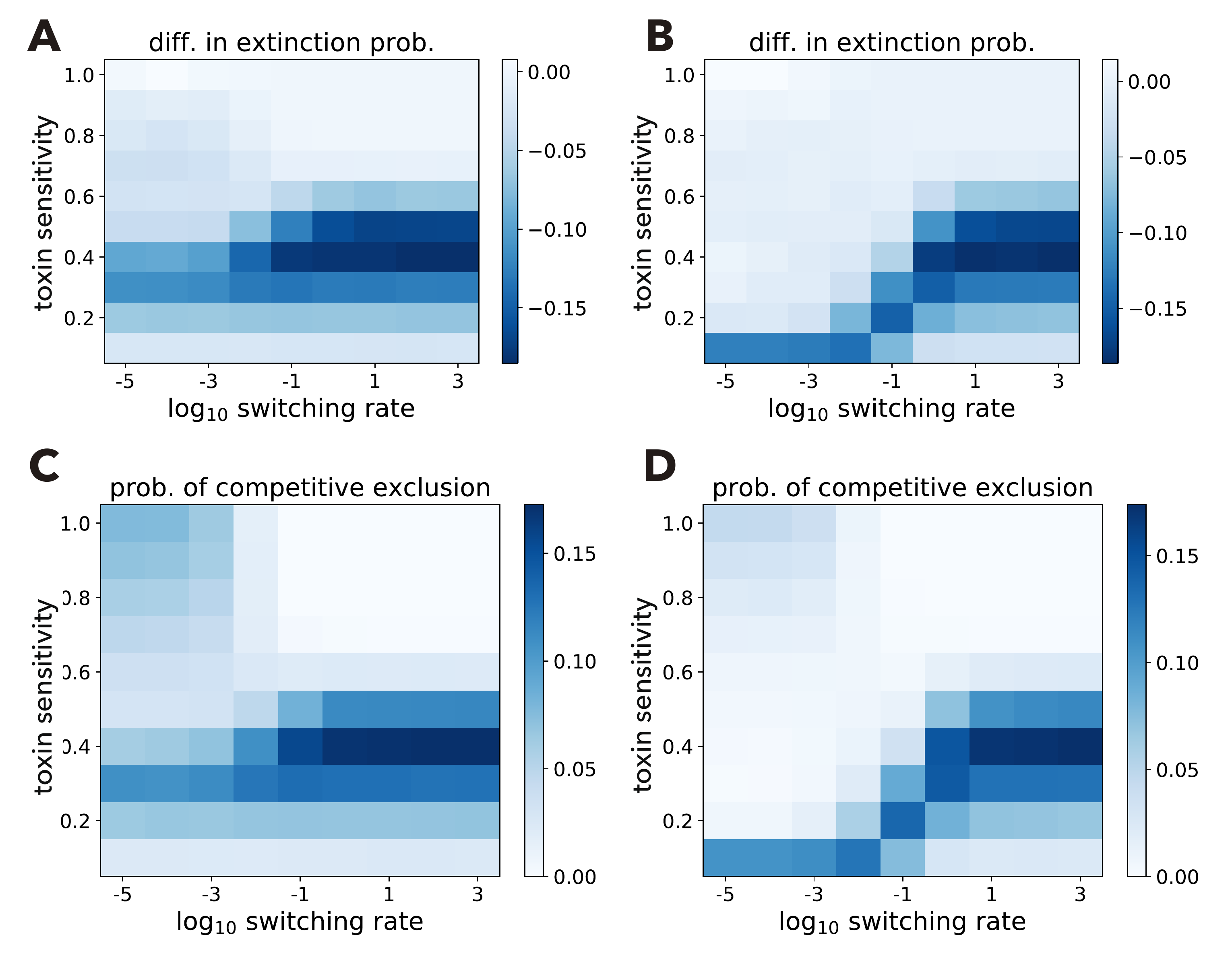}
    \caption{Effects of the environmental switching rate in alternative scenarios}
    \label{fig:alternative_scenario}
    \begin{flushleft}
    {\small
    Examples of the effect of switching rate in alternative scenarios. In the left column (A and C), toxin supply is switching (scenario 2), while both resource and toxin supplies switch and are negatively correlated (scenario 3) in the right column (B, D).
    A and B: difference between extinction probabilities in absence and presence of species 2. C and D: competitive exclusion probability. Parameter values: $R_1^+=200$, $R_1^-=50$, $T^+_1=200$, and $T^-_1=50$.}
    \end{flushleft}
\end{figure}

In both scenarios 2 and 3, the difference in species 1's extinction probability $\Delta P\left(s_1(\sigma_{end})=0\right)$ is very similar to the negative value of the probability of exclusion of the fittest when the sign of $\Delta P\left(s_1(\sigma_{end})=0\right)$ is negative (Fig. \ref{fig:alternative_scenario}). This once again confirms that we can use one measure for the other. 
\par
In addition, in both scenarios, the probability of exclusion of the fittest are bi-modal across toxin sensitivities at very slow environmental switching $\nu=10^{-5}$, but uni-modal at very fast environmental switching ($\nu=10^3$) (Fig. \ref{fig:alternative_scenario}C, D). As explained in the main text, when $\nu\to 0$, there are no switches and the environmental state is randomly allocated to harsh or mild conditions at $t=0$ with the same probability (the mean of $\xi$ is zero), yielding a bi-modal distribution at low switching rate. In the limit $\nu\to \infty$, there are so many switches that environmental noise averages out, i.e. $\xi$ is replaced by its mean (that is zero). This results in a uni-modal distribution when $\nu\gg 1$. 
\par
\begin{figure}
    \centering
    \includegraphics[scale=0.6]{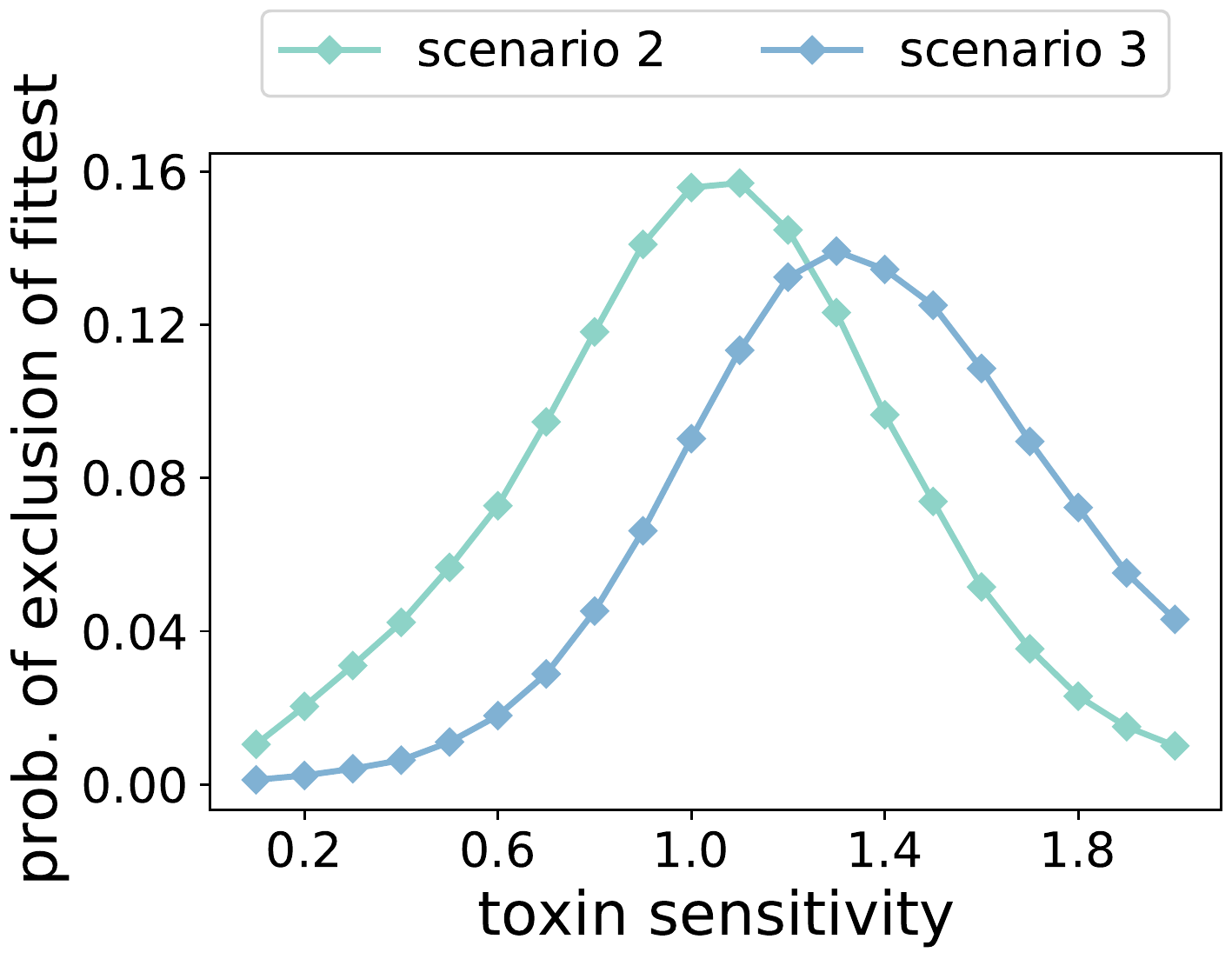}
    \caption{Critical toxin sensitivities under mild environments}
    \label{fig:critical_appendix}
    \begin{flushleft}
        {\small
        The critical toxin sensitivities (i.e., toxin sensitivity that maximizes the  probability of exclusion of the fittest in the absence of environmental switching) under the mild environments (scenario 2 :scarce toxin supply $T_1^{-}=50$, and scenario 3: abundant resource supply $R_1^+=200$ and scarce toxin supply) are $>1$. 
        }
    \end{flushleft}
\end{figure}

 In scenario 2, however, we do not observe any non-monotonic changes over the switching rate (Fig. \ref{fig:alternative_scenario}C). This is because the critical toxin sensitivity under abundant toxin supply ($\delta=0.3$) is close to that under mean toxin supply ($\delta=0.4$). In contrast, we do observe non-monotonic effects of the switching rate in scenario 3 (Fig. \ref{fig:alternative_scenario}D): when $\delta=0.2$, an intermediate switching rate ($\nu=10^{-1}$) shows the minimum difference in extinction probability, and the maximum probability of competitive exclusion. Although in scenario 3 the same intermediate switching rate minimizes the probability of competitive exclusion at toxin sensitivity $0.6$, the same non-monotonic effect is not observed in the difference in extinction probabilities (Fig. \ref{fig:alternative_scenario}B). Note that the critical toxin sensitivities under the mild environments in scenarios 2 (scarce toxin supply) and 3 (abundant resource supply and scarce toxin supply) are slightly larger than 1.0 (Fig. \ref{fig:critical_appendix}) and therefore not visible in Fig. \ref{fig:alternative_scenario}D. Table \ref{tab:sum_criticall} shows the critical toxin sensitivities in each scenario.

\section{Effects of resource supply}\label{sec:ChangeResourceSupply}
In this section, we again focus on environmental switching scenario 1 (changing only resource supply). We will see that the amount of resource supplies $R_1(\xi)$ changes the critical toxin sensitivities under scarce, mean, and abundant, resource supplies, affecting the likelihood of the non-monotonic effect of the environmental switching rate. 
\par
\begin{figure}
    \centering
    \includegraphics[scale=0.5]{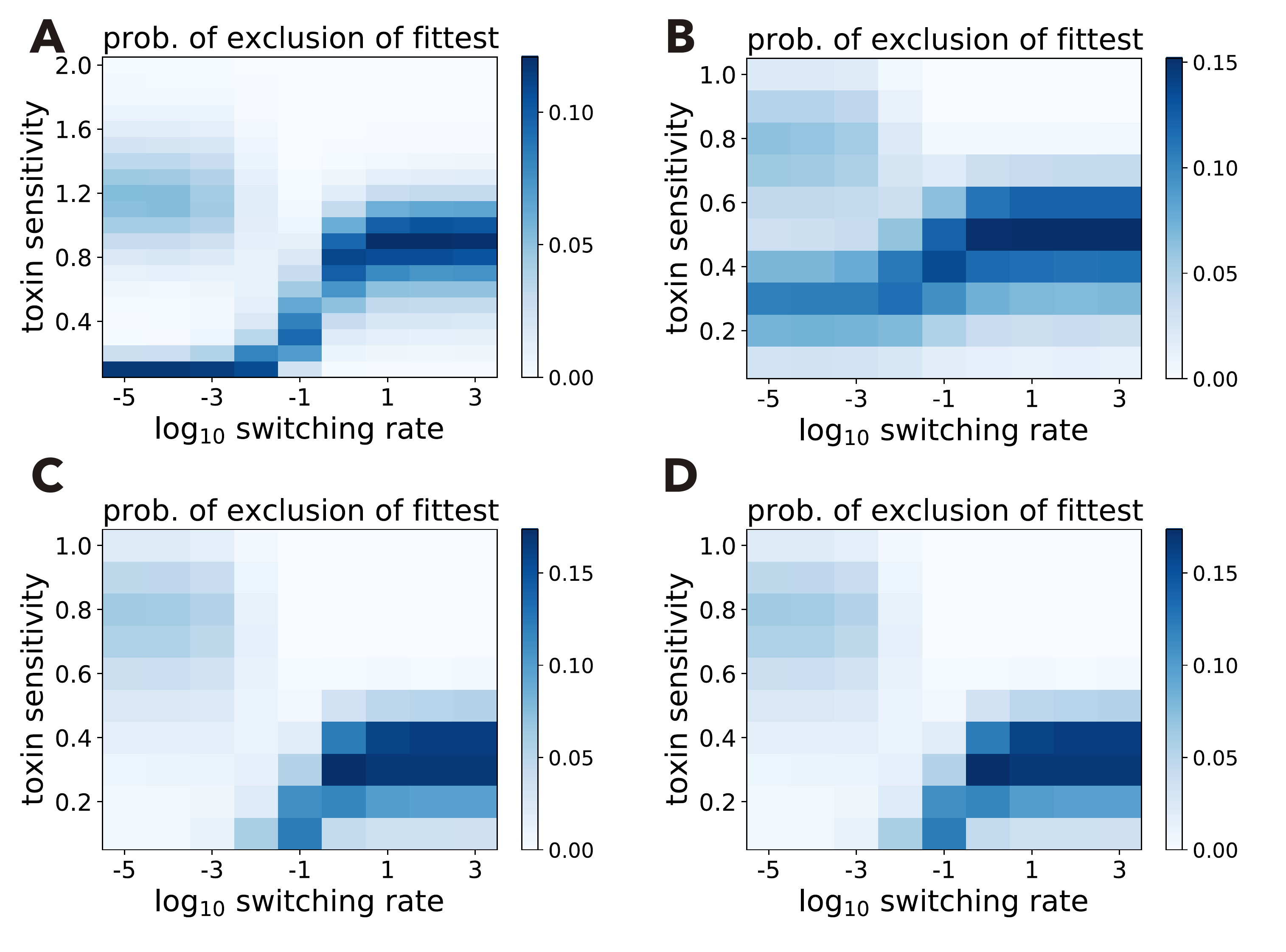}
    \caption{Effects of resource supplies on exclusion of the fittest}
    \label{fig:ChangeResourceSupply}
    \begin{flushleft}
    {\small
    Top: abundant resource supply becomes twice (A) or half (B) of $R_1^+=200$, i.e. $R_1^+=400$ in panel (A) and $R_1^+=100$ in panel (B). Bottom: scarce resource supply becomes twice (C) or half (D) of $R_1^-=50$, i.e. $R_1^-=100$ in panel (C) and $R_1^-=25$ in panel (D).
    }
    \end{flushleft}
\end{figure}
\begin{figure}
    \centering
    \includegraphics[scale=0.5]{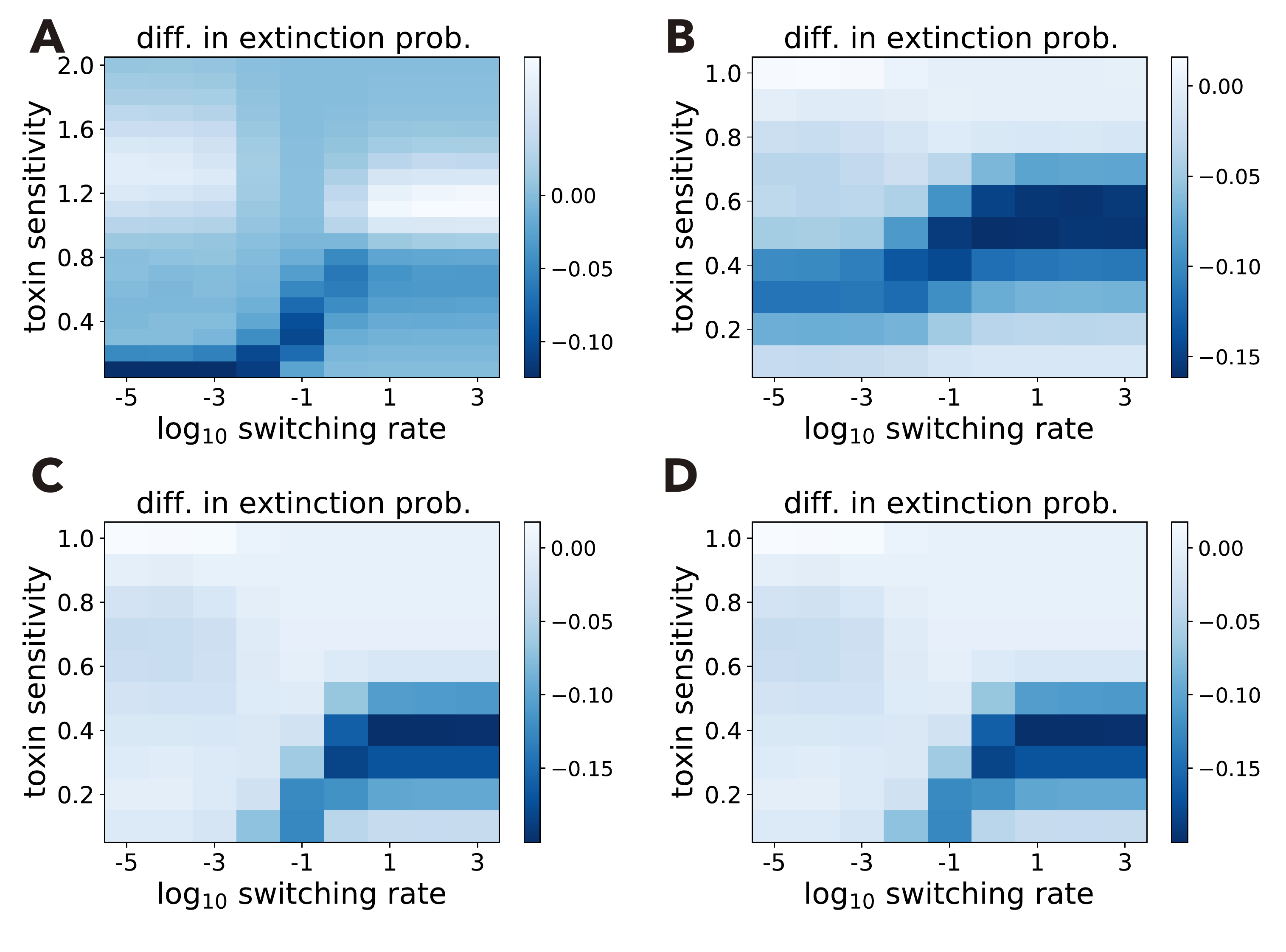}
    \caption{Effects of resource supplies on difference in extinction probability}
    \label{fig:ChangeResourceSupply-interaction}
    \begin{flushleft}
    {\small
    Similar to Fig. \ref{fig:ChangeResourceSupply} but showing species 2's effect on species 1's extinction probability. Top: abundant resource supply becomes twice (A) or half (B) of $R_1^+=200$, i.e. $R_1^+=400$ in panel (A) and $R_1^+=100$ in panel (B). Bottom: the scarce resource supply becomes twice (C) or half (D) of $R_1^-=50$, i.e. $R_1^-=100$ in panel (C) and $R_1^-=25$ in panel (D). We plotted toxin sensitivity from 0.1 to 2.0 in panel A to see non-monotonic positive species interactions (see also Fig.\ref{fig:exception} )
    }
    \end{flushleft}
\end{figure}

By increasing the abundant or decreasing the scarce resource supply (Figs. \ref{fig:ChangeResourceSupply}A and D, respectively), the distance between the critical toxin sensitivities under scarce and mean (or mean and  abundant) resource supplies becomes larger (Table \ref{tab:sum_criticall}).
Conversely, decreasing abundant or increasing scarce resource supply (Figs. \ref{fig:ChangeResourceSupply}B and C, respectively), decreases the distance between the critical toxin sensitivities under scarce and mean (or mean and  abundant) resource supplies (Table \ref{tab:sum_criticall}). Once again, changes in competitive  exclusion probability (\ref{fig:ChangeResourceSupply}) match changes in species 2's effect on species 1 (Fig. \ref{fig:ChangeResourceSupply-interaction}).

\par
\begin{figure}
    \centering
    \includegraphics[scale=0.4]{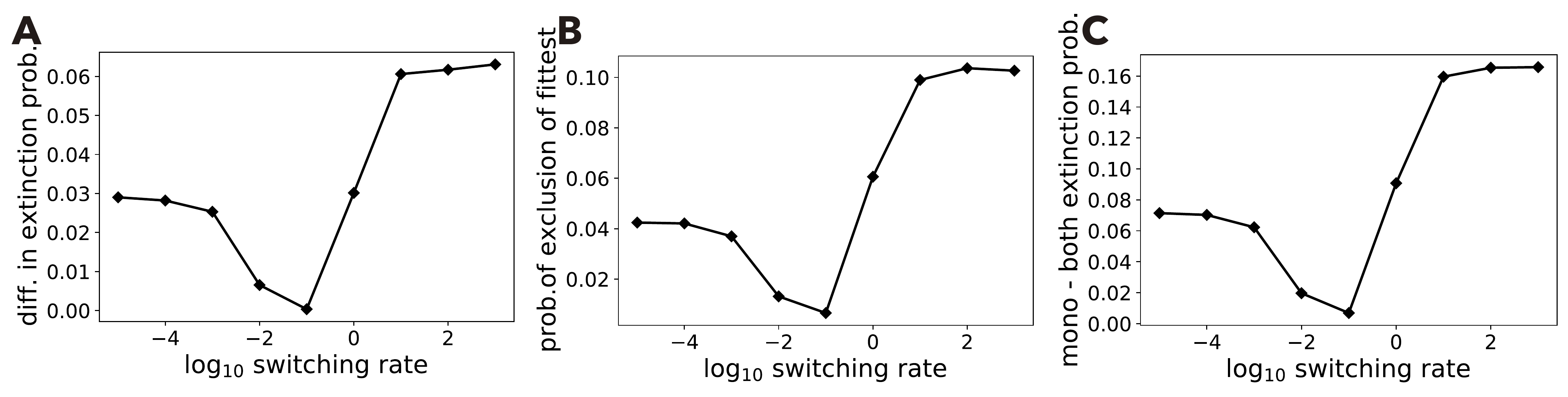}
    \caption{Positive interaction strength varies non-monotonically with switching rate}
    \label{fig:exception}
    \begin{flushleft}
    {\small  A: Difference in species 1's extinction probability with positive sign at $\delta=1.0$ and $R_1^+=400$, showing a non-monotonic effect of the environmental switching rate. B: Probability of exclusion of the fittest. C: Difference between the probability of species 1 going extinct alone and both species going extinct.}
    \end{flushleft}
\end{figure}

Analyzing the probability of exclusion of the fittest instead of the difference in extinction probability is valid only when the difference in extinction probability is negative: if it is positive, the second line in \eqref{eq:decomposedDiff} cannot be ignored. The sign of the difference can, however, become positive at $\delta=1.0$. Indeed, when $R_1^+=400$ and $\delta=1.0$, species 2 has a positive effect on species 1 whose strength varies non-monotonically with the rate of environmental switching (\ref{fig:exception}A). In this case, we analyze both (i) the probability of exclusion of the fittest (the first line in Eq \eqref{eq:decomposedDiff}, Fig. \ref{fig:exception}B) and (ii) the difference in species 1's extinction probability in mono-culture and both species extinction in co-culture (the second line in Eq \eqref{eq:decomposedDiff}, Fig. \ref{fig:exception}C).  The effects of the environmental switching rate on (i) and (ii) are similar, leading to similar non-monotonic effects of species 2 on species 1. In sum, non-monotonic effects of environmental switching rates on species interactions can be observed whether these interactions are positive or negative. Although the main text explains why non-monotonic effects of environmental switching rates on species interactions happen when interactions are negative, it remains unclear why such non-monotonic changes happen when species interactions are positive.

\section{Other forms of environmental fluctuations}\label{sec:variousflucutuation}
 
In this section, we analyze environmental fluctuations other than symmetric switching between two states. Our goal is here to show that our main findings qualitatively still hold and can therefore traced back to the generic interdependence of EFs and DN rather than detail of their coupling.

As in the main text, we assume that the environmental fluctuations change only the resource supply while the toxin supply is constant. First, we investigate asymmetric switching between two resource supply conditions. Second, we increase the number of environmental states and introduce a cyclic  change of the resource supply. 

Under asymmetrically switching or cyclically fluctuating environments, we find similar patterns of how species interactions change over species' toxin sensitivities and a rate of environmental fluctuations.

\subsection{Asymmetric switching}
In the main text and appendices other than this section, for the sake of simplicity, we assume symmetric switching rates between two states by Eq. \eqref{eq:transition-switch} (see \citet{Taitelbaum2020}). Here, we relax this assumption and introduce asymmetric switching rates because perfectly symmetric switching environments are very unlikely in nature; in gut microbiota, for example, duration that their host is starving would be longer than that the host is eating food. We implemented asymmetric switching as follows:
\begin{subequations}
\begin{align}
    \xi=1  &\xrightarrow{\nu_{1}} \xi=-1 \\
    \xi=-1 &\xrightarrow{\nu_{2}} \xi=1.
\end{align}
\end{subequations}
Without loss of generality, we define the two switching rates as follows:
\begin{subequations}
\begin{align}
    \nu_1&=\beta_1 \nu \\
    \nu_2&=\beta_2 \nu
\end{align}
\end{subequations}
where $\nu$ is the basal switching rates. In extreme cases ($\beta_1 \gg \beta_2$), for  asymmetric switching scenarios can correspond to systems with rare disturbances in nature. In this extremely asymmetric case, the
sojourn time in the harsh environment exceeds greatly that in the mild environment, which results in a strong effect of DN. In contrast, DN effects are less important when 
$\beta_2 \gg \beta_1$ and the population experiences more frequently the mild than the harsh environment\citep{Taitelbaum2020}. We recover a symmetric environmental switching when $\beta_1=\beta_2$.
\par
\begin{figure}
    \centering
    \includegraphics[scale=0.5]{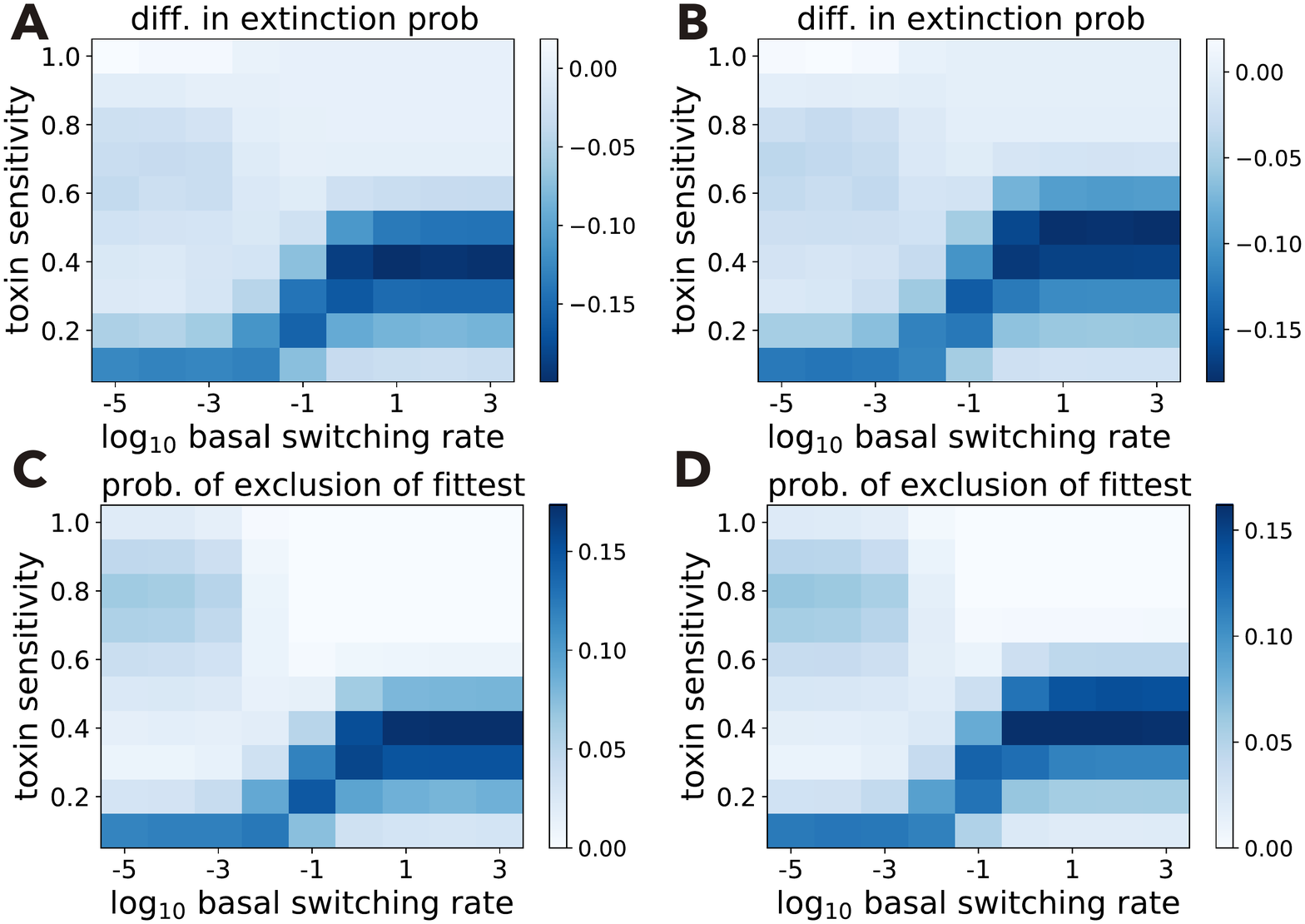}
    \caption{Extinction probabilities under asymmetrically switching environments}
    \label{fig:asymmetric}
    \begin{flushleft}
    {\small  Difference of species 1's extinction probability and the probability of exclusion of the fittest under asymmetric switching rates over the baseline of the switching rate $\nu$ and toxin sensitivity. Here, environmental fluctuations change the amounts of resource supplies. A and B: Difference of species 1's extinction probabilities in mono-culture minus co-culture wit species 2 when (A) the harsh environment continues longer than the mild environment ($\beta_1=1.2$ and $\beta_2=1$), or when (B) the mild environment continues longer than the harsh environment ($\beta_1=1$ and $\beta_2=1.2$), respectively. C and D: Probabilities that species 2 excludes species 1 when the environmental switching rates are identical to panels A and B, respectively. Other parameter values are shown in Table \ref{tab:prameter}.}
    \end{flushleft}
\end{figure}

Fig. \ref{fig:asymmetric} summarizes the two species interactions under resource supply fluctuations with asymmetric environmental switching rates when the initial environmental condition is $\xi(0)=1$ with probability of 0.5 (otherwise $\xi(0)=-1$). In Figs. \ref{fig:asymmetric} A and C, $\nu_1>\nu_2$ and therefore the sojourn time of $\xi=-1$ (an harsh environment) is longer than that of $\xi=1$ (a mild environment). On the other hand, Figs. \ref{fig:asymmetric} B and D shows the cases when the sojourn time the mild environment is longer than that of the harsh environment because $\nu_1<\nu_2$. In both cases, species 1's difference in extinction probabilities and the probability of exclusion of the fittest show monotonically increasing, monotonically decreasing, or non-monotonic changing with a minimum or maximum value at an intermediate switching rate, although they quantitatively differ from Figs. \ref{fig:interaction} A and C.

\subsection{Cyclic changes}

Here, we analyze cases when the number of environmental states in terms of resource supply is greater than two, but remains discrete and finite.
This simply reflects that natural environments do not always fluctuate between two states. In this subsection, an environmental state is given by $\xi=1,2, \ldots, n$ and environments cyclically fluctuate with rate $\nu$ as follows:
\begin{align}
    \xi \xrightarrow{\nu} \left\{\begin{array}{cl}
    \xi+1&\mbox{if }\xi=1, \ldots, n-1\\
    1& \mbox{otherwise}.
    \end{array}\right.
    \label{eq:season}
\end{align}
This is a natural extension of Eq \eqref{eq:transition-switch} by increasing the number of environmental conditions: $n=2$ recovers a symmetrically switching environment between two conditions ($\xi=1 \rightarrow 2 \rightarrow 1 \rightarrow \ldots$ although we use the notation $\xi=\pm 1 \rightarrow \mp 1 \rightarrow \pm1 \rightarrow \ldots$ in the main text).
\par
\begin{figure}
    \centering
    \includegraphics[scale=0.5]{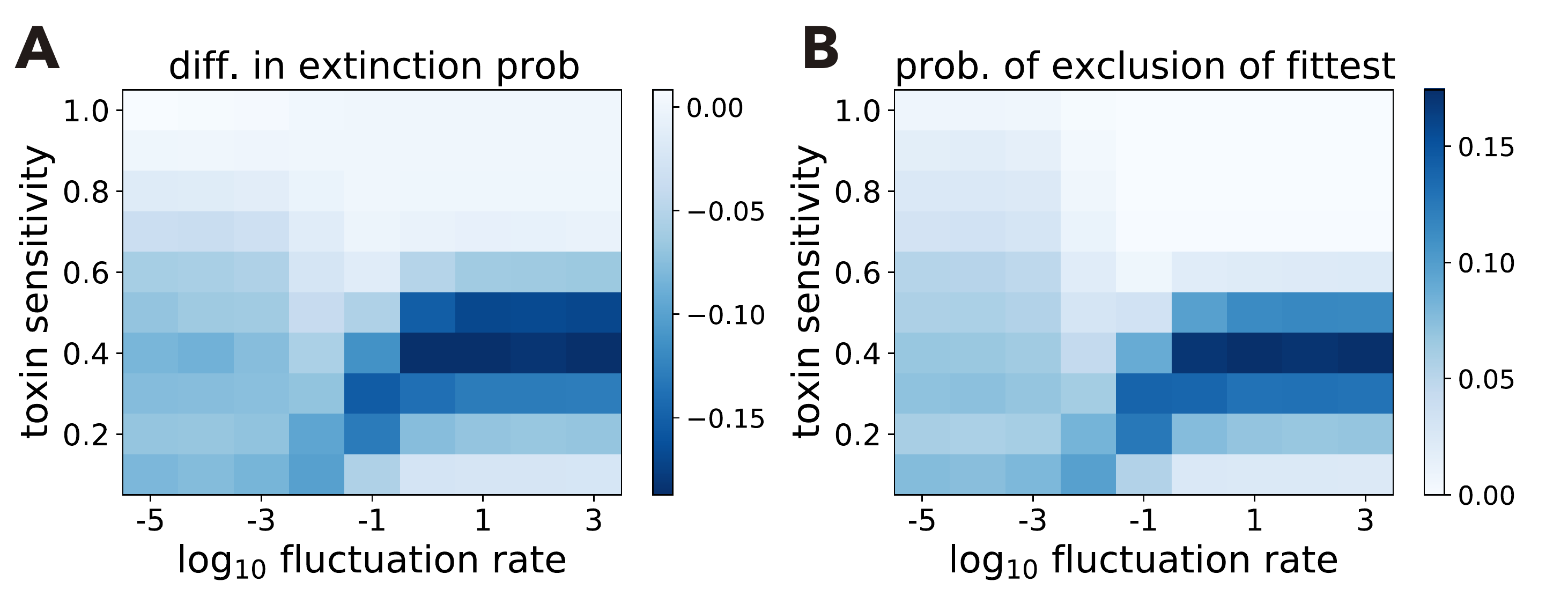}
    \caption{Extinction probabilities under cyclic environmental changes}
    \label{fig:seasonal}
    \begin{flushleft}
   Difference in species 1's extinction probability and the probability of exclusion of the fittest under under cyclically fluctuating resource supplies among four states: $R_1\left(\xi=1\right)=200$, $R_1\left(\xi=2\right)=150$, $R_1\left(\xi=3\right)=100$, and $R_1\left(\xi=4\right)=50$. A: Difference of species 1's extinction probabilities in mono-culture minus co-culture with species 2. B: Probability that species 1 goes extinct but species 2 survives. Other parameter values are shown in Table \ref{tab:prameter}.
    \end{flushleft}
\end{figure}

Fig. \ref{fig:seasonal} shows how species interactions between two species change when $n=4$ and the resource supply fluctuates such that $R_1\left(\xi=1\right)>R_1\left(\xi=2\right)>R_1\left(\xi=3\right)>R_1\left(\xi=4\right)$. In this analysis, an initial environmental condition $\xi(0)$ is one of four conditions (probability of 0.25 for each). As in Figs. \ref{fig:interaction} A and C, the rate of cyclical environmental change affects species 1's extinction probability and the probability that species 2 excludes species 1. We frequently observe that the probability of exclusion of the fittest non-monotonically changes (toxin sensitivity: $0.1$ -- $0.6$).
\par
 In this work, the fluctuating environment has been modeled
as randomly switching between a finite number of environmental states $\xi$.
This choice is particularly convenient as it allows us to deal with
bounded noise, and hence $R_i$ and $T_j$ to always remain positive, and are straightforward to simulate using the standard Gillespie algorithm.
The case of 
environmental noise varying continuously in time
is also of great interest, as it allows  
$R_i$ and $T_j$ to  take any values in a domain.
For instance, the environmental noise can be an Ornstein-Uhlenbeck process ($\xi_{OU}$), e.g. by letting $R_i=R_i(\xi_{OU})=\bar{R_i}(1+k\xi_{OU}) $, where $\bar{R_i}$ is constant, $\xi$ varies in time,  and $k>0$,
see, e.g., \citet{Assaf2013}. 
This poses a number of challenges since, $\xi_{OU}$ being unbounded, $R_i$ can  take negative (unphysical) values. Furthermore, there are no general methods to simulate 
exactly birth-death processes subject to continuous external noise, see, e.g., \citet{Berrios-Caro2020}.
Here, while the assumption of discrete environmental noise is a simplification of many real situations,
we think that our main findings are generic and shall hold also under continuous environmental noise. In fact, since our results stem chiefly  from the coupling of EFs and DN, a feature shared by discrete and continuous noise,
they  are expected to qualitatively hold also in the case of continuous external noise.  


\section{Diversity in communities of increasing species number}\label{sec:app-variouscom}
\begin{figure}
\rotatebox{90}{
    \begin{minipage}{1.0\textwidth}
    \centering
    \includegraphics[scale=0.45]{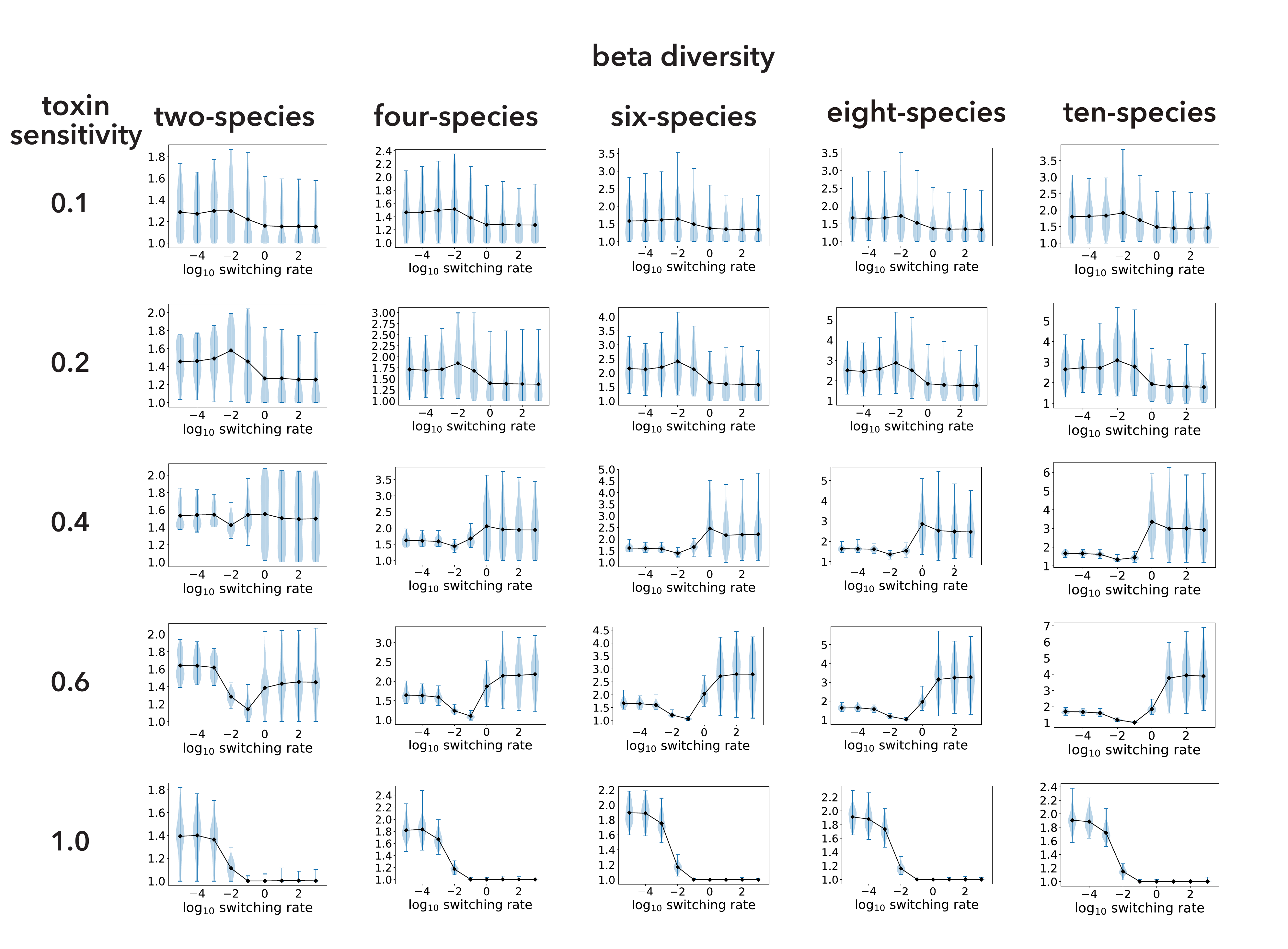}
    \caption{Beta diversity with increasing initial number of species in a community}
    \label{fig:beta-diversity-supp}
\end{minipage}
}
\begin{flushleft}
    {\small 
     Beta diversities with increasing initial numbers of species $N$ and mean toxin sensitivities $\bar{\delta}$. The black lines show the means and blue areas represent the probability distributions calculated by 10'000 simulations (100 beta diversity measurements using different parameter sets, each from 100 replicate runs).}
    \end{flushleft}
\end{figure}

In the main text, we show the distributions of beta diversity and species richness in communities when the initial number of species $N$ is two or ten. This section shows the results of intermediates values of $N=4, 6, 8$. The number of initial species does not change how beta diversity changes over the environmental switching rate, except when the mean toxin sensitivity $\bar{\delta}=0.4$ (Fig. \ref{fig:beta-diversity-supp}). These results indicate that beta diversity and the probability of competitive exclusion change similarly over the switching rate when the initial number of species is larger than two.
\par
\begin{figure}
\rotatebox{90}{
\begin{minipage}{1.0\textwidth}
    \centering
    \includegraphics[scale=0.45]{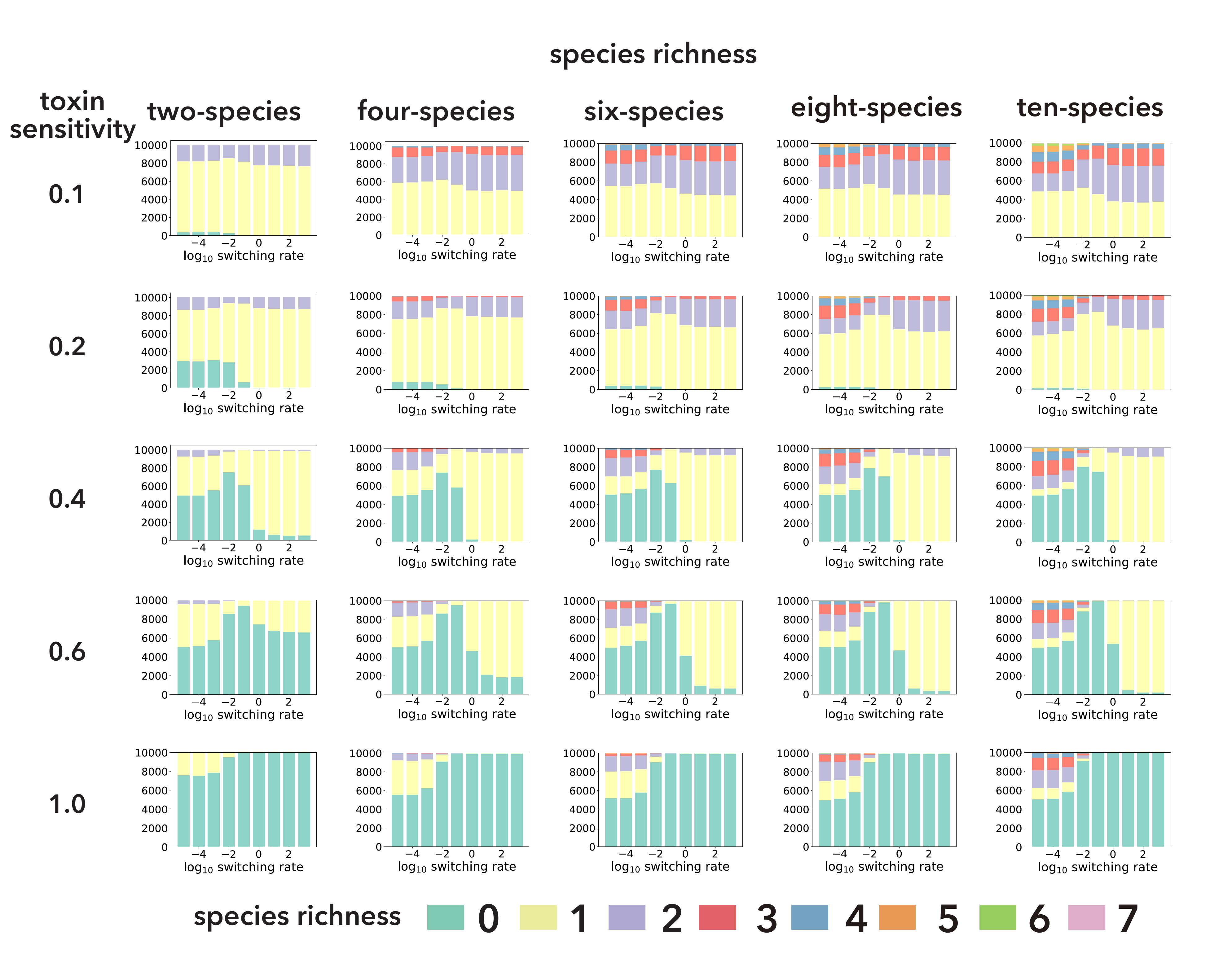}
    \caption{Species richness with increasing initial number of species in a community}
    \label{fig:richness-supp}
\end{minipage}
}
\begin{flushleft}
    {\small 
     Species richness with increasing initial numbers of species $N$ and mean toxin sensitivities $\bar{\delta}$. Each bar plot represents results of 10'000 simulations.}
    \end{flushleft}
\end{figure}

The initial number of species $N$ in a community affects the maximum values of species richness (Fig. \ref{fig:richness-supp}). However, how the switching rate and the mean toxin sensitivity affect the distribution of species richness was consistent for different values of $N$. In particular, increasing the mean toxin sensitivity decreases species richness. The effects of the environmental switching rate also depend on the mean toxin sensitivity: at mean toxin sensitivity $1.0$, species in all cases  are more likely to go extinct as the switching rate increases, while the likelihood of all species going extinct consistently shows a humped shape at toxin sensitivity $0.4$ or $0.6$.

\section{Quantified the similarity between the exclusion of fittest and beta diversity}\label{sec:qunatify_similarity}
In this section, we first quantify the similarity between the exclusion of the fittest and beta diversity in two-species communities. Then, we investigate how many species pairs we should analyze to predict the patterns of beta diversity in ten-species communities.
\par
\begin{figure}
    \centering
    \includegraphics[scale=0.35]{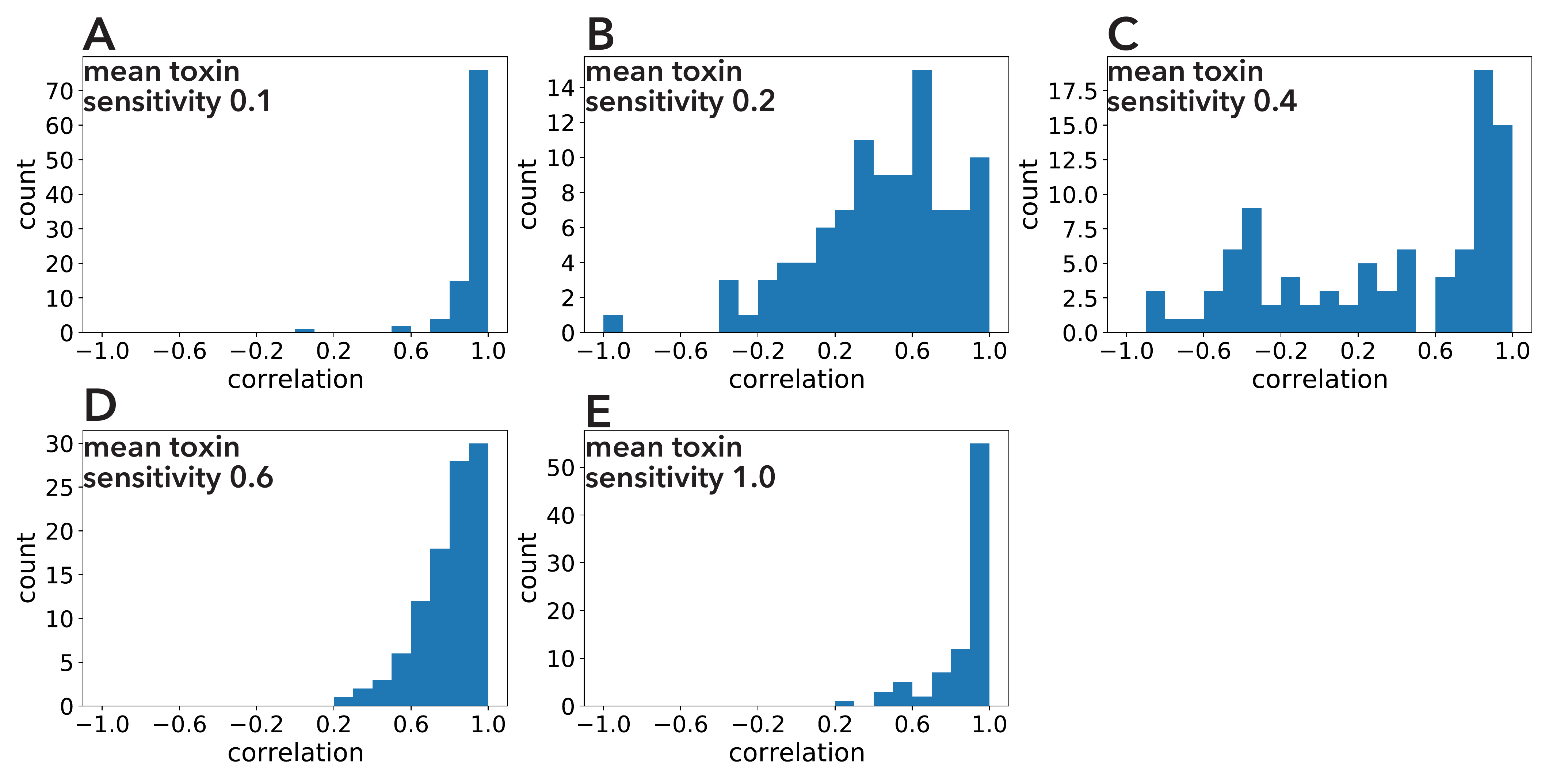}
    \caption{Correlations between exclusion of the fittest and beta diversity in two-species communities}
    \label{fig:corr_BetaExcl2}
\begin{flushleft}
    {\small 
     The Pearson correlation coefficients between probability of exclusion of the fittest (first column of Fig. \ref{fig:diversity}) and beta diversity  (second column of Fig. \ref{fig:diversity}) in two-species communities are shown. Each panel differs in mean toxin sensitivity (A: 0.1, B: 0.2, C: 0.4, D: 0.6, and E: 1.0) and we analyzed 100 two-species communities in each case.}
    \end{flushleft}
\end{figure}

In the main text, we show that the probability of exclusion of the fittest and beta diversity exhibit similar patterns (see columns A and B in Fig. \ref{fig:diversity}). We quantified the similarities with Pearson's correlation coefficients. Fig. \ref{fig:corr_BetaExcl2} shows the distributions of the correlation coefficients of 100 two-species communities at each mean toxin sensitivity (see also \ref{sec:detail-simulation2}). Except for the case that mean toxin sensitivity is 0.4 (where the probability of exclusion of the fittest and beta diversity do not match), the correlation coefficients are large positive ($>0.6$). Therefore, a large correlation coefficients indicates the similarity between the probability of the exclusion of the fittest and beta diversity.
\par
\begin{figure}
\rotatebox{90}{
    \begin{minipage}{1.0\textwidth}
    \centering
    \includegraphics[scale=0.45]{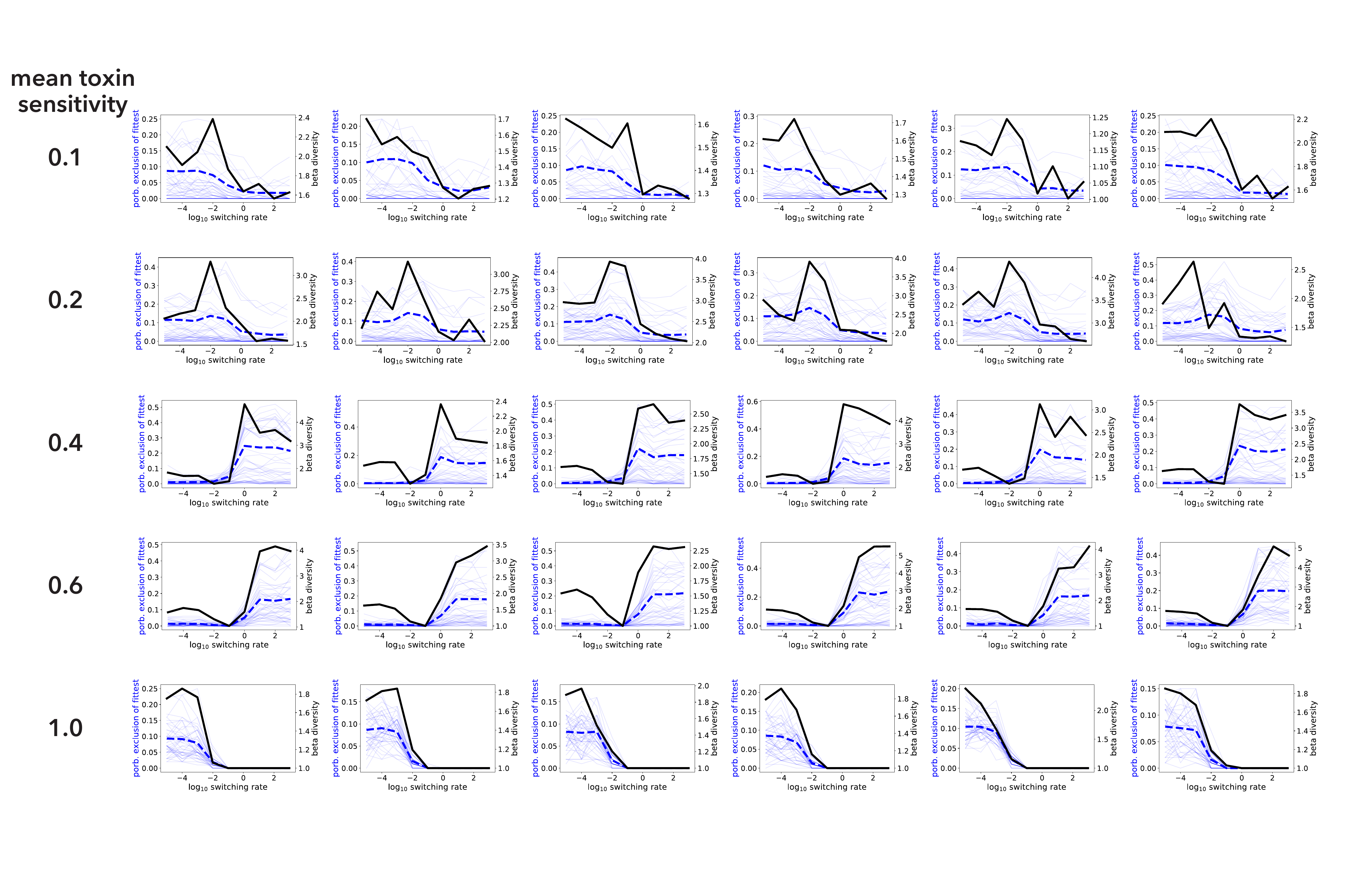}
    \caption{Beta diversity and exclusion of the fittest in ten-species communities}
    \label{fig:subsample}
\end{minipage}
}
\begin{flushleft}
    {\small 
     Each panel represents the beta diversity of a ten-species community (black lines), probabilities of exclusion of the fittest in 45 species pairs (solid blue lines) and the mean probability of exclusion of the fittest over the 45 pairs (dashed blue lines). Each row corresponds to mean toxin sensitivity $\bar{\delta}$. One can see six examples of ten-species communities at each mean toxin sensitivity.}
    \end{flushleft}
\end{figure}
\begin{figure}
    \rotatebox{90}{
    \begin{minipage}{1.0\textwidth}
    \centering
    \includegraphics[scale=0.45]{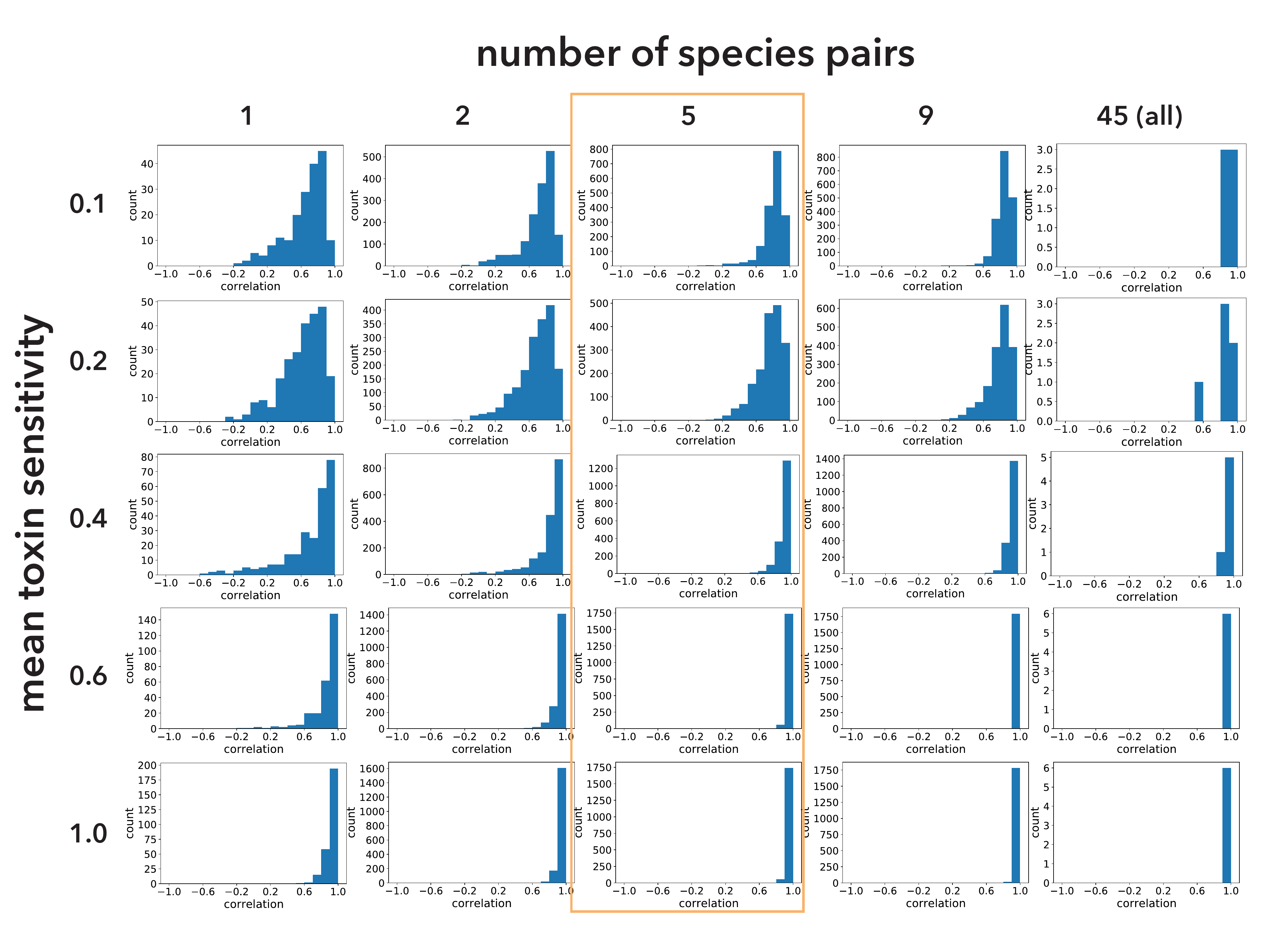}
    \caption{Correlations between exclusion of the fittest and beta diversity in ten-species communities}
    \label{fig:corr_BetaExcl10}
    \end{minipage}
    }
\begin{flushleft}
    {\small 
     The Pearson correlation coefficients between mean probability of exclusion of the fittest and beta diversity (Fig. \ref{fig:subsample}) in ten-species communities are shown. Each row represents the different value of mean toxin sensitivity $\bar{\delta}$ while each column represents the different number of sampled species pairs $m$.}
    \end{flushleft}
\end{figure}

We continued the analysis in ten-species communities. In these cases, we have 45 pairs of species in each community and we calculated the probabilities of the exclusion of the fittest by running 100 replicates in each species pair of thirty  ten-species communities (six communities at five mean toxin sensitivities). Fig. \ref{fig:subsample} shows that the exclusion of the fittest in some pairs match the patterns of beta diversity of whole communities but other pairs do not. Then, we investigated the number of species pairs $m$ that is necessary to predict a pattern of beta diversity over the environmental switching rate (i.e., a large correlation between probability of exclusion of the fittest and beta diversity). When $m \ge 2$, we calculated Pearson's correlation coefficient between beta diversity and mean probability of exclusion of the fittest within $m$ pairs. However, we have many possible choices of $m$ pairs when $1<m<45$. In these cases, we randomly chose 300 sets of $m$ pairs in each of community and thus we obtained 1800 correlation coefficients  at each mean toxin sensitivities. When $m=1$ or $45$, we analyzed all possible choices of $m$ species pairs and we obtained 45 or 1 correlation coefficient(s) in each community, respectively. Fig. \ref{fig:corr_BetaExcl10} indicates that the correlation coefficients can be large even when $m=1$ or $2$, but larger $m$ increases correlation coefficients. We suggest that $m=5$ is the best because the $m=5$ and $m=9$ show little difference in the distributions of the correlation coefficients and we are very likely to obtain a large correlation coefficient with $m=5$. In conclusion, we do not have to analyze the exclusion of the fittest for all species pairs to
predict how the environmental switching rate affect beta diversity.  
\newpage

\begin{figure}
    \centering
    \includegraphics[scale=0.6]{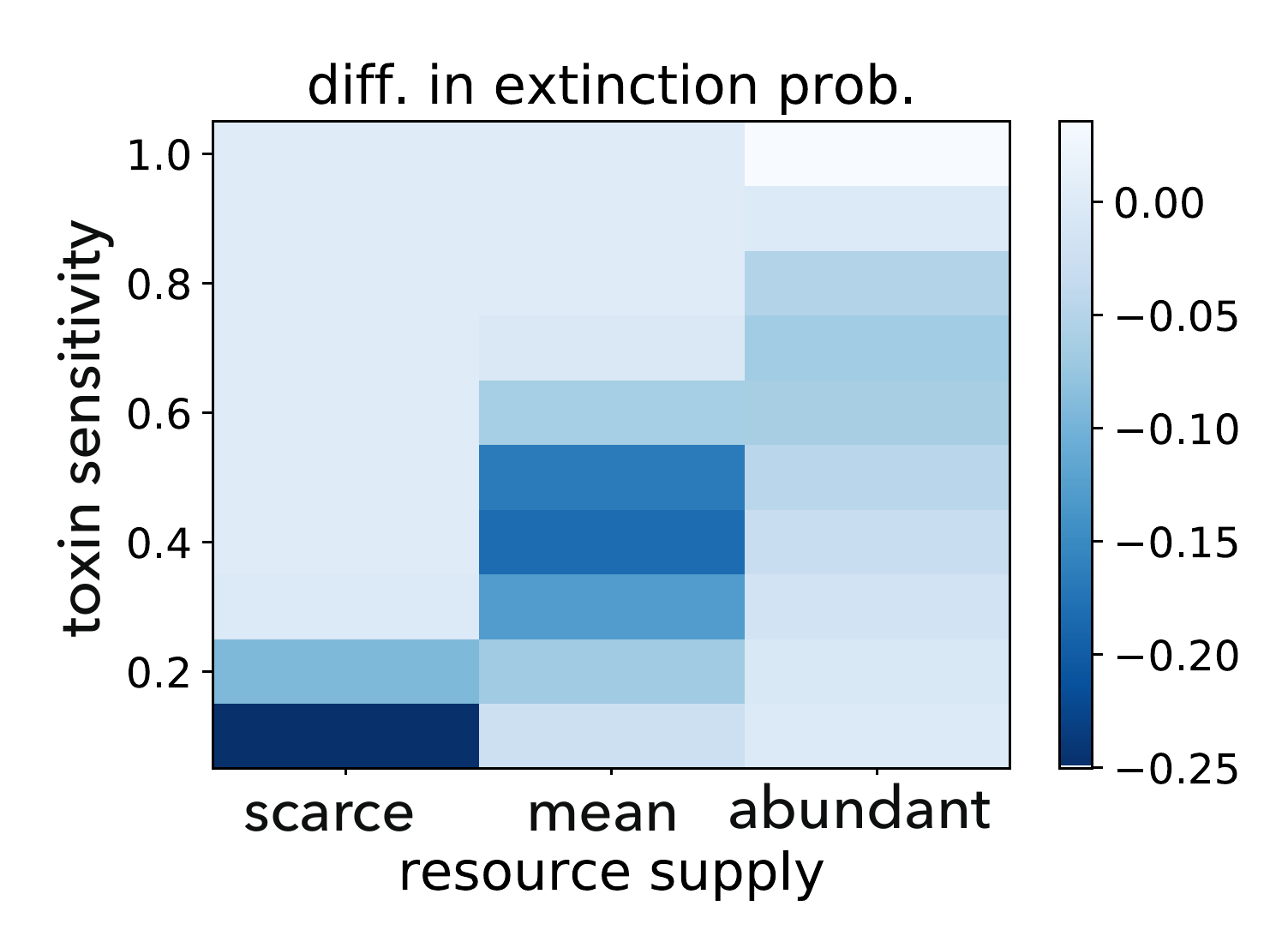}
    \caption{Species 2's effect on species 1 in the absence of environmental switching}
    \label{fig:diff_constant}
    \begin{flushleft}
    \small{Species 2's effect on species 1 when the resource supply is fixed to be scarce ($R_1^-$), mean ($\left<R_1\right>$), or abundant ($R_1^+$). The toxin sensitivities that minimize species 2's effect on species 1 correspond to the peak sensitivities in Fig. \ref{fig:CompExcl}.}
    \end{flushleft}
\end{figure}

\begin{figure}
    \centering
    \includegraphics[scale=0.7]{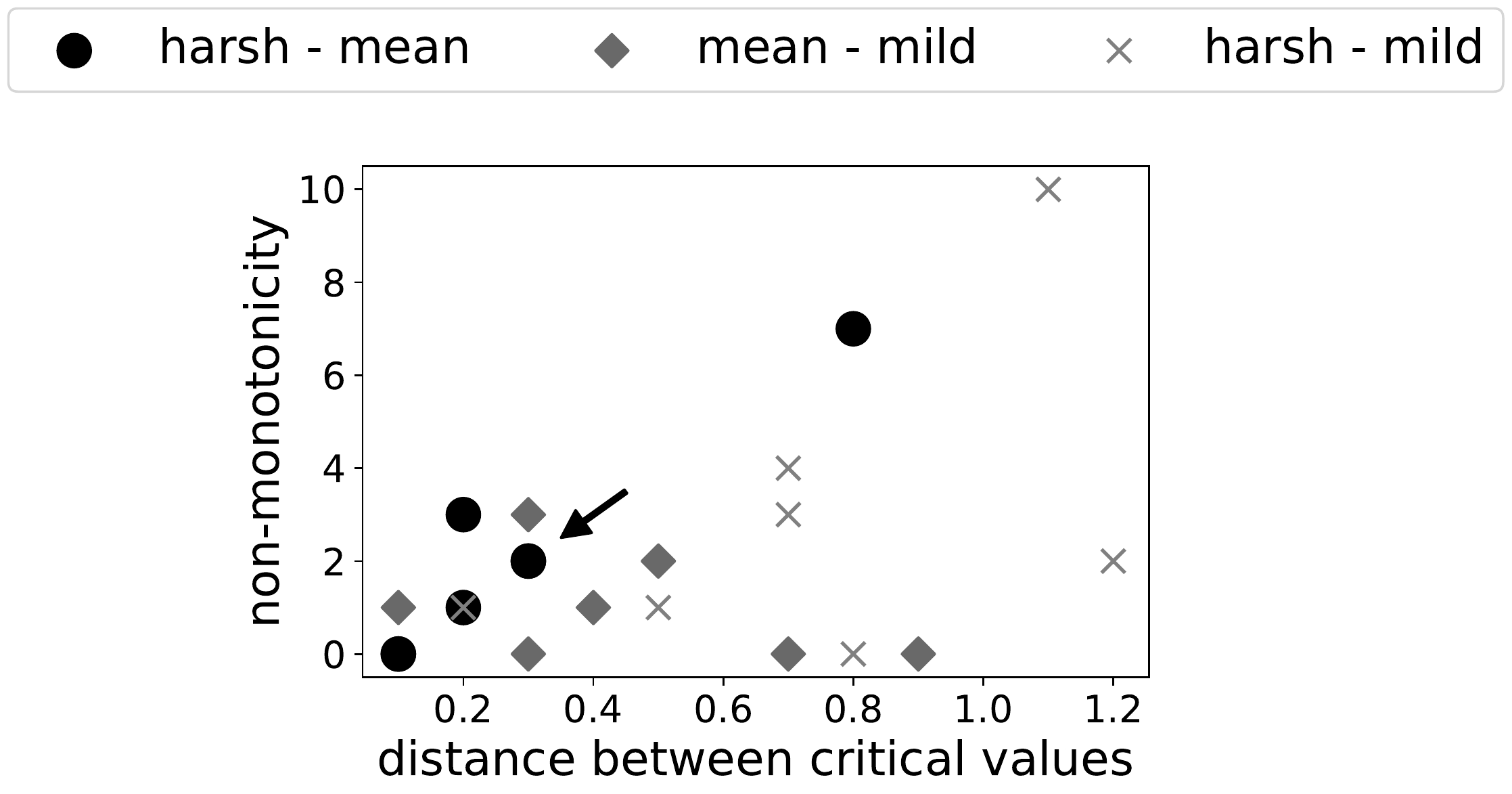}
    \caption{Non-monotonicity and critical toxin sensitivities}
    \label{fig:monotonic_toxin_crti}
    \begin{flushleft}
    {\small
    The number of times we observe non-monotonic changes of species 1's difference in extinction probability across the explored parameter range varies with the distance between the two critical toxin sensitivities, depending on where distances are measured (between harsh and mean environments: dots, between mean and mild: environment diamonds, and between harsh and mild environments: crosses). These three distances were measured in each of the following seven scenarios: three different scenarios of environmental switching (Table \ref{tab:resoure-toxin-switch} and \ref{sec:alternative}) and four environmental switching scenario 1s with changing amounts of resource supplies (\ref{sec:ChangeResourceSupply}). The correlation is only significantly positive for the distance between scarce resource or abundant toxin supplies (i.e., harsh environments) and mean resource/toxin supplies (Spearman's $\rho=0.77$, P-value: $0.043$). The dot indicated by the arrow
    corresponds to the scenario analyzed in the main text.
    
    }
    
    \end{flushleft}
\end{figure}

\begin{figure}
    \centering
    \includegraphics[scale=0.6]{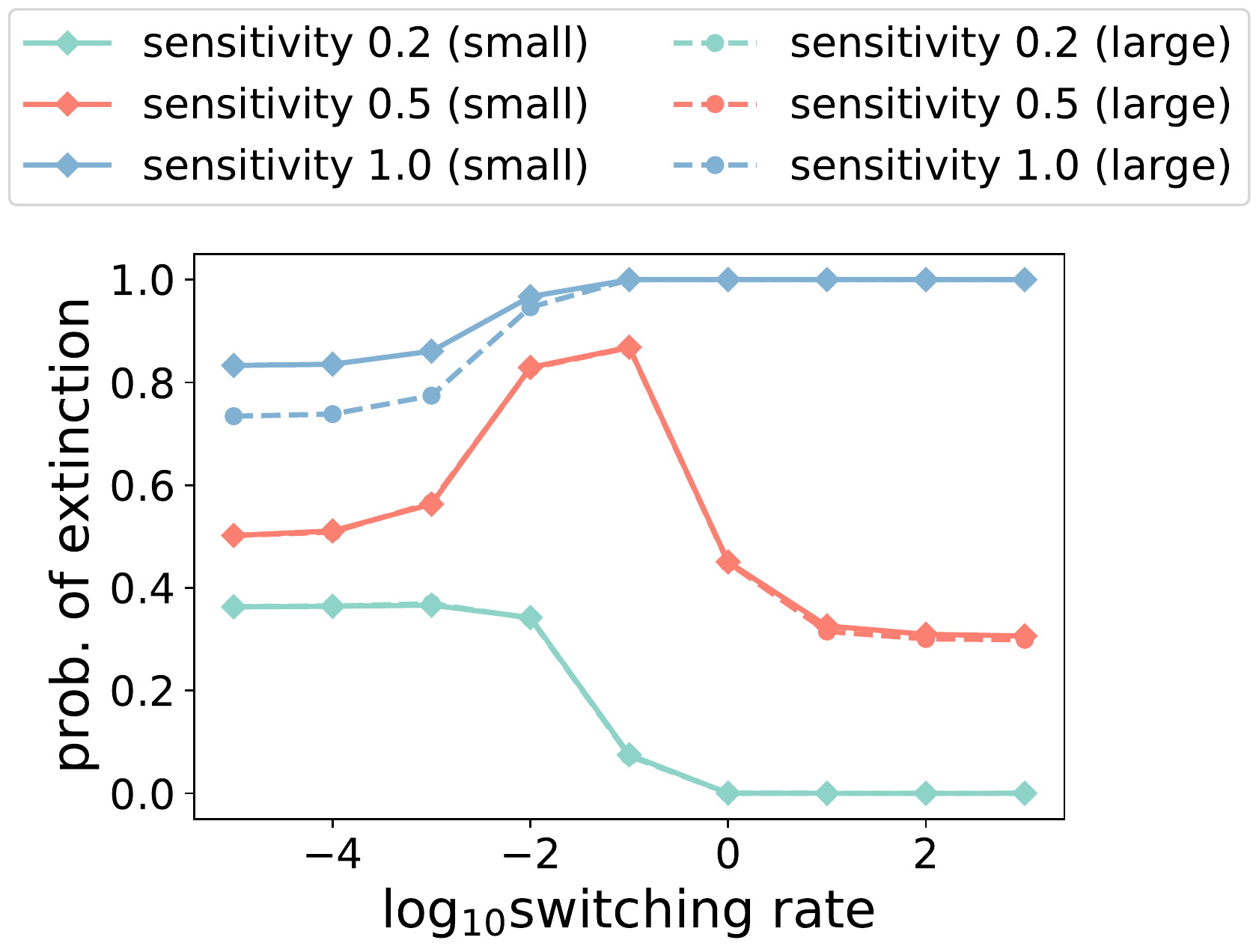}
    \caption{Initial Population size's effect on extinction probabilities}
    \label{fig:init_size}
    \begin{flushleft}
    {\small Extinction probability of species 1 in mono-culture when the initial species abundance is default (small: $s_1(0)=10$) or larger (large: $s_1(0)=20$). When the switching rate is small and the toxin sensitivity is $1.0$, the extinction probability is lower with the larger initial species abundance. In the rest cases, the extinction probabilities are not affected by the initial species abundance. The parameter values are as shown by Table \ref{tab:prameter}.}
    \end{flushleft}
\end{figure}

\begin{figure}
    \centering
    \includegraphics[scale=0.4]{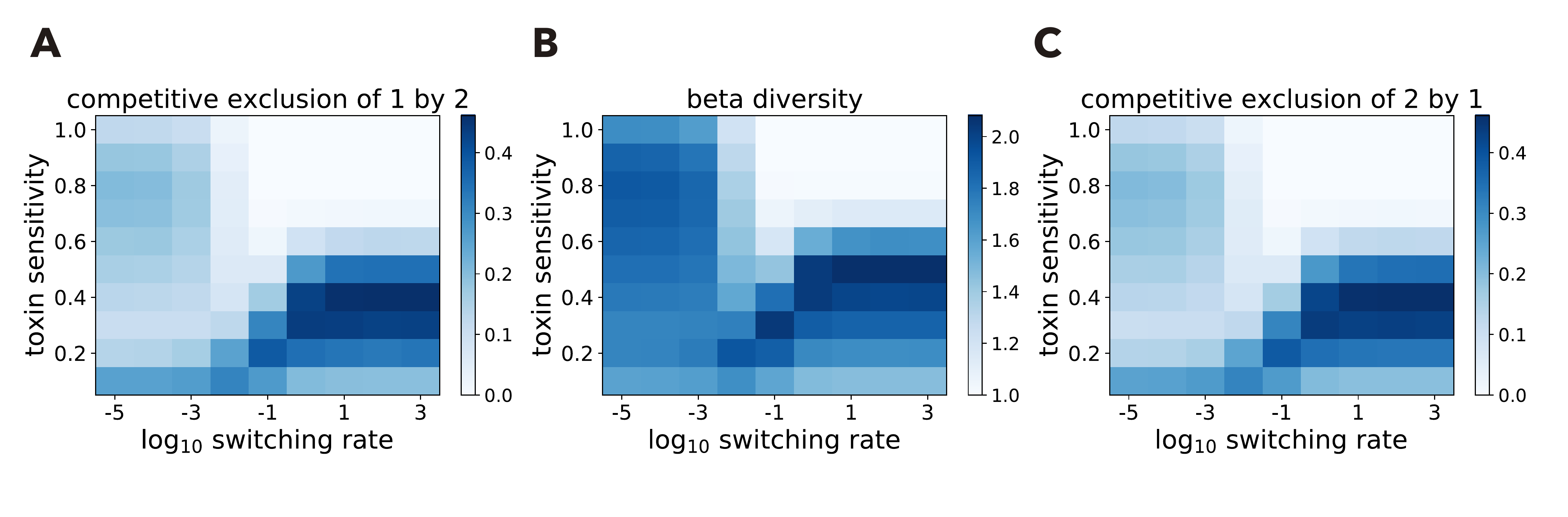}
    \caption{Exclusion probabilities and beta diversity in neutral cases}
    \label{fig:neutral}
    \begin{flushleft}
    {\small Two species dynamics under neutral cases (i.e., two species differ only in their labels). In this case, the probabilities that species 2 outcompete species 1 (A) are identical to those that species 1 outcompete species 2 (C). Without loss of generality,  we can call exclusion of either species as exclusion of the fittest in the neutral scenarios. As in Fig. \ref{fig:diversity}, there are similarities between how the exclusion of the fittest (A or C) and  beta diversity (B) changes over the switching rate at each toxin sensitivity.  The parameter values are as shown by Table \ref{tab:prameter}.}
    \end{flushleft}
\end{figure}

\begin{figure}
    \centering
    \includegraphics[scale=0.35]{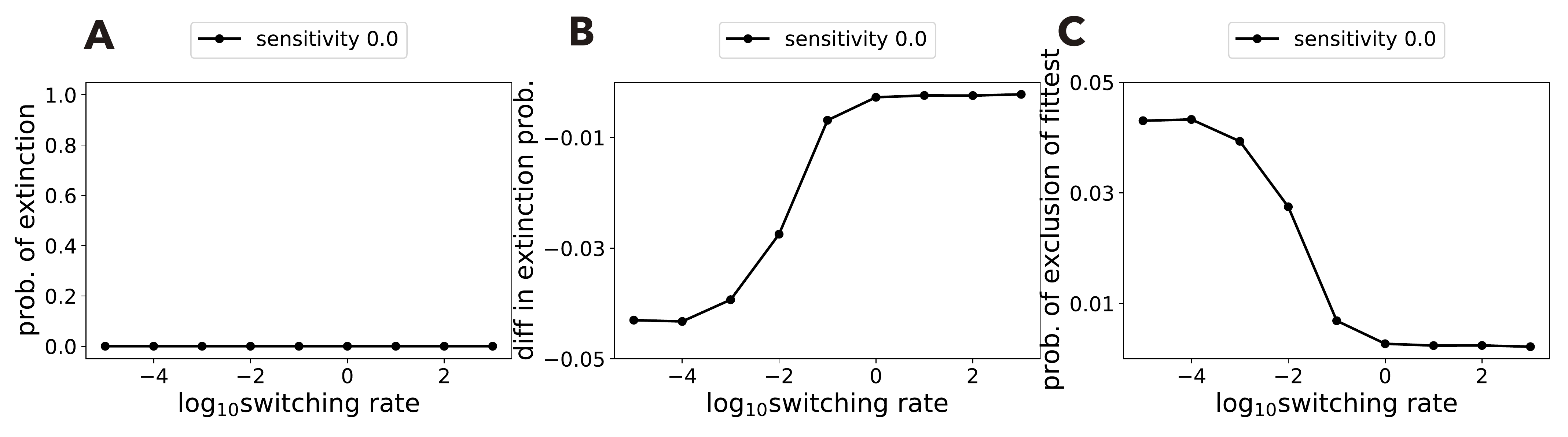}
    \caption{Extinction probabilities with zero toxin sensitivity}
    \label{fig:notoxin}
    \begin{flushleft}
    {\small When species's toxin sensitivities are zero, the dynamics are equivalent with the case without toxins because species die only due to dilution. A: Species 1 does not go extinct in mono-culture. B: Difference of species 1's extinction probabilities in mono-culture minus co-culture. C: Probability that species 2 outcompetes species 1 in co-culture. Panels B and C show qualitatively similar results with the case of toxin sensitivity 0.1, see Figs. \ref{fig:interaction}A and C because species 2 can outcompete species 1 under the harsh environment with low switching rates, due to DN.}
    \end{flushleft}
\end{figure}

\clearpage
\begin{table}[h]
    \centering
    \caption{List of fixed parameters in the interaction analysis}
    \begin{tabular}{ccl}\hline
         Symbol&Value&Description  \\ \hline
         $\alpha$&$0.1$& dilution rate of the chemostat \\ 
         $R_1^\pm$&$R_1^+=200, R_1^-=50$ & abundant or scarce  resource supply concentration\\ 
         $T_1^\pm$&$T_1^+=200, T_1^-=50$ & abundant or scarce toxin supply concentration\\ 
         $Y^r_{1k}$&$Y^r_{1k}=1$ for $k=1,2$&species $k$'s biomass yields of resource\\ 
         $Y^t_{1k}$&$Y^t_{1k}=1$ for $k=1,2$&species $k$'s biomass yields of toxin \\ 
         $\mu_{11}$ & $1.0$ & maximum growth rate of species 1 on resource $1$\\
         $\mu_{12}$ & $0.91$ & maximum growth rate of species 2 on resource $1$ \\
         $\delta_{1k}$ &$ \left[0.1,\dots,1.0\right]$  and $\delta=\delta_{11}=\delta_{12}$ &sensitivity of species  $k$ to toxin $1$.\\
         $K^r_{1k}$ &$100$ & amount of resource $1$ that gives half-max growth rate of species $k$ \\ 
         $K^t_{1k}$ &$100$ & amount of toxin $1$ that gives half-max death rate of species $k$ \\ \hline
    \end{tabular}
    \label{tab:prameter}
\end{table}

\begin{table}[htb]
    \centering\caption{Summary of critical toxin sensitivities and number of times non-monotonicity is observed}
    \begin{tabular}{cc|ccccccc}
    Switching scenario& &1&2&3&\multicolumn{4}{c}{1} \\ \hline
    Amounts of supplies& &\multicolumn{3}{c}{base line $^*$}&
    $\uparrow R_1^+$&$\downarrow R_1^+$ &  $\uparrow R_1^-$ & $\downarrow R_1^-$ \\ \hline
    Critical toxin& harsh& 0.1 &0.3 & 0.1 & 0.1 & 0.1& 0.3 & 0.1 \\
    sensitivities& mean & 0.4 & 0.4 & 0.4 &0.9 & 0.2 & 0.5 & 0.3 \\
    & mild &0.8&1.1 &1.3&1.2 & 0.3 & 0.8 & 0.8 \\\hline
    number of  & mean and harsh & 2 &0 &2&7 &0&1&3 \\
    non-monotonic changes& mean and mild & 1 & 0 & 0 & 3 & 0 & 0 &2\\
    observed between&mild and harsh& 3&0& 2& 10 & 0 & 1 &4 $^\dagger$
    \end{tabular}
    \\
    \begin{flushleft}
    \small{
    $^*$ For exact parameter values, see Table \ref{tab:prameter}. \\
    $^\dagger$ At the critical toxin sensitivity corresponding to the mean environment ($\delta=0.3$), the species interaction non-monotonically changes over the switching rate: the frequency of non-monotonicity is $3+2-1=4$.}
    \end{flushleft}
    
    \label{tab:sum_criticall}
\end{table}

\end{document}